\documentclass[journal]{IEEEtran}
\usepackage{cite}
\usepackage{amsmath,amssymb,amsfonts}
\usepackage{bbm}
\usepackage{tikz}
\usepackage{graphicx}
\usepackage{textcomp}
\usepackage{circledsteps}
\usepackage{xcolor}
\usepackage{soul,color}
\usepackage{kotex}
\usepackage{algorithmicx,algpseudocode}
\usepackage{subfigure}
\usepackage{booktabs}
\usepackage[linesnumbered,ruled]{algorithm2e}
\usepackage{tabularx}
\usepackage{multicol}
\usepackage{multirow}
\definecolor{color1}{RGB}{228,26,28}
\definecolor{color2}{RGB}{55,126,184}
\definecolor{color3}{RGB}{77,175,74}
\definecolor{color4}{RGB}{152,78,163}
\definecolor{color5}{RGB}{255,127,0}
\definecolor{color6}{RGB}{241,241,241}
\definecolor{color7}{RGB}{156,156,156}
\definecolor{color8}{RGB}{96,96,96}
\definecolor{color0}{RGB}{162, 20, 47}
\usetikzlibrary{patterns,arrows}
\usepackage[normalem]{ulem}
\usepackage{lettrine}

\newcommand{\argmaxD}{\arg\!\max} 

\newcommand*\circled[1]{\Circled[inner color=white, fill color= color0, outer color=color0]{\footnotesize{#1}}} 
 
\newcommand{\BfPara}[1]{{\noindent\bf#1.}\xspace}

\newcommand{\tBOSS}[1]{{\color{black}{#1}}}

\begin{document}

\title{Multi-Agent Reinforcement Learning for Cooperative Air Transportation Services in City-Wide Autonomous Urban Air Mobility}

\author{
    Chanyoung Park, 
    Gyu Seon Kim,
    Soohyun Park, 
    Soyi Jung,~\IEEEmembership{Member, IEEE}, and 
    \\
    Joongheon Kim,~\IEEEmembership{Senior Member, IEEE}\thanks{Preliminary version of this paper was accepted to \textit{IEEE International Conference on Communications (ICC)}, Rome, Italy, May/June 2023~\cite{icc2023park}.}
    \thanks{This work was supported in part by the National Research Foundation of Korea (NRF-Korea) under Grant 2021R1A4A1030775 and in part by the Institute of Information and Communications Technology Planning and Evaluation (IITP) Grant through the Korea Government [Ministry of Science and Information and Communications Technology (MSIT)], Intelligent 6G Wireless Access System, under Grant 2021-0-00467. \textit{(Corresponding authors: Soohyun Park, Soyi Jung)}}
    \thanks{Chanyoung Park, Gyu Seon Kim, Soohyun Park, and Joongheon Kim are with the Department of Electrical and Computer Engineering, Korea University, Seoul 02841, Republic of Korea (e-mails: \{cosdeneb,kingdom0545,soohyun828, joongheon\}@korea.ac.kr).}
    \thanks{Soyi Jung is with the Department of Electrical and Computer Engineering, Ajou University, Suwon 16499, Republic of Korea (e-mail: sjung@ajou.ac.kr).}
}
\maketitle

\begin{abstract}
The development of urban-air-mobility (UAM) is rapidly progressing with spurs, and the demand for efficient transportation management systems is a rising need due to the multifaceted environmental uncertainties. Thus, this paper proposes a novel air transportation service management algorithm based on multi-agent deep reinforcement learning (MADRL) to address the challenges of multi-UAM cooperation. Specifically, the proposed algorithm in this paper is based on communication network (CommNet) method utilizing centralized training and distributed execution (CTDE) in multiple UAMs for providing efficient air transportation services to passengers collaboratively. 
Furthermore, this paper adopts actual vertiport maps and UAM specifications for constructing realistic air transportation networks. 
By evaluating the performance of the proposed algorithm in data-intensive simulations, the results show that the proposed algorithm outperforms existing approaches in terms of air transportation service quality. Furthermore, there are no inferior UAMs by utilizing parameter sharing in CommNet and a \textit{centralized critic} network in CTDE. 
Therefore, it can be confirmed that the research results in this paper can provide a promising solution for autonomous air transportation management systems in city-wide urban areas.
\end{abstract}

\IEEEpeerreviewmaketitle

\section{Introduction}

\IEEEPARstart{C}{ountless} hours are squandered daily due to road congestion in numerous global metropolises. Inhabitants of cities like Los Angeles and Sydney experience an annual commuting duration equivalent to seven full working weeks, during which two weeks are consumed by traffic delays~\cite{Uber_2}. The resulting time inefficiency contributes to decelerated economic expansion, diminished productivity, and unwarranted carbon emissions. To address this issue, expanding mobility from the horizontal plane to the vertical dimension is imperative. Vertical transportation modalities, such as urban-air-mobility (UAM), promise to mitigate these pervasive urban dilemmas~\cite {goodrich2015demand}.
\begin{figure}[t]
    \centering
    \includegraphics[width=0.99\linewidth]{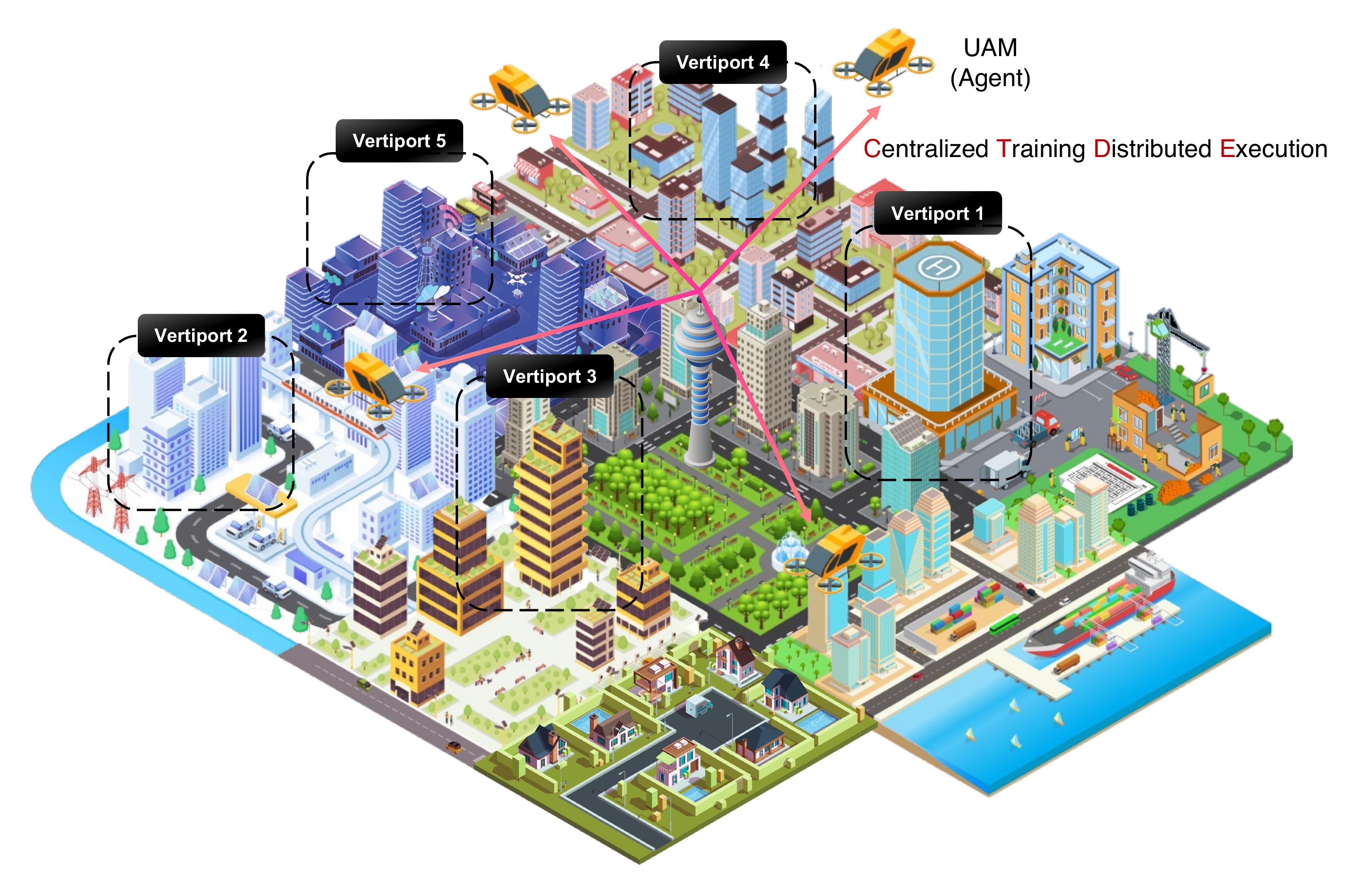}
    \caption{Reference air transportation service management network.}
    \label{fig:overall_architecture}
\end{figure}
UAM has garnered considerable interest in contemporary times, owing to its capacity for mitigating traffic bottlenecks and facilitating effective transportation services~\cite{9447255,neto2021trajectory,thipphavong2018urban,kohlman2019urban}.
{At present, a multitude of companies worldwide are engaged in the development of UAMs to commercialize them, such as the \textit{Volocopter 2X} by \textit{e-Volo GmbH}, \textit{Neva AirQuadOne} by \textit{Neva Aerospace}, \textit{Boeing PAV} by \textit{Aurora Flight Sciences}, \textit{Joby S2 VTOL\,/\,Cruise Configuration} by \textit{Joby Aviation}, and \textit{Opener Blackfly} by \textit{Opener}~\cite{straubinger2020overview}.
Nonetheless, there remains a need for more concrete evidence regarding the feasibility of air transportation in relation to the possibilities and demands, both in the presence and absence of autonomous vehicles.
As for delivering telecommunications, cellular networks emerge as viable options, owing to their widespread accessibility and substantial capacity, particularly in urban environments~\cite{straubinger2020overview}.
Furthermore, the 5G standard offers services such as ultra-low latency reliable communications and vehicle-to-everyting (V2X) communication, which serve as intriguing components for constructing cellular-based drone operations~\cite{americas2018new,ullah20195g}.
Next, from the perspective of feasibility in implementing pricing policies to convert existing transportation users to UAM users, the work in~\cite{choi2022exploring} used actual ground access transportation data collected from Incheon International Airport (ICN) and a multinomial logit model (MNL) to estimate the fare range for UAM services between Seoul Station and ICN. Assuming that UAM services can reduce travel time by 30-40 minutes compared to conventional ground taxis, the fare range was estimated to be between 96 and 108 US dollars. When comparing these estimated fares with those from expert institutions, it was demonstrated that such pricing policies are reasonable.
Lastly, the work in~\cite{courtin2018feasibility} demonstrates that UAM aircraft presents a viable option for air transportation networks, offering significant advantages with respect to vehicle payload and certification risks. Although the required runway lengths for UAMs range from 100 to 300 feet, which may appear short, they are achievable with near-future technology. In fact, a 300-foot runway length can be easily attained for UAM aircraft even with current technology, making it feasible from an urban infrastructure development perspective as well.}

{The primary advantage of UAM compared to traditional ground-based transportation lies in its potential to significantly reduce overall travel durations.} Additionally, UAMs address the issue of air contamination resulting from greenhouse gas emissions (GGE), which adversely affects the well-being of proximate inhabitants~\cite{soltani2020eco}. This is achievable because UAM operates on the basis of an eco-friendly electric propulsion system~\cite{kim2022effects}.
Despite the potential benefits, the integration of UAMs within prevailing transportation frameworks encounters a multitude of difficulties, such as the synchronization and collaboration between multiple UAMs~\cite{9447255}. Therefore, this paper puts forth a novel multi-agent deep reinforcement learning (MADRL) strategy with the objective of devising a trustworthy and effective aerial transportation system.
Within the suggested MADRL-focused methodology, a communication network (CommNet)\cite{sukhbaatar2016learning} is employed, incorporating centralized training and decentralized execution (CTDE)\cite{foerster2018counterfactual}.
CommNet facilitates inter-communication among multi-UAM to achieve coordination, while CTDE employs a \textit{centralized critic} to enhance the training efficacy of all \textit{actors} commensurate with the agent's policy.
The proposed MADRL-centric method tackles numerous obstacles in UAM functioning, including collision circumvention, trajectory formulation, and passenger routing. The employment of a multi-agent framework facilitates optimized utilization of airspace and resources, concurrently guaranteeing the security of passengers and UAM vehicles.
This paper evaluates the suggested MADRL-centric approach using data-driven simulations within realistic environment settings. A diverse set of findings demonstrates that the proposed technique surpasses current methodologies with respect to efficiency while maintaining safety. This proposed strategy holds the potential to substantially advance the development of a resilient and effective air transportation framework.

\subsection{Contributions}

The main contributions of this work are as follows.
\begin{itemize}
    \item \BfPara{Multi-UAM Cooperation and Coordination using MADRL} This paper utilizes a novel CommNet/CTDE-based MADRL strategy to manage air transportation networks efficiently under the concept of the cooperation and coordination of multiple UAMs. This advanced training structure helps multiple UAMs perform given common tasks more efficiently without explicit central coordination rules after the policy training phase.
    \item \BfPara{Realistic City-Wide Environment Design and Performance Evaluation} A vertiport map and UAM model are designed with the actual reference model to construct a realistic autonomous air transportation system. The energy charging/discharging model is accurately modeled with the specifications of the adopted UAM model, and even the passenger boarding and alighting system are designed in detail. Therefore, the justification of the proposed air transportation system becomes more potent than when an arbitrary system is assumed. Furthermore, in order to validate the MADRL supremacy of the proposed algorithm, data-intensive performance evaluation is conducted with the latest DRL algorithms from various aspects. Furthermore, this paper analyzes the results in detail with the subdivided groups according to the learning strategy.
\end{itemize}

\subsection{Organization}
The remainder of the paper is structured as follows. Sec.~\ref{sec:II} reviews related work in air transportation management systems and Sec.~\ref{sec:III} explains the system model considered in this paper.
Sec.~\ref{sec:IV} presents the proposed management algorithm for efficient autonomous air transportation networks employing CommNet/CTDE-based MADRL. Sec.~\ref{sec:V} demonstrates the performance of the proposed algorithm via various data-intensive evaluations. Finally, Sec.~\ref{sec:VI} concludes this paper.

\section{Related Work}\label{sec:II}
Self-governing vehicular have increasingly garnered interest from scholarly and commercial domains due to their potential to address enduring issues in transportation, such as safety enhancement, traffic alleviation, energy conservation, and other related concerns~\cite{wang2022metavehicles,cao2022future,huang2022survey}.
Advanced computational methodologies, such as deep learning (DL) and machine learning (ML) approaches, are employed to augment the efficacy of autonomous mobility systems~\cite{grigorescu2020survey,chen2017learning}.
{In urban settings, traffic data exhibit consistent characteristics concerning location and time within logistical systems. Consequently, learning-based algorithms demonstrate resilience, as they are capable of capturing the recurrence of similar patterns occurring at identical locations and times.}
A joint K-mean clustering and Gaussian mixture model-based expectation maximization~\cite{lu2020machine} and the attention mechanism are utilized for trajectory optimization~\cite{kong2022trajectory}.
{
However, the strategies utilized in the aforementioned traditional optimization and rule-based research, including those in~\cite{mozaffari2016optimal,kalantari2016number} that suggest methods for optimizing trajectories in autonomous aerial vehicles, predominantly concentrate on centralized optimization concerns.
This limited scope results in hindrances when attempting to provide real-time solutions for complex, dynamic, interrelated, and widespread transportation systems~\cite{sayyadi2020integrated}.
Furthermore, their approaches in massive environment may cause pseudo-polynomial computational complexity based on dynamic programming~\cite{bellman2010dynamic}.
In the context of deep reinforcement learning (DRL) algorithms, the computational complexity does not exhibit a general pseudo-polynomial time complexity; instead, it primarily depends on factors such as the size of the state space and the size of the action space.
}
{In addition,} imitation learning with advanced sensors is also employed for smart cruise control and lane-keeping systems~\cite {ijcai2019shin}. {However, imitation learning necessitates the provision of expert demonstrations and may suffer from covariate shift or suboptimal performance due to the limited exploration.}

Besides, among various DL and ML algorithms, DRL techniques have demonstrated the most effective performance by adaptively executing sequential decision-making processes in dynamic autonomous driving environments~\cite{kiran2021deep,grigorescu2020survey}.
Furthermore, the scalability of RL allows for the application of MADRL to govern the operation of multiple mobilities.
Among MADRL algorithms, the Q-Mix algorithm incorporates a comprehensive action-value function derived from the amalgamation of individual agents' action-value functions~\cite{rashid2020monotonic}. This algorithm is applied for the trajectory optimization of multiple electric vertical takeoff and landings (eVTOLs)~\cite{goodrich2015demand} in the context of aerial drone-taxi applications~\cite{icte202103yun}.
In addition, there are MADRL algorithms founded upon their inherent communication protocols, which utilize a neural network structure designed to facilitate inter-agent data exchange. Examples of these algorithms include differentiable inter-agent learning (DIAL)~\cite{foerster2016learning}, bilateral complementary network (BiCNet)~\cite{hou2021bicnet}, and CommNet.
DIAL-based agents use differentiable communication channels within their neural networks to learn cooperative strategies through end-to-end backpropagation, leading to improved coordination and cooperation among agents in complex, dynamic environments.
Next, BiCNet introduces a cutting-edge approach to address complex coordination challenges in MADRL by utilizing a twin neural network architecture, which enables agents to learn complementary policies collaboratively.
Finally, as elucidated in~\cite{sukhbaatar2016learning}, agents employing the CommNet framework acquire communication capabilities in tandem with a unified, centralized single deep neural network to process local observations for multiple agents.
Subsequently, each agent's decisions are influenced by its individual observations and the mean of other agents' observations~\cite{zhu2022survey}.
The CommNet architecture is extensively applied in diverse multi-agent systems in a distributed manner, such as management for electric vehicle charging station~\cite{tii202005shin}, charging scheduling in UAV networks~\cite{tvt202106jung}, and autonomous surveillance system~\cite{tii202210yun}.

While the previous study in~\cite{icc2023park} also suggests the CommNet-centric MADRL algorithm for establishing an autonomous multi-UAM network, several distinctions exist between the present and aforementioned works.
Primarily, the algorithm proposed herein further utilizes centralized critic, in contrast to only focusing on CommNet in~\cite{icc2023park}.
Moreover, this work evaluates the performance of the proposed MADRL algorithm while considering various communication statuses compared to previous studies in~\cite{icc2023park}. Additionally, an additional inference step is conducted to validate the effectiveness of the learned policy.

\section{Realistic City-Wide Autonomous Air Transportation System Design}\label{sec:III}

This paper constructs a realistic air transportation environment and UAM model based on actual vertiports and aircraft. This realistic design suggests a direction in which UAM can be realized in various bustling metropolitan areas, including Dallas, Texas, and Bedtown Frisco, USA.


\subsection{Realistic Air Transportation Environment}\label{sec:III-A}
\begin{figure}
    \centering
    \includegraphics[width=\linewidth]{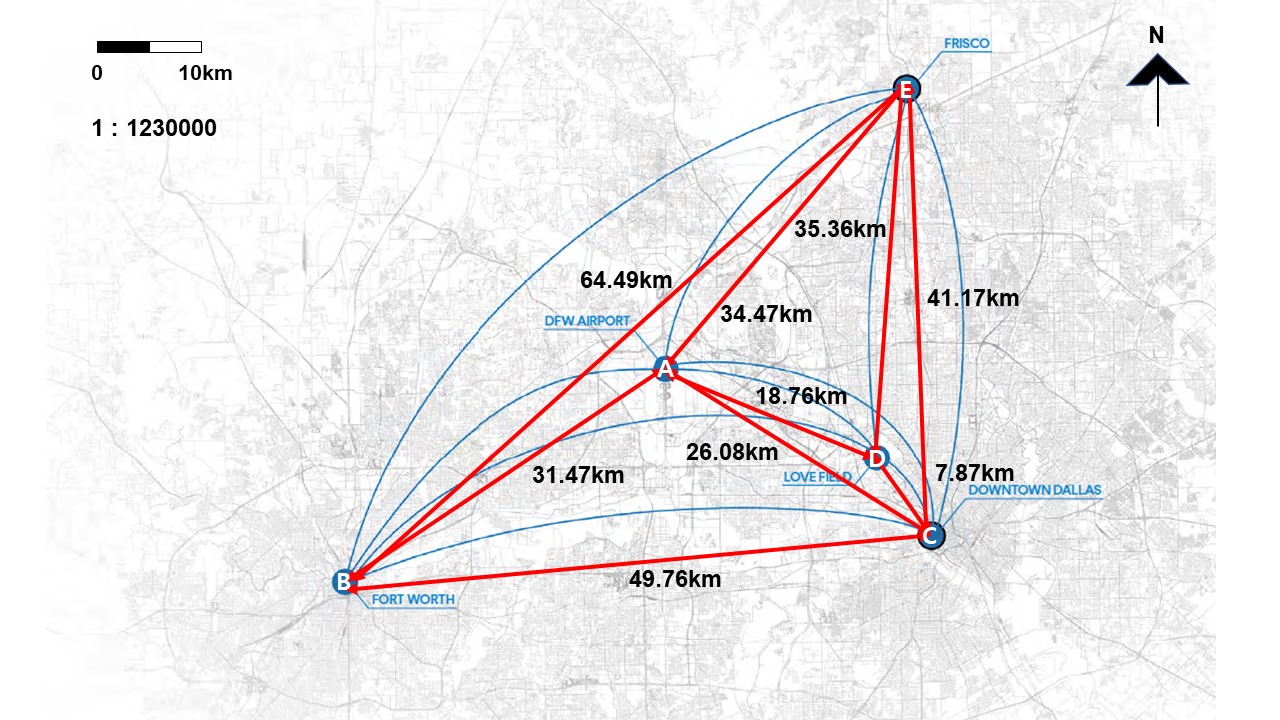}
    \caption{Vertiport map considered in this paper for realistic design~\cite{Uber_1}.}
    \label{fig:vertiport_map}
\end{figure}

Uber will provide UAM service to various metropolitan areas such as Sao Paulo and Los Angeles within 10 years~\cite{Uber_2}. Dallas, Texas, a central US metropolitan area, is no exception. Uber Air will begin commercial operations in Dallas in 2023, making it the first city to provide flights. The vertiport environment with the highest possibility of realization is Dallas, Texas. The actual background of the vertiport map used in the experiment corresponds to a movement zone connecting Downtown Dallas, Texas, USA, and Bedtown Frisco to the north. The actual detailed vertiport map information is illustrated in Fig.~\ref{fig:vertiport_map}~\cite{Uber_1}.
Fig.~\ref{fig:vertiport_map} shows the central transportation network of the United States, which connects downtown Dallas, Texas, with Frisco, centered on Dallas Fort Worth International(DFW) Airport. In this transportation network, Uber will build a total of five vertiports at DFW AIRPORT ($A$), FORT WORTH ($B$), DOWNTOWN DALLAS ($C$), LOVE FIELD ($D$), and FRISCO ($E$). The black number means the distance between vertiports. The length of the bar on the top left represents about $10\,km$ in real space. The scale is $1:1,230,000$, thus $1\,cm$ in Fig.~\ref{fig:vertiport_map} corresponds to $12.3\,km$.

\subsection{Realistic UAM Model}\label{sec:III-B}
\tBOSS{It is vital to take the energy consumption/remains of UAM devices into account because they are energy-constrained and power-hungry~\cite{9447255,mobisys2010paek}.
In contrast to the aircraft powered by internal combustion engines, which produce energy by burning fossil fuels, UAM is an electrified aircraft that relies on batteries. Accordingly, the energy model of UAM can be represented by aerodynamic power calculations independent to the specific fuel consumption (SFC). UAM's hovering power expenditure $P_h$ when take-off or landing with passengers can be mathematically expressed as follows~\cite{equation_sjung},
\begin{equation}
P_h=\underbrace{\frac{C_{d} }{8}\rho sA\Omega^3R^3}_{\textit{blade\,profile}, \,P_{o}}+\underbrace{(1+k)\frac{W^{\frac{3}{2}}}{\sqrt{2\rho A}}}_{\textit{induced}, \,P_{i}},
\label{eq:hovering energy}
\end{equation}
where} $C_{d}$ is the drag coefficient, which is a dimensionless coefficient that quantifies the drag force of a body in a fluid, $\rho$ is the density of air that decreases exponentially with altitude, and $s$ is the rotor solidity that means the ratio of the rotor blade area to the rotor disk area. In addition, $A$, $\Omega$, $R$, $W$ stand for the rotor disc area, blade angular velocity, rotor radius, and total weight considering passenger payload, respectively. $k$ is an induced drag coefficient that is inversely proportional to the efficiency factor ($e$) and aspect ratio ($AR$). UAM must rotate the rotor to overcome the drag increased by the induced drag, so $k$ must be considered in~\eqref{eq:hovering energy}. When a UAM carrying a passenger rises to an altitude of $600m$ by eVTOL and propels forward, the force on the x-axis is added. Thus, the propulsion power consumption $P_p$ must be considered. Here, the energy expenditure during a round trip or propulsion can be mathematically expressed as follows~\cite{zeng2017energy},


\begin{table}[t!]
\centering
\scriptsize
\caption{Specification of UAM Model.~\cite{jobyaviation}}
\renewcommand{\arraystretch}{1.0}
\begin{tabular}{l||r}
\toprule[1pt]
\textsf{\textbf{Notation}} & \textsf{\textbf{Value}} \\ \midrule
Maximum number of passengers, $\Lambda$ & 4 \\
Flight speed, $v$ & 73.762\,[$\mathrm{m/s}$] \\
Aircraft mass including battery and propellers, $m$ & 1,815\,[$\mathrm{kg}$]\\
Aircraft weight including battery and propellers, ${W}={mg}$ & 17,799\,[$\mathrm{N}$]\\ 
Rotor radius, $R$ & 1.45\,[$\mathrm{m}$] \\
Rotor disc area, $A=\pi R^{2}$ & 6.61\,[$\mathrm{m^{2}}$] \\
Number of blades , $b$ & 5 \\
Rotor solidity, $s=\frac{0.2231b}{\pi R}$ & 0.2449 \\
Blade angular velocity, $\Omega$ & 78\,[$\mathrm{radius/s}$]\\
Tip speed of the rotor blade , $U_{tip}=\Omega R^{2}$ & 112.776\,[$\mathrm{m/s}$] \\
Air density, $\rho$ & 1.225\,[$\mathrm{kg/m^{3}}$] \\
Fuselage drag ratio, $d_{0}=\frac{0.0151}{sA}$ & 0.01 \\
Mean rotor-induced velocity in hovering, $v_{0}=\sqrt\frac{W}{s\rho A}$ & 26.45\,[$\mathrm{m/s}$] \\
Profile drag coefficient, $C_{d}$ & 0.045 \\
Incremental correction factor to induced power, $k$ & 0.052 \\
\bottomrule[1pt]
\end{tabular}
\label{tab:parameters of uam}
\end{table}

\begin{multline}
P_p= \underbrace{P_i\left(\sqrt{1+\frac{v^4}{4v_0^4}}-\frac{v^2}{2v_0^2}\right)^{0.5}}_{\textit{induced}}\\+ \underbrace{P_0\left(1+\frac{3v^2}{U_{tip}^2}\right)}_{\textit{bladeprofile}}+\underbrace{\frac{1}{2}d_0\rho sAv^3}_{\textit{parasite}},
\label{eq:round trip traveling energy}
\end{multline}
where $v$ is the cruising flying speed of UAM at which UAM reaches an altitude of $600\,m$ and cruises after picking up passengers. $v_0$ is mean rotor-induced velocity which is the flow's average speed as caused by the wing tip vortex. $U_{tip}$ is tip speed of the rotor blade and $d_0$ is fuselage drag ratio. The values of the aerodynamic parameters used in~\eqref{eq:hovering energy} and~\eqref{eq:round trip traveling energy} are for JOBY AVIATION's S4 and are displayed in Table~\ref{tab:parameters of uam}~\cite{jobyaviation}.


According to Joby Aviation's analyst day presentation~\cite{jobyaviation}, S4 ascends to an altitude of $600\,m$ in around $30$ seconds after carrying passengers. Since the turnaround time is about $6$ minutes, UAM landing on a vertiport must unload passengers and recharge its battery as much as possible in the meantime. Here, the realistic actual charging time is only $5$ minutes, excluding the time to turn on/off the electric engine and plug in/out the charging cap. However, S4 can charge its battery quickly via a charger plugged directly behind the inboard tilt electric motor nacelle. It is equipped with four battery packs, two on both sides of the main wing inboard and the others in the nacelle at the rear of the main wing inboard electric motor. To charge each battery with $30\,kWh$ in $5$ minutes, the charger's supply power must be at least $360\,kW$.
Under this condition, S4's current rate (C-rate) indicating the battery charge/discharge rate becomes $2.4$ per hour when a $150\,kWh$ battery is charged with a $360\,kW$ charger. Finally, the state of charge (SOC) that is the time to charge the battery to $100\%$ takes $25$ minutes, and S4 can charge $20\%$ of the total battery capacity per journey. Table~\ref{tab:specification_of_UAM_battery} summarizes a detailed description of the battery specifications.


\begin{table}[t!]
\centering
\caption{Specification of S4's Battery.~\cite{jobyaviation}}
\renewcommand{\arraystretch}{1.0}
\begin{tabular}{l||r}
\toprule[1pt]
\textsf{\textbf{Notation}} & \textsf{\textbf{Value}} \\ \midrule
Battery capacity & $150\,[kWh]$ \\
Battery charge capacity per journey & $30\,[kWh]$ \\
Charging time per journey & $5\,[min]$ \\
Charger supply power & $360\,[kW]$ \\
C-rate & $2.4\,[per\,hour]$ \\
SOC & $25\,[min]$ \\
Charge rate per journey & $20\,\%$ \\
\bottomrule[1pt]
\end{tabular}
\label{tab:specification_of_UAM_battery}
\end{table}

\section{Algorithm Design}\label{sec:IV}

\subsection{RL Formulation for UAM Networks}\label{sec:IV-A}
\tBOSS{In order to consider the physical capabilities of UAM, this paper formulates UAM networks as a decentralized partially observable MDP (Dec-POMDP)~\cite{pajarinen2011periodic, cui2022multi}. According to the nature of RL-based algorithms, each UAM with Dec-POMDP can sequentially make action decisions founded on partial environmental information. The considering reference air transportation model consists of $J$, $N$, and $\Xi$ numbers of UAMs, vertiports, and passengers, respectively. Thus, the sets of UAMs, vertports, and passengers are defined as $\forall u_j \in \mathcal{U}$ where $\mathcal{U}\triangleq\{u_1,\cdots,u_j,\cdots,u_J\}$, 
    $\forall \varrho_\xi \in \mathcal{G}$ where $\mathcal{G}\in\{\varrho_1,\cdots,\varrho_\xi,\cdots,\varrho_\Xi\}$, and 
    $\forall \nu_n \in \mathcal{V}$ where $\mathcal{V}\in\{\nu_1,\cdots,\nu_n,\cdots,\nu_N\}$. Here, the air transportation service provided by UAMs in this paper can be defined as the passenger delivery using UAM from one vertiport to the other target vertiport.}

The POMDP of the proposed air transportation networks with $J$ UAMs can be modeled as $\langle J, \mathcal{S}, \mathcal{O}, \mathcal{A}, \mathcal{R}, \mathcal{P}, \mathcal{Z}, \gamma\rangle$ where $s\in\mathcal{S}$ is a set of ground truth states. Next, ${\mathrm{o}_j}\in \mathcal{O}_j$ and ${\mathrm{a}_j}\in \mathcal{A}_j$ stand for a set of $j$-th UAM's observations and actions, respectively. Note that these two sets can be jointly denoted as ${O}_j\subset\mathcal{O}$ and ${A}_j\subset\mathcal{A}$. At every time step, each UAM gets reward $\mathrm{r}_j$ with reward function $\mathcal{R}(\mathrm{s},\mathbf{a},\mathrm{s}')$ by selecting joint action $\mathbf{a}$ while observing joint observation information $\textbf{o}$ based on conditional observation probability $\mathcal{Z}(s',\textbf{a}, \textbf{o})=\mathcal{S}\times\mathcal{A}\rightarrow\mathcal{O}$. Then, the global state $s$ is transited to the next state $s'$ with state transition probability function $P(s'\,|\,s,\textbf{a})=\mathcal{S}\times\mathcal{A}\times\mathcal{S}\rightarrow\mathcal{S}$. Lastly, $\gamma$ is a discount factor that weights current rewards versus future rewards. The below subsections materialize the MDP delineating observation, state, action, reward, and objective of the proposed UAM networks via mathematical description.

\subsubsection{Observation}\label{sec:IV-A-1}
Due to the physical limitation of the considered UAM model, every UAM can recognize other UAMs or vertiports within its coverage. Here, the eyesight of the $j$-th UAM is dependent on its absolute position denoted as $p_j\in\{x_j,y_j,z_j\}$ corresponding to Cartesian coordinates. The $j$-th agent can recognize distances with other UAMs and vertiports in its observation scope $D_{th}$, which information can be mathematically represented as follows,
\begin{equation}
    d(u_j,u_{j'}) = \begin{cases}
        \|u_j - u_{j'}\|_2,& \textit{if.}~~~\|u_j - u_{j'}\|_2 \leq D_{th}, \\
        -1,& \textit{(otherwise)},
    \end{cases}
\end{equation}
\begin{equation}
    d(u_j,\nu_{n'}) = \begin{cases}
        \|u_j - \nu_{n}\|_2,& \textit{if.}~~~\|u_j - \nu_{n}\|_2 \leq D_{th}, \\
        -1,& \textit{(otherwise)},
    \end{cases}
\end{equation}
where $d(\cdot)$ and $\|\cdot\|_2=(\sum_{i=1}^N|\cdot|^2)^{\frac{1}{2}}$ stand for the function that outputs the distance between two inputs and L2-norm, respectively. UAMs can also carry up to $\Lambda$ passengers, and also know the status of their seats denoted as $\Phi_j=\{\phi^1_j,\cdots,\phi^\lambda\cdots,\phi^\Lambda_j\}$, where $\phi^\lambda_j$ is defined as follows,
\begin{equation}
    \phi^\lambda_j =
    \begin{cases}
        \varrho_\xi,& \textit{if.}~~\text{$\varrho_\xi$ boards the $\lambda$-th seat on $u_j$}, \\
        -1,& \textit{(otherwise)},
    \end{cases}
\end{equation} Lastly, each $j$-th UAM has to observe its energy state $e_j$ to avoid the battery's full discharge for a safe air transportation system denoted as follows,
\begin{equation}
    e_j = \begin{cases}
    e_{\mathrm{max}}-(P_p\times\frac{|\mathcal{A}_h|^2}{(vt)^2} + P_h\times\frac{|\mathcal{A}_v|^2}{(vt)^2}),& \textit{if.}~~~t=0, \\
    \max{(e_j-(P_p\times\frac{|\mathcal{A}_h|^2}{(vt)^2} + P_h\times\frac{|\mathcal{A}_v|^2}{(vt)^2}}),& \textit{(otherwise)},
    \end{cases}
\end{equation}
where the $P_h$ and $P_p$ stand for power consumption when UAM takes-off or lands on the vertiport defined in Sec.~\ref{sec:III-B}. In addition, $\mathcal{A}_h$ and $\mathcal{A}_v$ are the action UAM takes, which are specified in Sec.~\ref{sec:IV-A-2}. In brief, the partial information that $j$-th UAM can observe in the environment is connoted as follows, 
\begin{equation}
O_j \triangleq\{p_j, \bigcup_{j\neq j'}^{J}\{d(u_j,u_{j'})\}, \bigcup_{j\neq j'}^{J}\{d(u_j,\nu_{n})\}, \Phi_j, e_j\}.
\end{equation}

\subsubsection{State}\label{sec:IV-A-2}
The ground truth state includes the overall air transportation service information consisting of the number of passengers serviced by each UAM represented as $\bigcup^{\Xi}_{\xi=1}\{\mathbbm{1}_j(\varrho_\xi)\}$, where the $\mathbbm{1}(\cdot)$ is an indicator function differentiating serviced (one) or not (zero). The type of vertiports each UAM landed on is also contained in the ground truth state denoted as $\bigcup^{N}_{n=1}\{\mathbbm{1}_j(\nu_n)\}$, which represents visited (one) or not (zero). Additionally, the distance to the target vertiport of the passenger boarding the $j$-th UAM is one of the components of the state space, which is $d'(u_j,\nu_{n}^{\mathrm{target}})$. Here, the distance between $u_j$ and $\nu_\mathrm{target}^{\lambda}$ dependent on seating state can be defined as follows,
\begin{equation}
d'(u_j, \nu_\mathrm{target}^{\lambda}) = 
    \begin{cases}
        \|u_j - \nu_{\mathrm{target}}^\lambda\|_2,& \textit{if.}~~~\phi_j^\lambda\neq-1, \\
        -1,& \textit{(otherwise)}.
    \end{cases}
\end{equation}
To recap, the ground truth state can be organized as follows,
\begin{equation}
\begin{split}
& \mathcal{S}\triangleq \\
& \{\bigcup^{J}_{j=1}\bigcup^{\Xi}_{\xi=1}\{\mathbbm{1}_j(\varrho_\xi)\},\bigcup^{J}_{j=1}\bigcup^{N}_{n=1}\{\mathbbm{1}_j(\nu_n)\},\bigcup_{j=1}^{J}\bigcup_{\lambda=1}^{\Lambda}\{d'(u_j,\nu_\mathrm{target}^{\lambda})\}.
\end{split}
\end{equation}

\begin{figure}[t!]
    \centering
    \includegraphics[width=0.4\columnwidth]{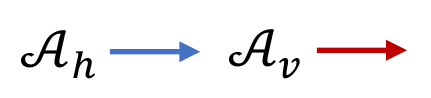}\\
    \subfigure[Top View.]
    {
        \includegraphics[width=0.465\columnwidth]{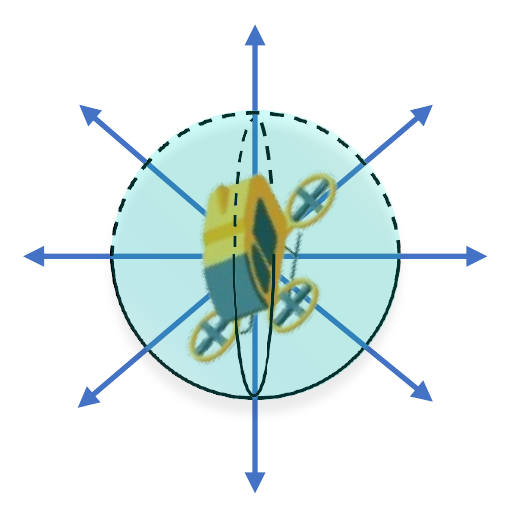}
        \label{fig:Top}
    }
    \subfigure[Forward View.]
    {
        \includegraphics[width=0.3333\columnwidth]{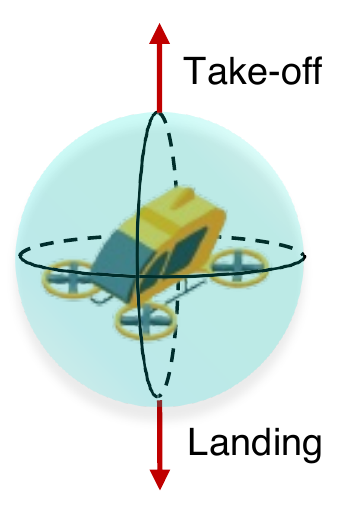}
        \label{fig:Forward}
    }
    \caption{Types of actions UAM can take.}
    \label{fig:Action_Type}
\end{figure}

\subsubsection{Action}\label{sec:IV-A-3}
In every time step $t$, every UAM can take two types of actions; \textit{i)} horizontal and \textit{ii)} vertical moving, where the set of actions is $\mathcal{A}\triangleq\{\mathcal{A}_h,\mathcal{A}_v\}$ as illustrated in Fig.~\ref{fig:Action_Type}. Note that $\mathcal{A}_h$ and $\mathcal{A}_v$ have an alternate relationship, thus $\mathcal{A}_h$ is zero when $\mathcal{A}_v$ is selected by UAM (vice versa).
The first type of action is horizontal moving from one vertiport to another vertiport in ordinal directions expressed in vector form as follows,
\begin{equation}
\mathcal{A}_h\in\{\langle\pm vt,0,0\rangle, \langle 0,\pm vt,0\rangle,\langle\pm\frac{1}{\sqrt{2}}vt,\pm\frac{1}{\sqrt{2}}vt,0\rangle\}.
\end{equation}
The other is moving for vertical take-off and landing in any vertiport, which can be indicated in vector form as follows,
\begin{equation}
\mathcal{A}_v\in\{\langle0,0,vt\rangle,\langle0,0,-vt\rangle\}.
\end{equation}

\subsubsection{Reward}\label{sec:IV-A-4}
Every UAM takes action at every time step, and then it gets a reward from reward function $\mathcal{R}(s,\mathbf{a},s')$ when the state $s$ is transited to the next state $s'$. There is a common goal of UAMs in MADRL, this paper divides the reward function into two elements; \textit{i)} individual reward, \textit{ii)} common reward. Thus, the reward function can be configured as $\mathcal{R}(\mathrm{s},\mathbf{a},\mathrm{s}')=\sum^{J}_{j=1}(\mathcal{R}^j_{\mathrm{Indiv}}(\mathrm{s},\mathbf{a},\mathrm{s}'))+\mathcal{R}_{\mathrm{Comm}}(\mathrm{s},\mathbf{a},\mathrm{s}')$.
First, each UAM receives individual rewards based on its actions. Accordingly, all UAMs have different individual reward values from each other, which are denoted as follows,
\begin{equation}
    \begin{split}
    & \mathcal{R}^j_{\mathrm{Indiv}}(\mathrm{s},\mathbf{a},\mathrm{s}') = 
    \prod_{j\neq j'}^{J}\mathbbm{1}(d(u_j,u_{j'})\geq C_{th})
    \\
    & \times\!\big[\sum_{\xi=1}^{\Xi}\mathbbm{1}_j(\varrho_\xi)\!+\!\sum_{n=1}^{N}\mathbbm{1}_j(\nu_n)\!-\!\sum_{\lambda=1}^{\Lambda}\frac{d'(u_j,\nu_\mathrm{target}^\lambda)}{\Gamma/2}\!+\!\frac{e_j}{e_\mathrm{max}}\big],
    \end{split}
\end{equation}
where $C_{th}$ and $\Gamma$ are the minimum distance to occur collision and considered environment size in this paper, respectively.
Next, all UAMs have the same common reward value simultaneously which helps them to cooperatively achieve a shared objective without preemption or competition. The common reward is established as follows,
\begin{equation}
    \mathcal{R}_{\mathrm{Comm}}(\mathrm{s},\mathbf{a},\mathrm{s}') = 
    \sum_{j=1}^{J}\frac{\mathbbm{1}_j(\varrho_\xi)}{J}.
\end{equation}
As a result, as seen in the above definition of the reward function, all UAMs try to maximize not only the quality components of air transportation service but also consider safety conditions. By shaping appropriate reward functions, UAMs can learn the intelligence to select the suitable strategy concerning shared objectives in any given state.

\subsubsection{Objective}\label{sec:IV-A-5}
All UAMs' main objectives can be mathematically expressed as follows,
\begin{equation}
    \pi^*_{\boldsymbol{\theta}} = 
    \argmaxD_{\boldsymbol{\theta}}\mathbb{E}_{\mathrm{s}\sim E,\,\mathrm{a}\sim\pi_{\boldsymbol{\theta}}}\left[\sum_{t=1}^T\gamma^{t-1}\!\cdot\!\mathcal{R}\left(\mathrm{s},\mathbf{a},\mathrm{s}'\right)\right],
    \label{eq:goal}
\end{equation}
where $\boldsymbol{\theta}$, $E$, and $T$ correspond to the parameters of \textit{actor}-network, environment, and episode length, respectively. In summary, it is obvious that the main objective of UAMs is to find the optimal policy maximizing the expected cumulative reward over a given finite step $T$.

\begin{figure*}[t!]
    \centering
    \includegraphics[width=\linewidth]{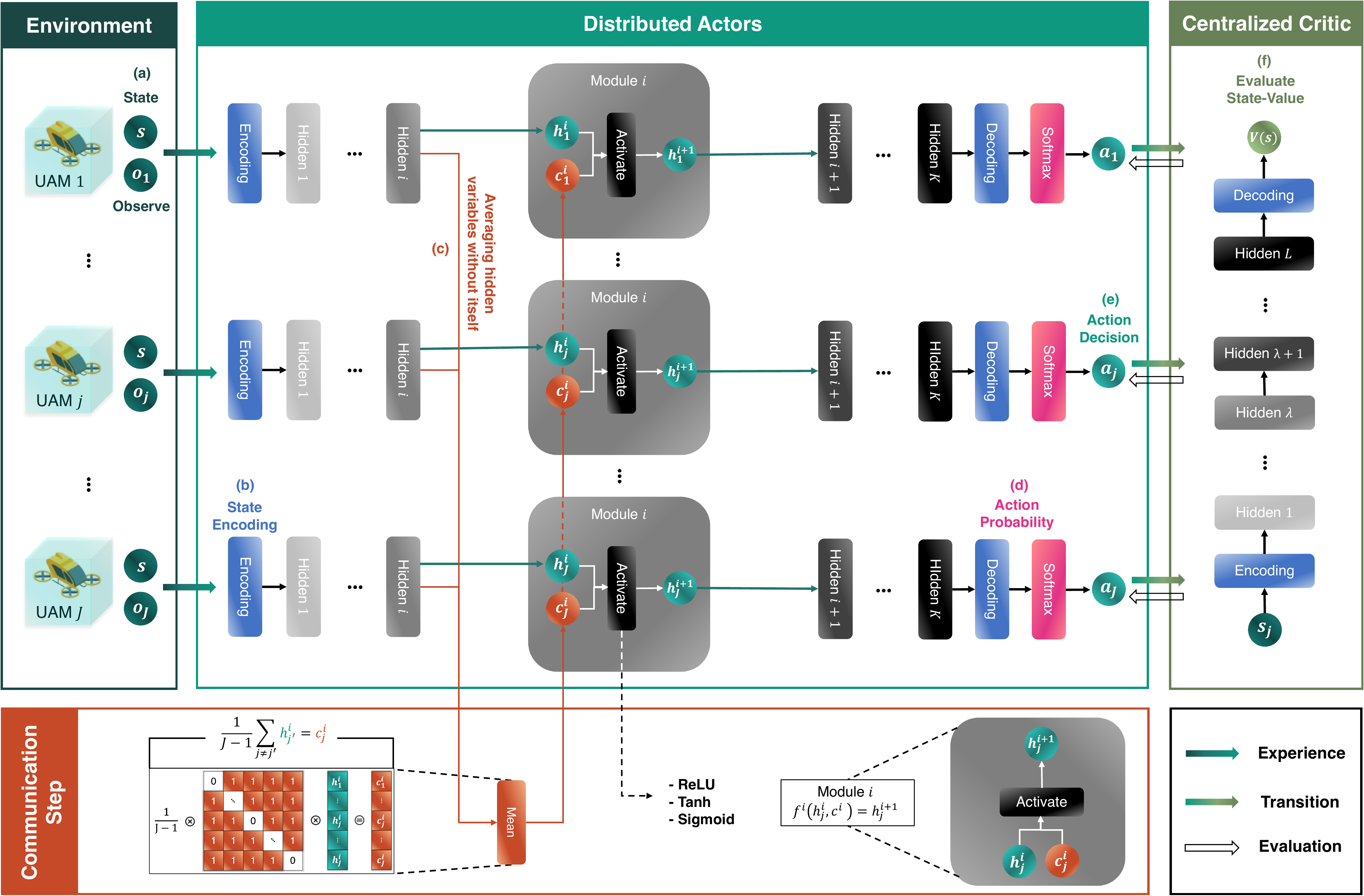}
    \caption{Overall CTDE-based neural network training architecture of CommNet algorithm.}
    \label{fig:CommNet}
\end{figure*}

\subsection{Information Sharing by CommNet}
The CommNet algorithm provides mutual communication between multiple UAMs with a communication step when multiple \textit{actors} proceed to learn their hidden variables as depicted in Fig.~\ref{fig:CommNet}. This process of sharing observation information with each other is particularly effective when multiple agents try to achieve a common goal in a Dec-POMDP environment where agents cannot observe the global state. First of all, every UAM explores the environment consisting of state and observation in Fig.~\ref{fig:CommNet}(a). Their experiences are encoded into hidden variables of the first layer $h^{1}_{j}$ with encoder function in Fig.~\ref{fig:CommNet}(b) as follows,
\begin{equation}
    h^{1}_{j} = \textit{Encoder}\,(\mathrm{s}, \mathrm{o}_j).
    \label{eq:encoding}
\end{equation}
When hidden variables are fed into the deeper hidden layers, the communication variables are entered simultaneously. The communication variable of $j$-th UAM in $i$-th hidden layer is acquired by averaging hidden variables of other UAMs in Fig.~\ref{fig:CommNet}(c) as follows,
\begin{equation}
    c^{i}_{j} = \frac{1}{J-1} \sum_{j \neq j'}^J\nolimits h^{i}_{j'}.
    \label{eq:comm}\\
\end{equation}
At the input of every $i$-th layer except for the first layer, $h^{i}_{j}$ and $c^{i}_{j}$ are feed-forwarded to the next layer with the single-agent module $f^{i}(\cdot)$, which returns output vector $h^{i+1}_j$ as follows,
\begin{equation}
    {h}^{i+1}_{j} = {f}^{i}({h}^{i}_{j}, {c}^{i}_{j}),
    \label{eq:hidden}
\end{equation}
subject to
\begin{equation}
    {f}^{i}(\cdot) = \textit{Activ}(\textit{Concat}({h}^{i}_{j}, {c}^{i}_{j})), 
    \label{eq:module}
\end{equation}
where $\textit{Activ}(\cdot)$ and $\textit{Concat}(\cdot)$ stand for a non-linear activation function (\textit{e.g.}, ReLU, hyperbolic tangent, or sigmoid) and concatenate function. The information sharing by inter-communication between UAMs comes about when averaging hidden variables of different UAMs. Finally, the action probabilities of $j$-th UAM are obtained by decoding the output of the last layer in Fig.~\ref{fig:CommNet}(d) as follows,
\begin{equation}
    p_{\boldsymbol{\theta}_j}(\mathcal{A}_j;\mathrm{o}_j) = \textit{Decoder}\,(h_{j}^{K}),
    \label{eq:output}
\end{equation}
where $p(\cdot)$ is the action distribution. To sum it up, the above sequential process of feed-forwarding can be implied as follows,
\begin{equation}
    p_{\boldsymbol{\theta}_j}(\mathcal{A}_j;\mathrm{o}_j) = Q(\mathrm{o}_j,\mathrm{a}\,;[\boldsymbol{\theta}_1,\cdots,\boldsymbol{\theta}_j,\cdots,\boldsymbol{\theta}_J]),
\end{equation}
where $Q(\mathrm{o}_j,\mathrm{a})$ is commensurate with the action-value function in Dec-POMDP. It is noteworthy that the output probabilities of $j$-th UAM are also dependent on the network parameters of other UAMs.

\begin{algorithm}[t]
\small
    Initialize parameters of \textit{actor} networks and \textit{centralized critic} network which are denoted as $[\boldsymbol{\theta}_1,\cdots,\boldsymbol{\theta}_j,\cdots,\boldsymbol{\theta}_J]$ and $\boldsymbol{\phi};$ \\
    Initialize replay buffer $\mathcal{D}=\{\}$ and mini-batch $\mathcal{B}=\{\};$ \\
    \For{Epoch = 1, MaxEpoch}{
        $\triangleright$ \textbf{Initialize Air Transportation Environments}, set $s_0$; \\
        \For{time step = 1, $T$}{
            \For{each $j$-th UAM}{
                $\triangleright$ Select the action $\mathrm{a}_j$ based on its policy $\pi_{\boldsymbol{\theta}_j}(\mathrm{a}_j^t\,|\,\mathrm{o}_j^t)$ at time step $t$; \\
            }
            $\triangleright$ $s^t \rightarrow s^{t+1}$, $\textbf{o}^t \rightarrow \textbf{o}^{t+1}$ with the reward $\textbf{r}^t;$

            $\triangleright$ Set $\xi= \{s^t, \textbf{o}^t, \textbf{a}^t, \textbf{r}^t, s^{t+1}, \textbf{o}^{t+1}\};$ \\
            
            $\triangleright$ Update $\mathcal{D}$: $Enqueue(\mathcal{D},\xi);$ \\
        }
        \If {$\mathcal{D}$ \textbf{is full enough to train}}{
        
        \For{each $j$-th UAM}{
        $\triangleright$ Get $V_{\boldsymbol{\phi}}$ by sampling mini-batch $\mathcal{B}$ from $\mathcal{D};$\\
        
        $\triangleright$ Update $\boldsymbol{\phi}$ by \textbf{gradient descent} to loss function of the \textit{centralized critic} network: $\nabla_{\boldsymbol{\phi}}\mathcal{L}(\boldsymbol{\phi});$ \\
        
        $\triangleright$ Update $\boldsymbol{\theta}_j$ by \textbf{gradient ascent} to objective function of $j$-th \textit{actor} network: $\nabla_{\boldsymbol{\theta}_j} J(\boldsymbol{\theta}_j);$
        }
    }
    }
    \For{Episode = 1, MaxEpisode}{
        $\triangleright$ \textbf{Initialize Air Transportation Environments}, set $s_0;$ \\
        \For{time step = 1, $T$}{
            \For{each $j$-th UAM}{
                    $\triangleright$ Select the action $\mathrm{a}_j$ based on its policy $\pi_{\boldsymbol{\theta}_j}(\mathrm{a}_j^t\,|\,\mathrm{o}_j^t)$ at time step $t;$ \\
                }
                $\triangleright$ $s^t \rightarrow s^{t+1}$, $\textbf{o}^t \rightarrow \textbf{o}^{t+1};$ \\
        }
    }
    \caption{Policy Training and Inference in CTDE}
    \label{alg:CTDE}
\end{algorithm}

\subsection{CTDE-based Parameterized Policy Training}
As motivated in~\cite{lowe2017multi}, multiple \textit{actors} distributedly get experiences by exploring environments, and the \textit{centralized critic} evaluates every global state's value in Fig.~\ref{fig:CommNet}(f). The \textit{centralized critic} may be a central server such as a control tower observed in Fig.~\ref{fig:overall_architecture}. This strategy helps all UAMs symmetrically to learn near-optimal policies. Additionally, below introduces the multi-agent policy gradient (MAPG) based on the \textit{temporal difference (TD) actor-critic}
method~\cite{fujimoto2018addressing} with \textit{Bellman optimality equation} to prevent the occurrence of high variance.

\BfPara{Centralized Critic}
To evaluate the value of parameterized policies of decentralized \textit{actors}, the \textit{centralized critic} tries to learn its network parameters $\boldsymbol{\phi}$ to approximate the optimal joint state-value function which is configured as follows,
\begin{equation}
    V_{\boldsymbol{\phi}}(\mathrm{s}) \!=\! 
    \mathbb{E}_{\mathrm{s}\sim E,\,\mathrm{a}\sim\pi_{\boldsymbol{\theta}}}\left[\,\sum_{u=t}^{T} \gamma^{u-t}\!\!\cdot \mathcal{R}(\mathrm{s}^u,\mathbf{a}^u,\mathrm{s}^{u+1})\,\right].
    \label{eq:state-value-function}
\end{equation}
With~\eqref{eq:state-value-function}, the \textit{centralized critic} learns its network parameters to minimize the loss function which is leveraged as follows,
\begin{equation}
    \nabla_{\boldsymbol{\phi}}\mathcal{L}(\boldsymbol{\phi}) = \sum^{T}_{t=1}\nabla_{\boldsymbol{\phi}}\left\|\delta^t_{\boldsymbol{\phi}}\right\|^2,
    \label{eq:l_critic}
\end{equation}
subject to
\begin{equation}
    \delta_{\boldsymbol{\phi}}^t = 
    \underbrace{\mathcal{R}(\mathrm{s}^t,\mathbf{a}^t,\mathbf{s}^{t+1})+\gamma V_{\boldsymbol{\phi}}(\mathrm{s}^{t+1})}_{\text{TD Target}}
    -V_{\boldsymbol{\phi}}(\mathrm{s}^t),
    \label{eq:delta}
\end{equation}
where $\delta^t_{\boldsymbol{\phi}}$ is the TD error based on \textit{Bellman optimality equation} in time step $t$. As seen in~\eqref{eq:delta}, the TD target is composed of the summation of current and future reward values. Based on~\eqref{eq:delta}, the \textit{centralized critic} trains its network parameters in the direction of minimizing the loss function by gradient descent as follows,
\begin{equation}
    \boldsymbol{\phi}^{t+1} \approx \boldsymbol{\phi}^{t} + \alpha_{\mathrm{critic}}\times[\,\delta^t_{\boldsymbol{\phi}} \cdot \nabla_{\boldsymbol{\phi}}V_{\boldsymbol{\phi}}(\mathrm{s}^{t})\,],
\end{equation}
where $\alpha_{\mathrm{critic}}$ stands for a learning rate of the \textit{centralized critic} network that decides the inclination of updating neural network parameters by policy gradient.

\BfPara{Multiple Actors}
Actors correspond to UAMs providing air transportation service to passengers in environments. They learn policy parameters $\boldsymbol{\theta}$ to approximate optimal policy for providing efficient air transportation service. At every time step $t$, they make sequential decision-making based on their parameterized strategy function as follows,
\begin{equation}
    \mathrm{a}_j^t = \argmaxD_{\mathrm{a}^t}\pi_{\boldsymbol{\theta}_j}(\mathrm{a}^t|\,\mathrm{o}_j^t).
\end{equation}
It can be seen that each UAM selects an action with high probability among all possible actions as presented in Fig.~\ref{fig:CommNet}(e). Here, $j$-th UAM's parameterized policy $\pi_{\boldsymbol{\theta}_j}$ is defined as follows,
\begin{equation}
    \pi_{\boldsymbol{\theta}_j}(\mathrm{a}^t\,|\,\mathrm{o}_j^t) \triangleq \textit{softmax}(p_{\boldsymbol{\theta}_j}(\mathcal{A}_j|\,\mathrm{o}_j^t)),
\end{equation}
subject to
\begin{equation}
    \textit{softmax}(\mathbf{x}) \triangleq \left[\frac{e^{x_1}}{\sum_{i=1}^N e^{x_i}},\cdots,\frac{e^{x_N}}{\sum_{i=1}^N e^{x_i}}\right],
\end{equation}
where $\textit{softmax}(\cdot)$ stands for exponential softmax distribution function to activate normalization of action probabilities.
Finally, the objective function that dispersed actors need to maximize is mathematically constituted as follows,
\begin{equation}
    \nabla_{\boldsymbol{\theta}_j}J(\boldsymbol{\theta}_j) = \mathbb{E}_{\mathrm{o}_j \sim E} \left[\,\sum\limits^{T}_{t=1} \delta^t_{\boldsymbol{\phi}}\!\cdot\!\nabla_{\boldsymbol{\theta}}\log\pi_{\boldsymbol{\theta}_j}(\mathrm{a}_j^t|\,\mathrm{o}_j^t)\right].
    \label{eq:l_actor}
\end{equation}
Using~\eqref{eq:l_actor}, UAMs learn their parameters toward maximizing the objective function by gradient ascent as follows,
\begin{equation}
    \boldsymbol{\theta}_j^{t+1} \approx \boldsymbol{\theta}_j^{t} + \alpha_{\mathrm{actor}}\times[\,\delta^t_{\boldsymbol{\phi}} \!\cdot\!\nabla_{\boldsymbol{\theta}}\log\pi_{\boldsymbol{\theta}_j}(\mathrm{a}_j^t|\,\mathrm{o}_j^t)\,],
\end{equation}
where $\alpha_{\mathrm{actor}}$ is a learning rate of the \textit{actor} network. The details of the CTDE-based policy training and inference are organized in Algorithm~\ref{alg:CTDE} in consecutive order.

\begin{table}[t!]
\centering
\caption{System Parameters for Performance Evaluation}
\renewcommand{\arraystretch}{1.0}
\begin{tabular}{l|r}
\toprule[1pt]
\textbf{\textsf{Notation}} & \textbf{\textsf{Value}} \\ \midrule
Size of environment, $\Gamma$ & $32,000\,[m]$ \\
Length of episode, $T$ & $60\,[min]$ \\
Number of UAMs, $J$ & $10$ \\
{Number of vertiports, $N$} & {$5$}\\
{Size of state space, $|\mathcal{S}|$} & {$67$}\\
{Size of action space, $|\mathcal{A}|$} & {$15$} \\
Size of mini-batch, $|\mathcal{B}|$ & $32$ \\
Size of experience replay buffer, $|\mathcal{D}|$ & $5\times10^4$ \\
Discount factor, $\gamma$ & $9.8\times10^{-1}$ \\
Initial value of epsilon, $\epsilon_{\mathrm{init}}$ & $0.275$ \\
Minimum value of epsilon, $\epsilon_{\mathrm{min}}$ & $0.01$ \\
Annealing epsilon & $5\times10^{-5}$ \\
Hidden layer dimension of \textit{actor} & $64$ \\
Hidden layer dimension of \textit{centralized critic} & $256$ \\
Learning rate of \textit{actor}, $\alpha_\mathrm{actor}$ & $1\times10^{-2}$ \\
Learning rate of \textit{centralized critic}, $\alpha_\mathrm{critic}$ & $2.5\times10^{-3}$ \\
Training epochs & $5,500$ \\
Activation function & ReLU \\
Optimizer & Adam \\
\bottomrule[1pt]
\end{tabular}
\label{tab:setup}
\end{table}

\section{Performance Evaluation}\label{sec:V}
\subsection{Experimental Setting}
The air transportation area has a size of $(2\times\Gamma)^2\,m^2$, illustrated in Fig.~\ref{fig:vertiport_map}. In that area, $J$-UAMs autonomously provide air transportation service to passengers by transporting them from one vertiport to the destination vertiport. As mentioned in Sec.~\ref{sec:III-B}, the realistic UAM model, named JOBY AVIATION's S4, transports passengers at an altitude of $600\,m$. In addition, this UAM model can carry up to four passengers based on the first-in-first-out (FIFO). UAMs start an episode at random vertiports, and destinations/departures of passengers also change from episode to episode. 
In addition, UAMs can recharge their batteries for five minutes at the vertiport while dropping off and picking up passengers as mentioned in Sec.~\ref{sec:III-B}.
{Lastly, note that this paper considers a total of five vertiports, as depicted in Fig.~\ref{fig:vertiport_map}. With merely five vertiport locations, the network effectively represents a city-wide, large-scale, and intricate system capable of realistically accommodating a significant number of passengers. In~\cite{Uber_1}, Uber has established five vertiports in the Dallas Metropolitan area, which is the closest region to the implementation of UAM-based air transportation service. Since Uber has already determined the number and location of vertiports based on actual urban traffic demand, a configuration with only five vertiports still results in a relatively complex and practical map.} The overall system parameter notations and corresponding values are arranged in Table~\ref{tab:setup}.

\subsection{Benchmarks}
\tBOSS{This paper conducts data-intensive experiments focusing on the validation of the proposed algorithms' performance,} which coincided with \textit{i)} CommNet and \textit{ii)} CTDE. For this purpose, this paper divides benchmarks into the following two groups.

\subsubsection{Benchmarks for CommNet}

To scrutinize the performance of CommNet in MADRL, an ablation study is conducted according to the number of UAMs participating inter-agent communications.

\begin{itemize}
    \item \BfPara{CommNet (Proposed)} All UAMs in this benchmark learn their policies with rich experiences (\textit{i.e.,} hidden variables) using information sharing via inter-communications.
    \item \BfPara{Hybrid} This benchmark has half CommNet-based UAMs and half DNN-based UAMs, where 'DNN' stands for the conventional neural network architecture. In other words, only half of UAMs communicate environmental information with each other.
    \item \BfPara{DNN} There are no mutual communications between UAMs. In other words, they behave as in a single-agent DRL environment.
    \item \BfPara{Monte Carlo} Since UAMs in this benchmark have no policies, they take actions randomly without sequential decision-making. Although it is not an ML/DRL algorithm, it can serve as a standard for evaluating the performance of other learning algorithms.
\end{itemize}

\subsubsection{Benchmarks for CTDE}

Benchmarks in the other group differ in their training methodology. The performance of the proposed CTDE-based training approach is compared with conventional DRL algorithms.

\begin{itemize}
    \item \BfPara{CTDE (Proposed)} A central server utilizes \textit{centralized critic} when evaluating the ground truth state made by decentralized \textit{actors}. 
    \item \BfPara{IAC} Instead of a central server, every UAM has its \textit{independent actor-critic} (IAC) networks. UAMs in this benchmark learn their policies based on the existing TD actor-critic algorithm~\cite{fujimoto2018addressing}.
    \item \BfPara{DQN} Deep Q-network (DQN) algorithm~\cite{mnih2013playing} only approximates Q-function without separate neural network approximating optimal state-value function (\textit{i.e., critic}).
    \item \BfPara{Monte Carlo} This benchmark is for the \textit{Monte Carlo} simulation as described above.
\end{itemize}

\subsection{Training Performance}\label{sec:V-C}

\subsubsection{Reward Convergence}\label{sec:V-C-1}
\begin{figure}[t!]
    \centering
    \includegraphics[width=\columnwidth]{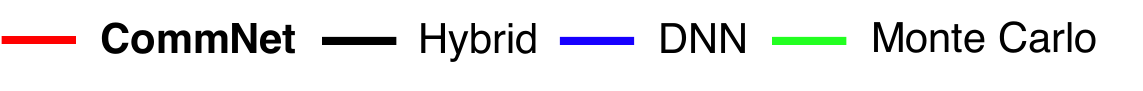}\\
    \subfigure[POMDP.]
    {
        \includegraphics[width=0.465\columnwidth]{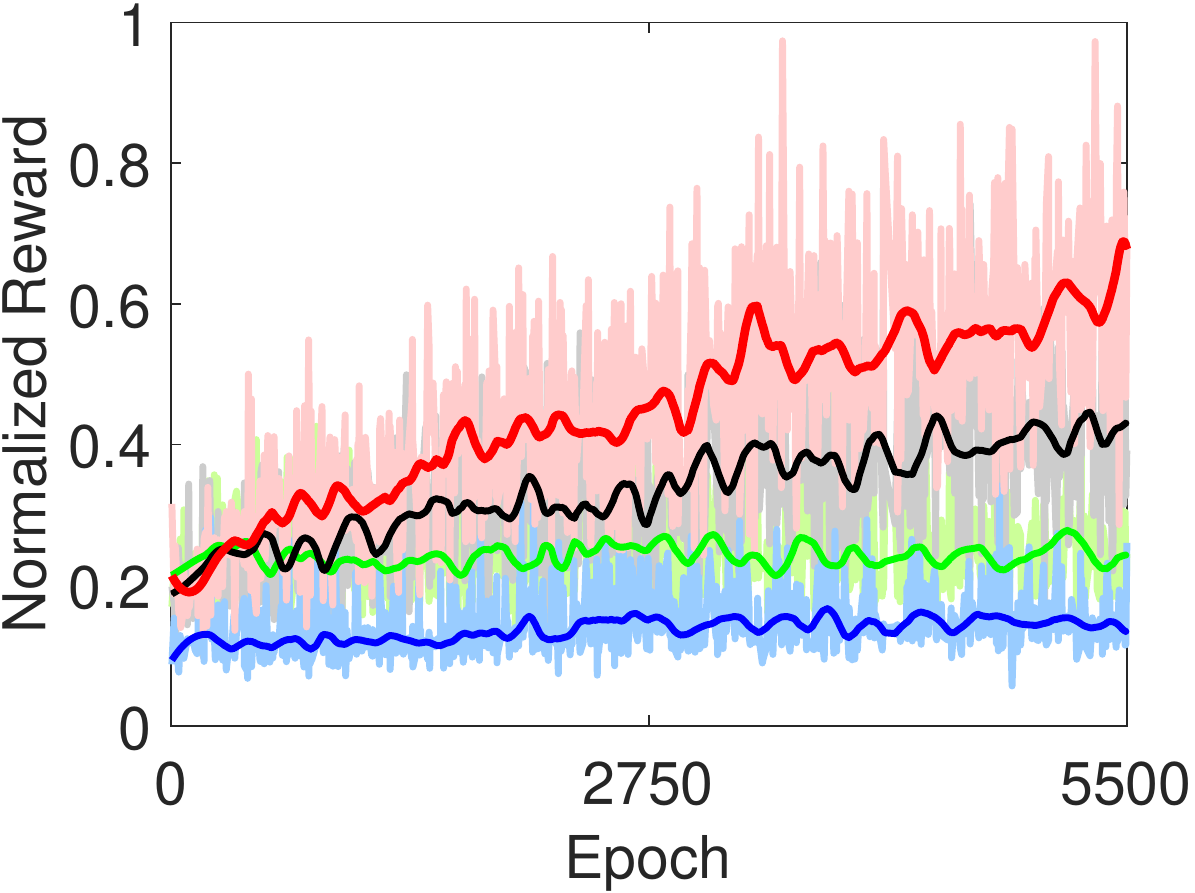}
        \label{fig:POMDP}
    }
    \subfigure[FOMDP.]
    {
        \includegraphics[width=0.465\columnwidth]{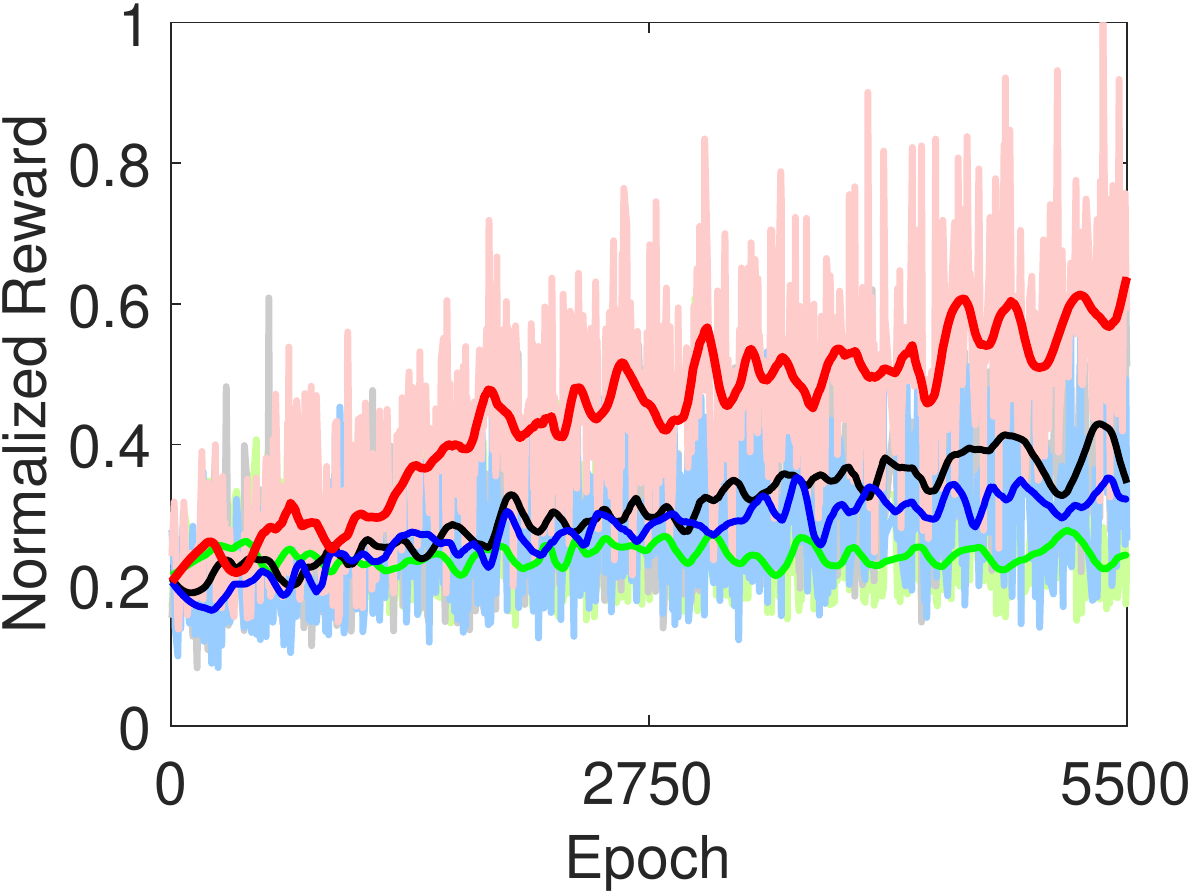}
        \label{fig:FOMDP}
    }
    \caption{Training performance in POMDP and FOMDP over training epochs.}
    \label{fig:Reward}
\end{figure}

\begin{figure}[t!]
    \centering
    \includegraphics[width=0.9\columnwidth]{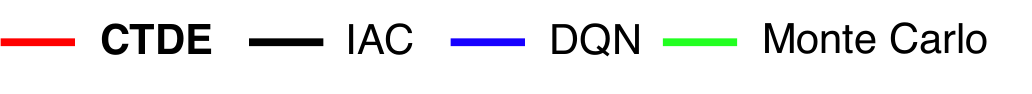}\\
    \subfigure[POMDP.]
    {
        \includegraphics[width=0.465\columnwidth]{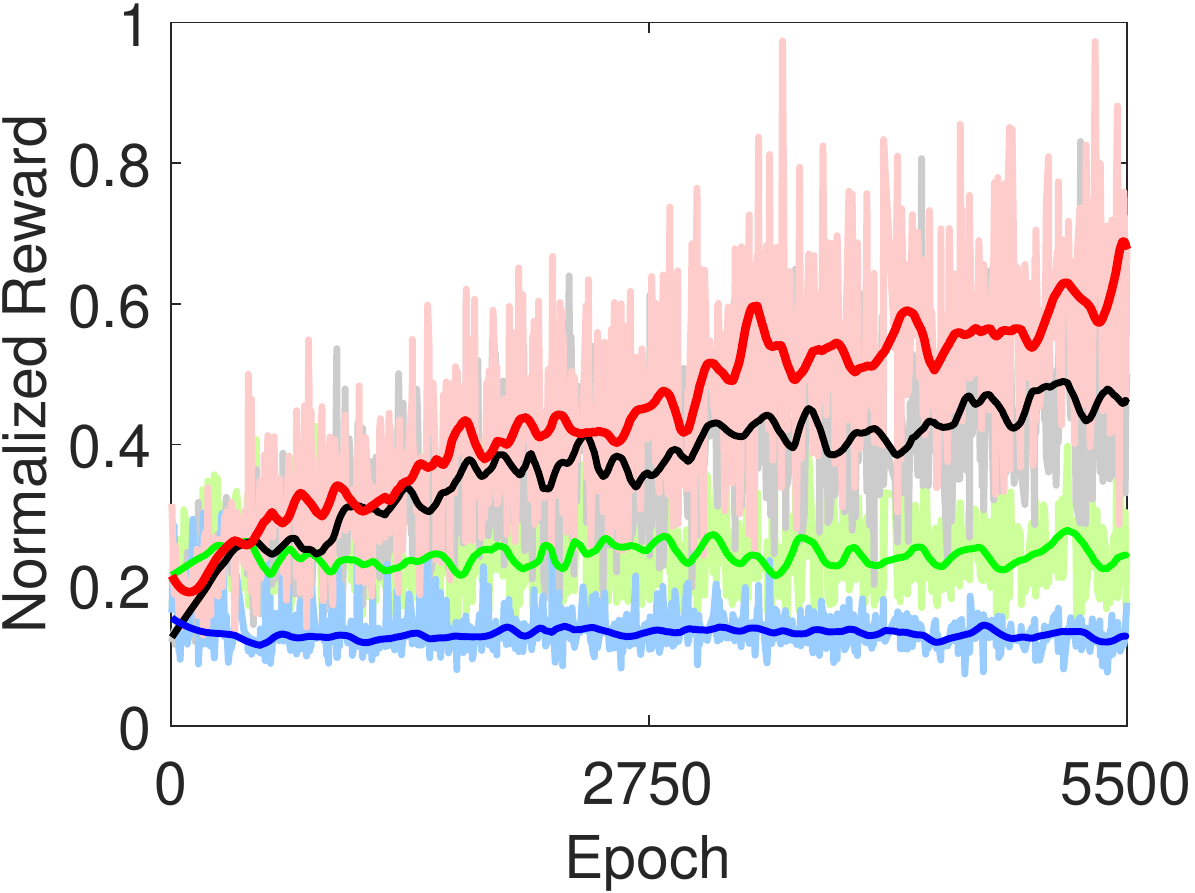}
        \label{fig:CTDE_POMDP}
    }
    \subfigure[FOMDP.]
    {
        \includegraphics[width=0.465\columnwidth]{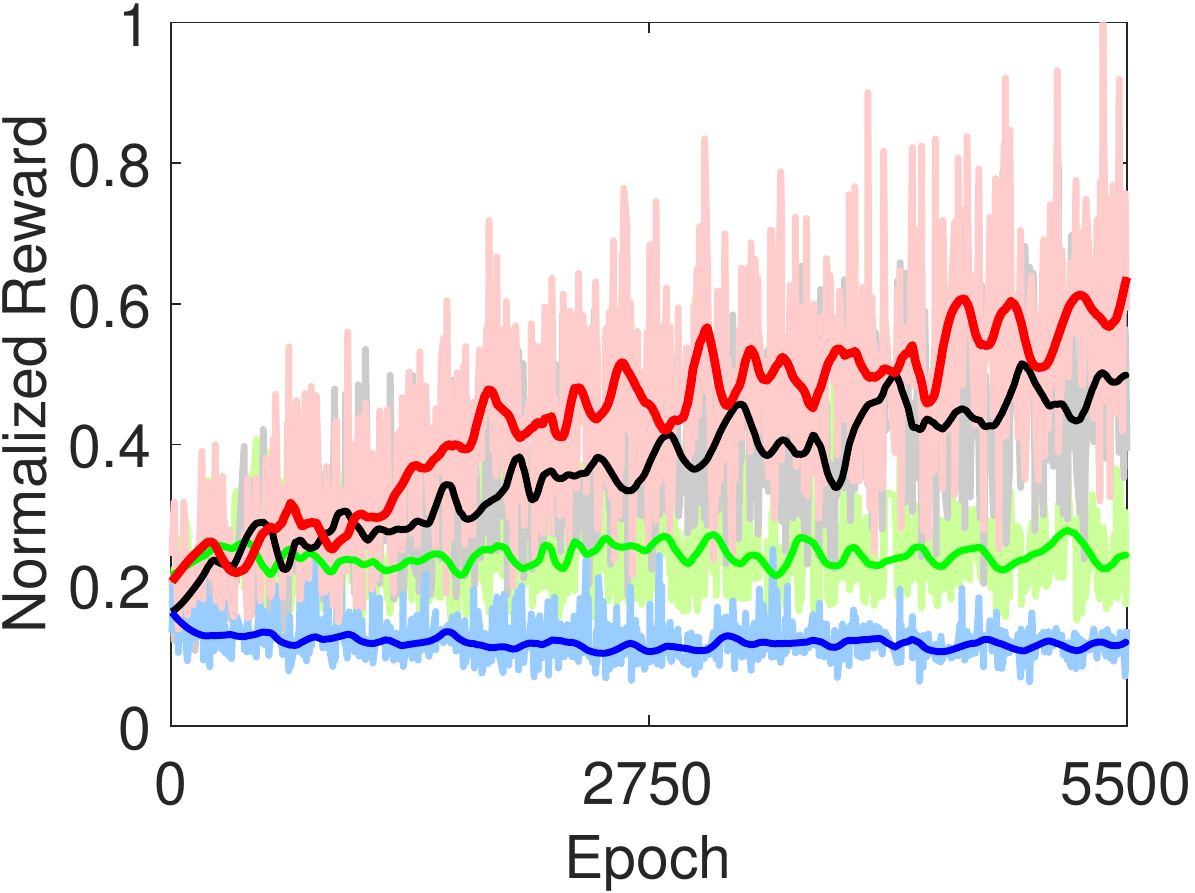}
        \label{fig:CTDE_FOMDP}
    }
    \caption{Training performance in POMDP and FOMDP over training epochs.}
    \label{fig:CTDE_Reward}
\end{figure}

Figs.~\ref{fig:Reward}--\ref{fig:CTDE_Reward} plot UAMs' obtained reward value while training their policies, and Table~\ref{table:reward}--\ref{table:CTDE_reward} summarize the final reward convergence value in both POMDP and FOMDP environments. Note that FOMDP is unrealistic due to UAM model's physical limitations.

\begin{table}[t!]
\centering         
\caption{Reward comparison to investigate the effect of CommNet.}
\resizebox{0.88\columnwidth}{!}{\begin{minipage}[h]{0.88\columnwidth}
\centering
\begin{tabularx}{1\linewidth}{l c c}

\toprule[1pt]

\multicolumn{3}{c}{\textbf{\circled{1} : Reward Convergence in POMDP and FOMDP}} \\
\midrule[.5pt]
\ \ \ \ \ \ \ \ Algorithm & POMDP  & FOMDP \\
\cmidrule(lr){1-1} \cmidrule(lr){2-2} \cmidrule(lr){3-3} 

\textbf{CommNet (Proposed)}
& $0.647$\; \tikz{
\draw[gray,line width=.3pt] (0,0) -- (1.1,0);
\draw[white, line width=0.01pt] (0,-2pt) -- (0,2pt);
\draw[black,line width=1pt] (0.572,0) -- (0.722,0);
\draw[black,line width=1pt] (0.572,-2pt) -- (0.572,2pt);
\draw[black,line width=1pt] (0.722,-2pt) -- (0.722,2pt);}
& $0.602$\; \tikz{
\draw[gray,line width=.3pt] (0,0) -- (1.1,0);
\draw[white, line width=0.01pt] (0,-2pt) -- (0,2pt);
\draw[black,line width=1pt] (0.527,0) -- (0.677,0);
\draw[black,line width=1pt] (0.527,-2pt) -- (0.527,2pt);
\draw[black,line width=1pt] (0.677,-2pt) -- (0.677,2pt);} \\

Hybrid
& $0.438$\; \tikz{
\draw[gray,line width=.3pt] (0,0) -- (1.1,0);
\draw[white, line width=0.01pt] (0,-2pt) -- (0,2pt);
\draw[black,line width=1pt] (0.363,0) -- (0.513,0);
\draw[black,line width=1pt] (0.363,-2pt) -- (0.363,2pt);
\draw[black,line width=1pt] (0.513,-2pt) -- (0.513,2pt);}
& $0.456$\; \tikz{
\draw[gray,line width=.3pt] (0,0) -- (1.1,0);
\draw[white, line width=0.01pt] (0,-2pt) -- (0,2pt);
\draw[black,line width=1pt] (0.381,0) -- (0.531,0);
\draw[black,line width=1pt] (0.381,-2pt) -- (0.381,2pt);
\draw[black,line width=1pt] (0.531,-2pt) -- (0.531,2pt);} \\

DNN
& $0.148$\; \tikz{
\draw[gray,line width=.3pt] (0,0) -- (1.1,0);
\draw[white, line width=0.01pt] (0,-2pt) -- (0,2pt);
\draw[black,line width=1pt] (0.073,0) -- (0.223,0);
\draw[black,line width=1pt] (0.073,-2pt) -- (0.073,2pt);
\draw[black,line width=1pt] (0.223,-2pt) -- (0.223,2pt);}
& $0.337$\; \tikz{
\draw[gray,line width=.3pt] (0,0) -- (1.1,0);
\draw[white, line width=0.01pt] (0,-2pt) -- (0,2pt);
\draw[black,line width=1pt] (0.262,0) -- (0.412,0);
\draw[black,line width=1pt] (0.262,-2pt) -- (0.262,2pt);
\draw[black,line width=1pt] (0.412,-2pt) -- (0.412,2pt);} \\

Monte Carlo
& $0.265$\; \tikz{
\draw[gray,line width=.3pt] (0,0) -- (1.1,0);
\draw[white, line width=0.01pt] (0,-2pt) -- (0,2pt);
\draw[black,line width=1pt] (0.19,0) -- (0.34,0);
\draw[black,line width=1pt] (0.19,-2pt) -- (0.19,2pt);
\draw[black,line width=1pt] (0.34,-2pt) -- (0.34,2pt);}
& $0.265$\; \tikz{
\draw[gray,line width=.3pt] (0,0) -- (1.1,0);
\draw[white, line width=0.01pt] (0,-2pt) -- (0,2pt);
\draw[black,line width=1pt] (0.19,0) -- (0.34,0);
\draw[black,line width=1pt] (0.19,-2pt) -- (0.19,2pt);
\draw[black,line width=1pt] (0.34,-2pt) -- (0.34,2pt);} \\

\bottomrule[1pt]
\end{tabularx}
\end{minipage}}
\label{table:reward}
\end{table}

Firstly, among benchmarks of the first group, UAMs in CommNet get the fastest reward-increasing rate in Fig.~\ref{fig:Reward}, and the most enormous reward value of $0.647$ in POMDP and $0.602$ in FOMDP as summarized in Table~\ref{table:reward}. Notably, only the proposed algorithm allows UAMs to get a higher reward in POMDP than FOMDP, regardless of information loss. On the other side, rewards of UAMs in Hybrid and DNN converge to $0.438$ and $0.148$, respectively. It means that these schemes are obviously susceptible to information loss in POMDP by getting $3.94\,\%$ and $56.1\%$ lower reward values than in FOMDP. In the case of DNNs utilizing only DNN-based UAMs, policy training failed with smaller rewards than \textit{Monte Carlo}. Indeed, the presence of a CommNet-based UAM helps serve robust air transportation service with limited environmental information in MADRL.

\begin{table}[t!]
\centering         
\caption{Reward comparison to investigate the effect of CTDE.}
\resizebox{0.83\columnwidth}{!}{\begin{minipage}[h]{0.83\columnwidth}
\centering
\begin{tabularx}{1\linewidth}{l c c}

\toprule[1pt]

\multicolumn{3}{c}{\textbf{\circled{2} : Reward Convergence in POMDP and FOMDP}} \\
\midrule[.5pt]
\ \ \ \ \ Algorithm & POMDP  & FOMDP \\
\cmidrule(lr){1-1} \cmidrule(lr){2-2} \cmidrule(lr){3-3} 

\textbf{CTDE (Proposed)}
& $0.647$\; \tikz{
\draw[gray,line width=.3pt] (0,0) -- (1.1,0);
\draw[white, line width=0.01pt] (0,-2pt) -- (0,2pt);
\draw[black,line width=1pt] (0.572,0) -- (0.722,0);
\draw[black,line width=1pt] (0.572,-2pt) -- (0.572,2pt);
\draw[black,line width=1pt] (0.722,-2pt) -- (0.722,2pt);}
& $0.602$\; \tikz{
\draw[gray,line width=.3pt] (0,0) -- (1.1,0);
\draw[white, line width=0.01pt] (0,-2pt) -- (0,2pt);
\draw[black,line width=1pt] (0.527,0) -- (0.677,0);
\draw[black,line width=1pt] (0.527,-2pt) -- (0.527,2pt);
\draw[black,line width=1pt] (0.677,-2pt) -- (0.677,2pt);} \\

IAC
& $0.440$\; \tikz{
\draw[gray,line width=.3pt] (0,0) -- (1.1,0);
\draw[white, line width=0.01pt] (0,-2pt) -- (0,2pt);
\draw[black,line width=1pt] (0.365,0) -- (0.515,0);
\draw[black,line width=1pt] (0.365,-2pt) -- (0.365,2pt);
\draw[black,line width=1pt] (0.515,-2pt) -- (0.515,2pt);}
& $0.481$\; \tikz{
\draw[gray,line width=.3pt] (0,0) -- (1.1,0);
\draw[white, line width=0.01pt] (0,-2pt) -- (0,2pt);
\draw[black,line width=1pt] (0.406,0) -- (0.556,0);
\draw[black,line width=1pt] (0.406,-2pt) -- (0.406,2pt);
\draw[black,line width=1pt] (0.556,-2pt) -- (0.556,2pt);} \\

DQN
& $0.131$\; \tikz{
\draw[gray,line width=.3pt] (0,0) -- (1.1,0);
\draw[white, line width=0.01pt] (0,-2pt) -- (0,2pt);
\draw[black,line width=1pt] (0.056,0) -- (0.206,0);
\draw[black,line width=1pt] (0.056,-2pt) -- (0.056,2pt);
\draw[black,line width=1pt] (0.206,-2pt) -- (0.206,2pt);}
& $0.111$\; \tikz{
\draw[gray,line width=.3pt] (0,0) -- (1.1,0);
\draw[white, line width=0.01pt] (0,-2pt) -- (0,2pt);
\draw[black,line width=1pt] (0.036,0) -- (0.186,0);
\draw[black,line width=1pt] (0.036,-2pt) -- (0.036,2pt);
\draw[black,line width=1pt] (0.186,-2pt) -- (0.186,2pt);} \\

Monte Carlo
& $0.265$\; \tikz{
\draw[gray,line width=.3pt] (0,0) -- (1.1,0);
\draw[white, line width=0.01pt] (0,-2pt) -- (0,2pt);
\draw[black,line width=1pt] (0.19,0) -- (0.34,0);
\draw[black,line width=1pt] (0.19,-2pt) -- (0.19,2pt);
\draw[black,line width=1pt] (0.34,-2pt) -- (0.34,2pt);}
& $0.265$\; \tikz{
\draw[gray,line width=.3pt] (0,0) -- (1.1,0);
\draw[white, line width=0.01pt] (0,-2pt) -- (0,2pt);
\draw[black,line width=1pt] (0.19,0) -- (0.34,0);
\draw[black,line width=1pt] (0.19,-2pt) -- (0.19,2pt);
\draw[black,line width=1pt] (0.34,-2pt) -- (0.34,2pt);} \\

\bottomrule[1pt]
\end{tabularx}
\end{minipage}}
\label{table:CTDE_reward}
\end{table}

Besides, in Fig.~\ref{fig:CTDE_Reward} and Table~\ref{table:CTDE_reward}, CTDE outperforms the other DRL benchmarks of the second group in POMDP/FOMDP. UAMs trained by IAC and DQN receive reward values of $0.440$ and $0.131$ in POMDP. Especially, DQN fails to learn UAMs in both MDP environments by attaining lower reward values of $0.131$ and $0.111$ than \textit{Monte Carlo}.
According to~\cite{malysheva2018deep}, a limitation of utilizing DQN-based DRL in multi-agent settings is the reduced effectiveness of experience replay. Unlike single-agent scenarios, the same action executed in an identical state may yield varying outcomes, contingent on other agents' actions. Consequently, agents that have trained neural network parameters through DQN in MARL may struggle to optimize their parameters efficiently. This occurrence leads to agents collectively exhibiting alike inappropriate behavior when encountering similar environmental information. As a result, adopting random actions, as observed in \textit{Monte Carlo}, could yield higher rewards than DQN due to the wider action decisions of agents.
Next, among DRL algorithms that successfully trained policies, only CTDE results in higher reward values in POMDP than in FOMDP. In a nutshell, the training performance of the proposed algorithm with CommNet and CTDE is corroborated by showing the powerful reward convergence ability despite information loss.

\begin{figure}[t!]
    \centering
    \includegraphics[width=\columnwidth]{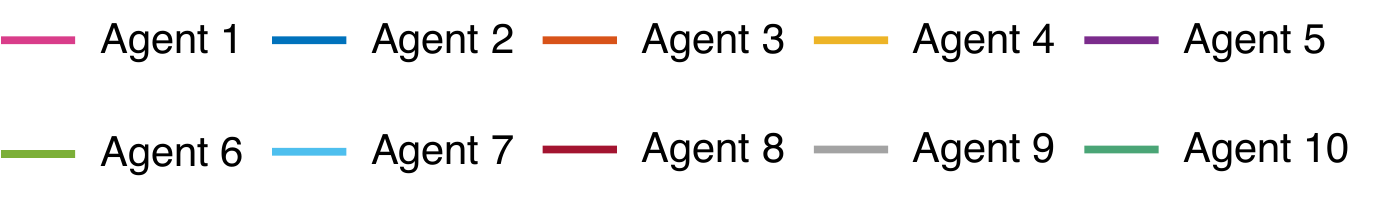}\\
    \includegraphics[width=\linewidth]{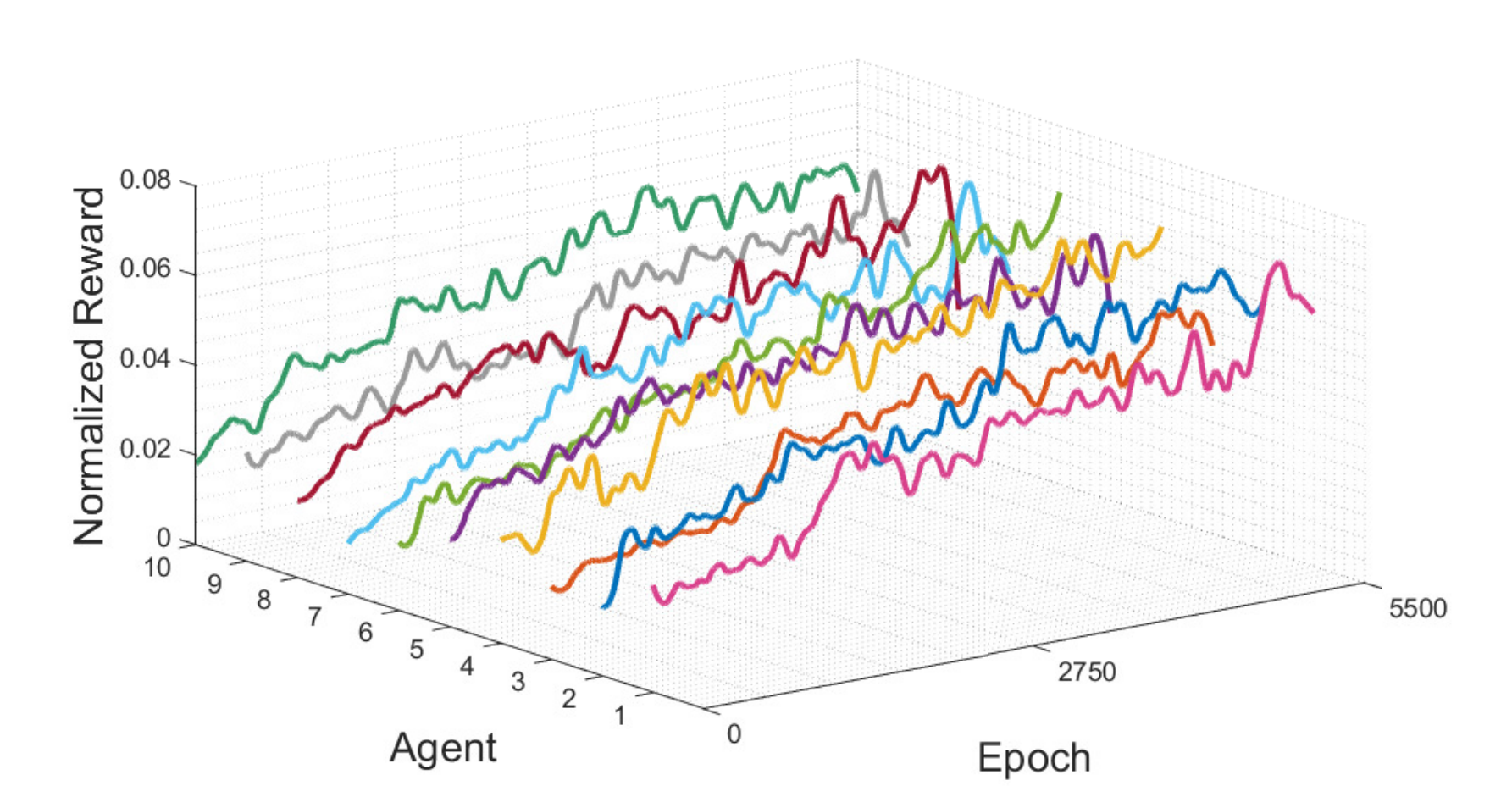}
    \caption{Every agent's learning progress with the proposed algorithm in the POMDP environment over training epochs.}
    \label{fig:Agentwise_reward}
\end{figure}

\subsubsection{Trained Trajectories}\label{sec:V-C-2}

\begin{figure*}[t!]
    \centering
    \subfigure[Agent 1.]
    {
        \includegraphics[width=0.166\columnwidth]{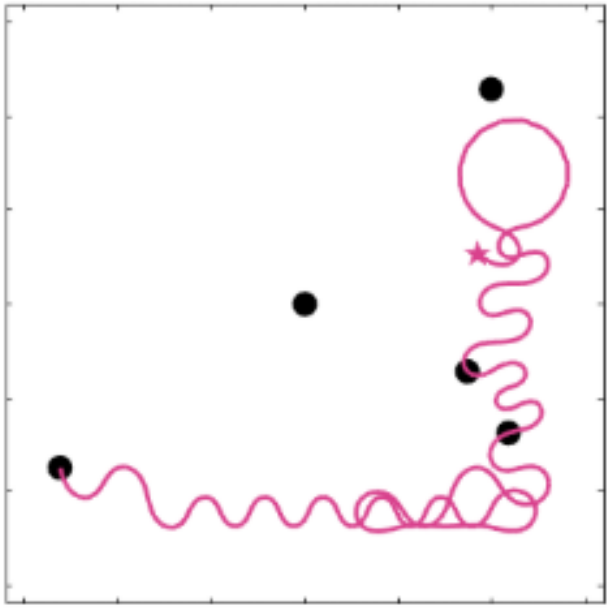}
        \label{fig:CommNet_agent1}
    }
    \subfigure[Agent 2.]
    {
        \includegraphics[width=0.166\columnwidth]{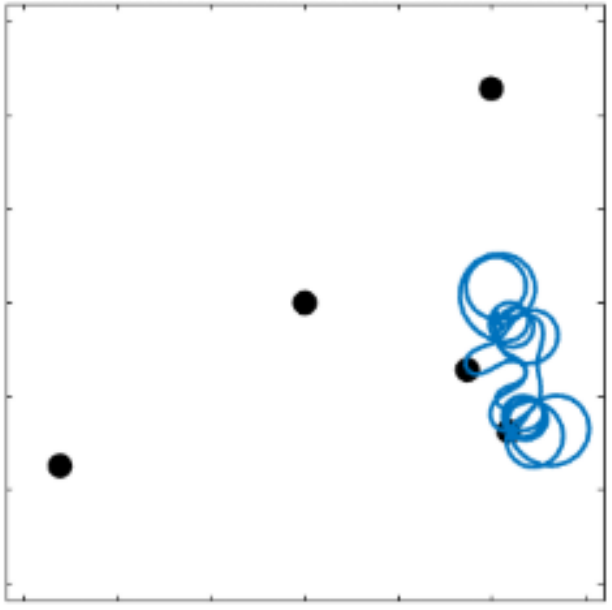}
        \label{fig:CommNet_agent2}
    }
    \subfigure[Agent 3.]
    {
        \includegraphics[width=0.166\columnwidth]{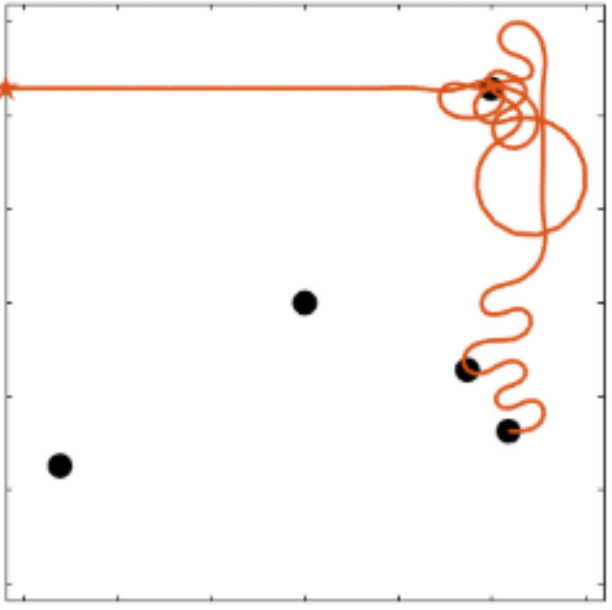}
        \label{fig:CommNet_agent3}
    }
    \subfigure[Agent 4.]
    {
        \includegraphics[width=0.166\columnwidth]{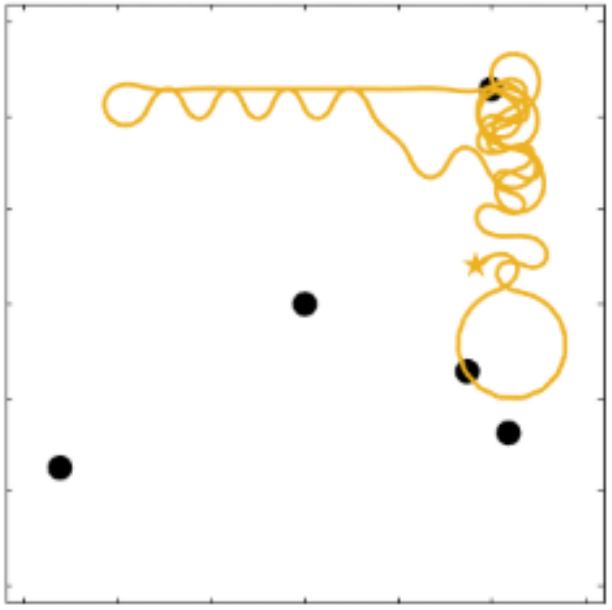}
        \label{fig:CommNet_agent4}
    }
    \subfigure[Agent 5.]
    {
        \includegraphics[width=0.166\columnwidth]{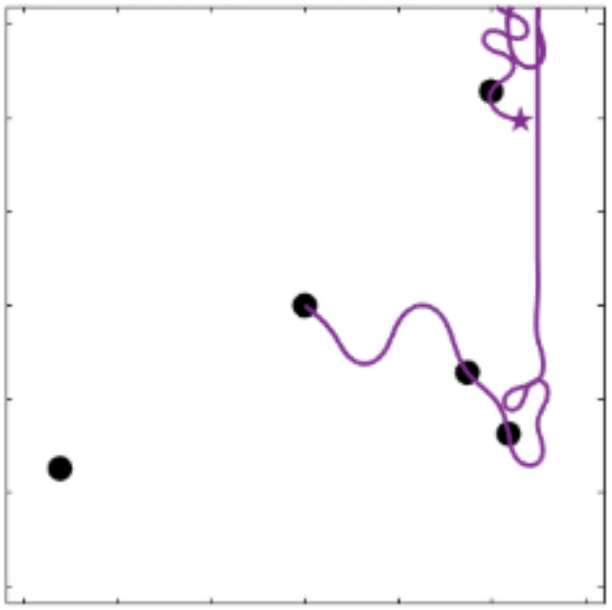}
        \label{fig:CommNet_agent5}
    }
    \subfigure[Agent 6.]
    {
        \includegraphics[width=0.166\columnwidth]{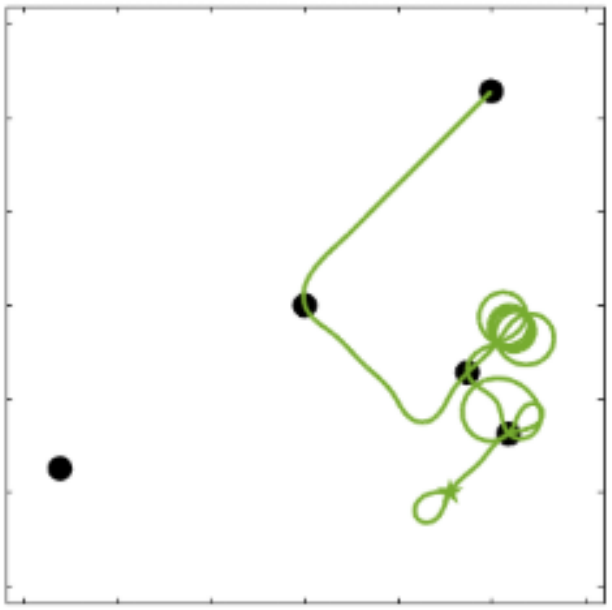}
        \label{fig:CommNet_agent6}
    }
    \subfigure[Agent 7.]
    {
        \includegraphics[width=0.166\columnwidth]{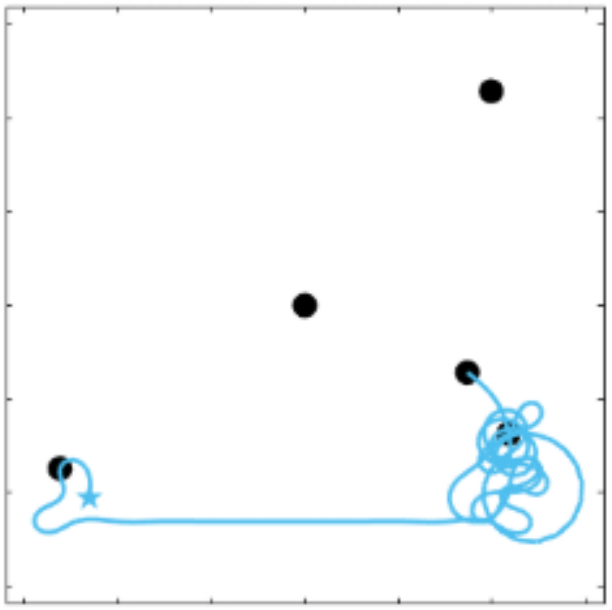}
        \label{fig:CommNet_agent7}
    }
    \subfigure[Agent 8.]
    {
        \includegraphics[width=0.166\columnwidth]{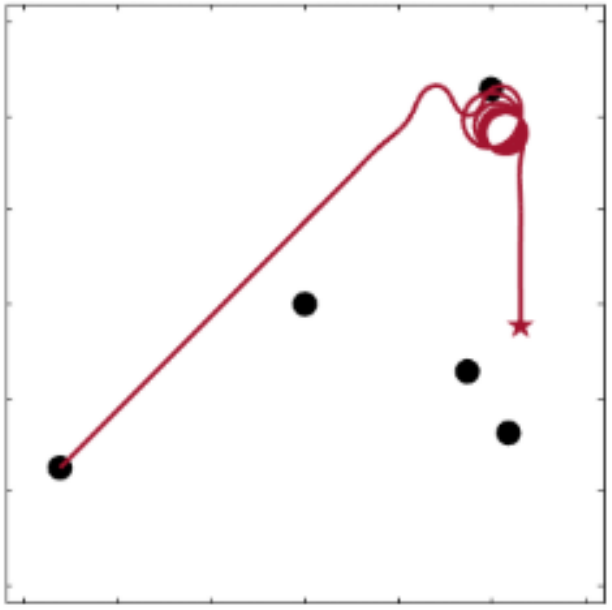}
        \label{fig:CommNet_agent8}
    }
    \subfigure[Agent 9.]
    {
        \includegraphics[width=0.166\columnwidth]{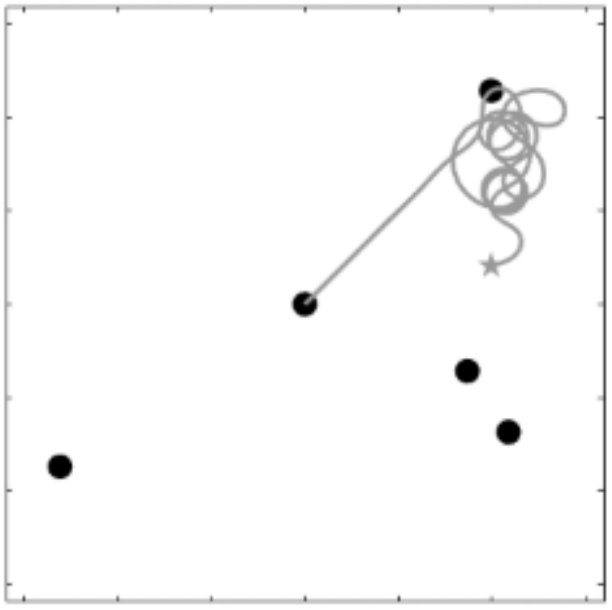}
        \label{fig:CommNet_agent9}
    }
    \subfigure[Agent 10.]
    {
        \includegraphics[width=0.166\columnwidth]{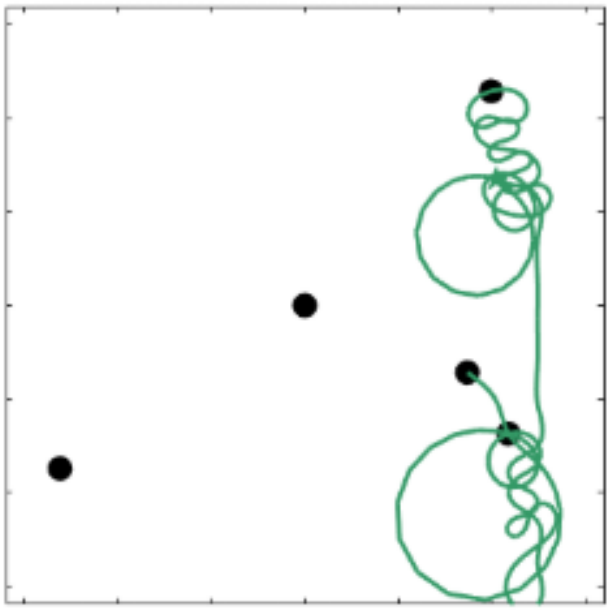}
        \label{fig:CommNet_agent10}
    }
    \caption{The trained trajectories of UAMs trained with \textbf{CommNet} in the progress of episodes at the end of the policy training.}
    \label{fig:Trajectory_CommNet}

    \subfigure[Agent 1.]
    {
        \includegraphics[width=0.166\columnwidth]{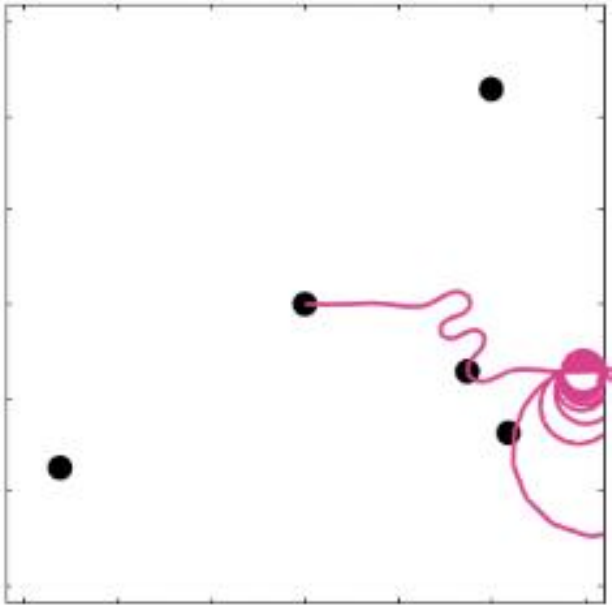}
        \label{fig:DNN_agent1}
    }
    \subfigure[Agent 2.]
    {
        \includegraphics[width=0.166\columnwidth]{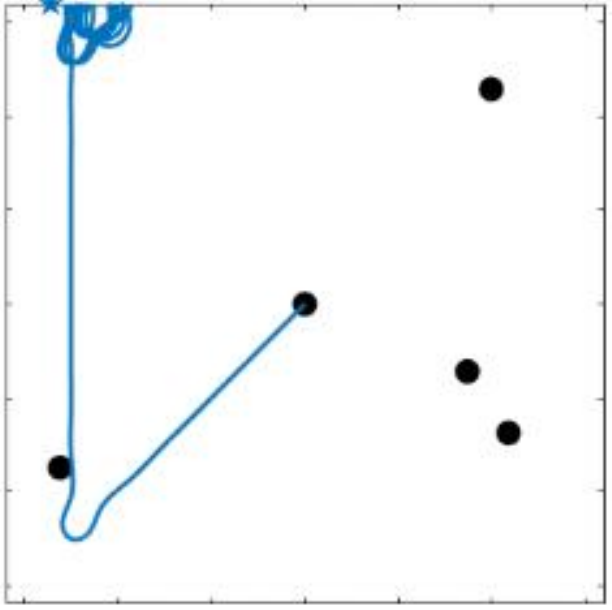}
        \label{fig:DNN_agent2}
    }
    \subfigure[Agent 3.]
    {
        \includegraphics[width=0.166\columnwidth]{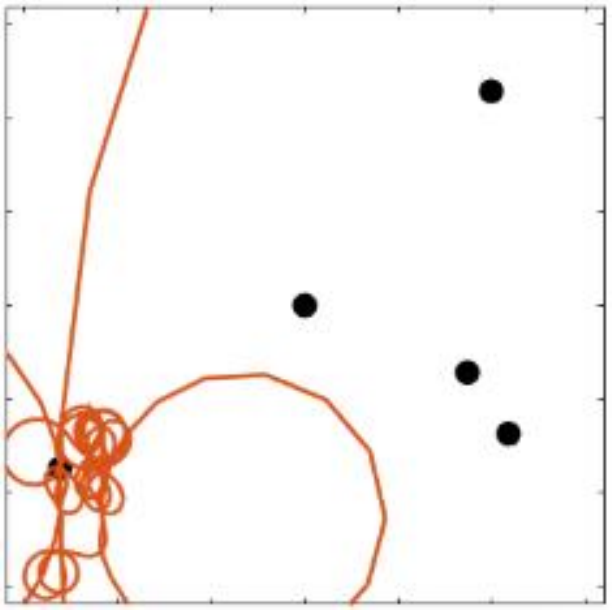}
        \label{fig:DNN_agent3}
    }
    \subfigure[Agent 4.]
    {
        \includegraphics[width=0.166\columnwidth]{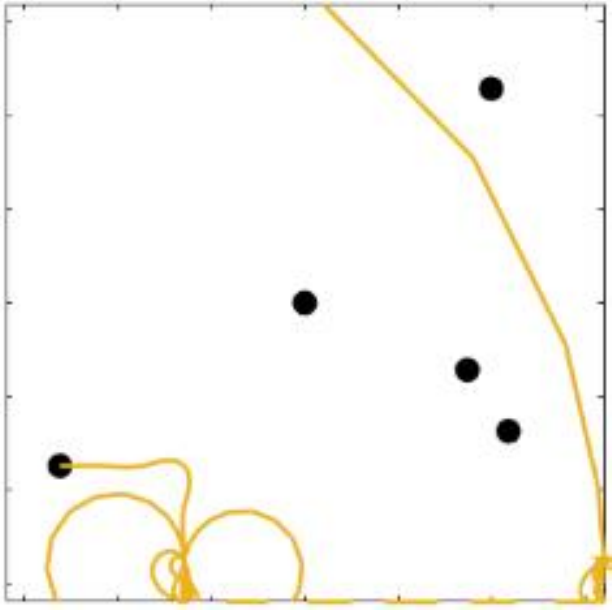}
        \label{fig:DNN_agent4}
    }
    \subfigure[Agent 5.]
    {
        \includegraphics[width=0.166\columnwidth]{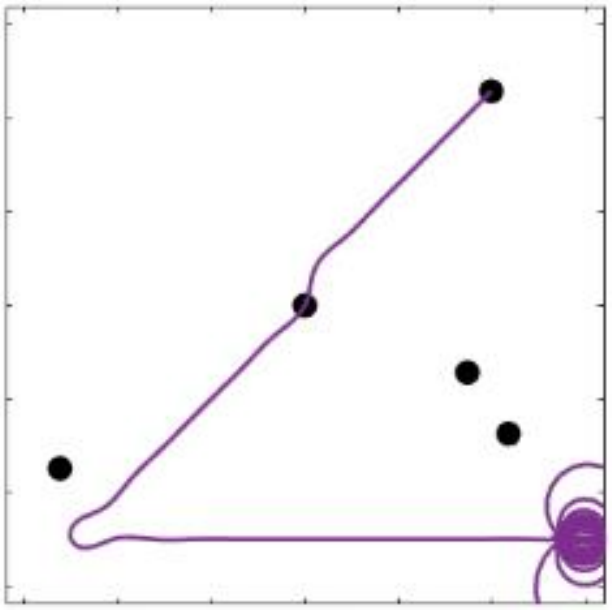}
        \label{fig:DNN_agent5}
    }
    \subfigure[Agent 6.]
    {
        \includegraphics[width=0.166\columnwidth]{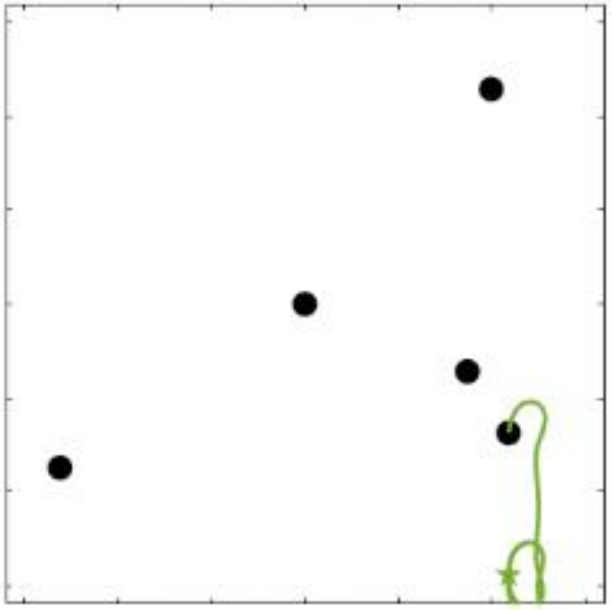}
        \label{fig:DNN_agent6}
    }
    \subfigure[Agent 7.]
    {
        \includegraphics[width=0.166\columnwidth]{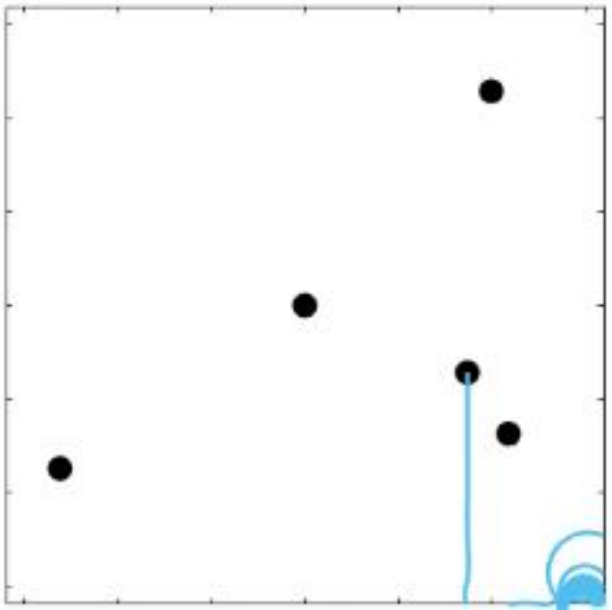}
        \label{fig:DNN_agent7}
    }
    \subfigure[Agent 8.]
    {
        \includegraphics[width=0.166\columnwidth]{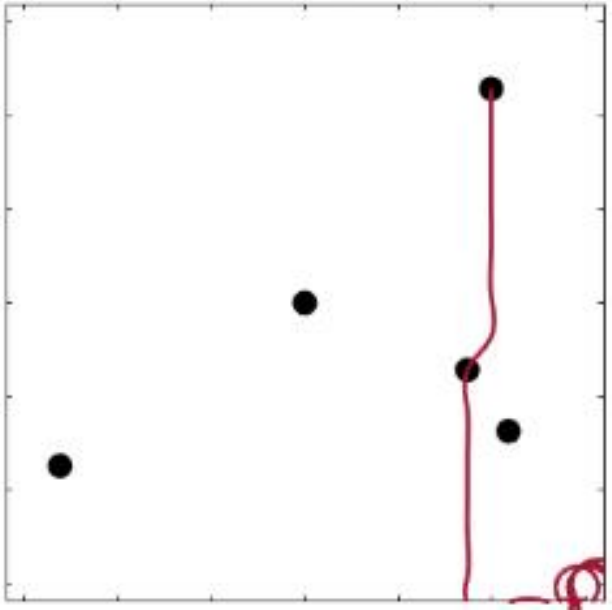}
        \label{fig:DNN_agent8}
    }
    \subfigure[Agent 9.]
    {
        \includegraphics[width=0.166\columnwidth]{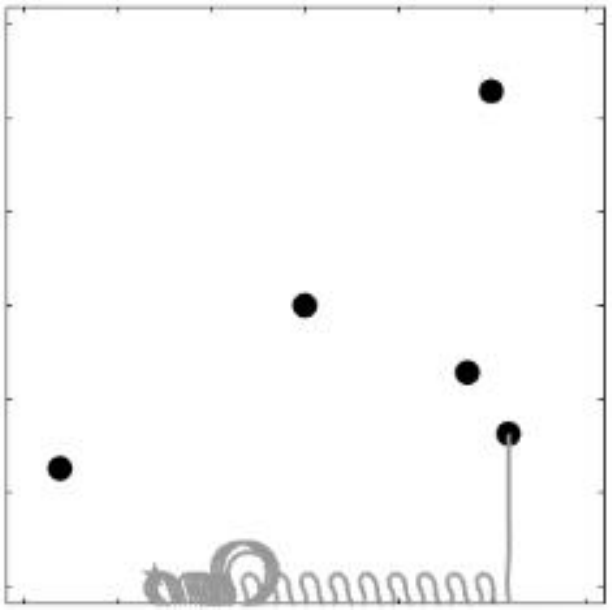}
        \label{fig:DNN_agent9}
    }
    \subfigure[Agent 10.]
    {
        \includegraphics[width=0.166\columnwidth]{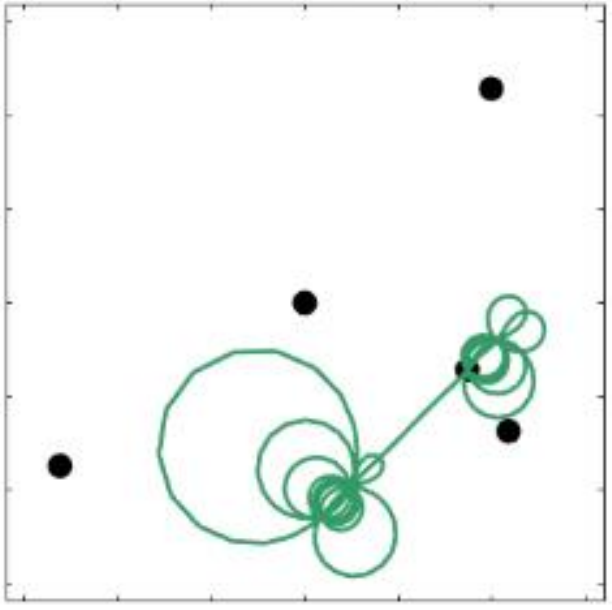}
        \label{fig:DNN_agent10}
    }
    \caption{The trained trajectories of UAMs trained with \textbf{DNN} in the progress of episodes at the end of the policy training.}
    \label{fig:Trajectory}
\end{figure*}

This section intuitively investigates the CommNet's training performance with Fig.~\ref{fig:Trajectory}, which exhibits the trajectories of UAMs trained by CommNet and DNN benchmarks in the POMDP environment. At the last training epoch, it can be seen that CommNet-based UAMs do not deviate much from the considered system map. In contrast, DNN-based UAMs have unnecessary trajectories in terms of air transportation service provision (Agents 4 and 10) or are isolated in a limited area (Agent 3). Some agents are even isolated in areas where vertiports do not exist (Agents 2, 5--6, and 7--9). In addition, it can be seen that UAMs have non-linear trajectories for transporting passengers to target vertiports. 
UAM's linear trajectories are likely to provide more ideal air transport services than non-linear trajectories, but not because of `safety'~\cite{weibel2011establishing}. In aviation systems, there are horizontal and vertical separations to prevent collisions between aircraft~\cite{ATC}. All aircraft must maintain a safe distance to the separation criteria. Especially in unmanned aerial system (UAS) operating without a pilot, the plane must detect and resolve potential collisions on its own. In the case of UAM, since many UAMs fly simultaneously at low altitudes in urban areas, separation must be managed more strictly than normal aircraft. In particular, near congested vertiport where many UAMs take off and landing, since they must fly while avoiding other UAMs, non-linear trajectories are more ideal for transporting large numbers of passengers with safety considerations.


In Fig.~\ref{fig:Trajectory_CommNet}, transportation services are relatively concentrated in $A$, $C$, $D$, and $E$ except for $B$ (FORT WORTH). This is the result of considering actual passenger demand. In Fig.~\ref{fig:vertiport_map}, the traffic circle connecting Dallas, Texas and Frisco is very busy. In 2018, 27.2 million people visited Dallas, and $40\%$ of Frisco Collin's residents commute there~\cite{Uber_1}. So traffic can achieve economies of scale, and there are DFW Airport ($A$) and Dallas Love Field Airport($D$), which are the hubs of air traffic in the heart of the United States, in the metropolitan area. Demand is concentrated at airports with high traffic volumes or at FRISCO between DOWNTOWN DALLAS, where commuting volume is high. In the vertiport map in Fig.~\ref{fig:vertiport_map}, the vertiport layout is designed based on these actual passenger demands, so the experimental results shown in Fig~.\ref{fig:Trajectory_CommNet} are also very reasonable.



\begin{figure}
    \centering
    \ \ \ \ \ \ \ \ \includegraphics[width=0.45\columnwidth]{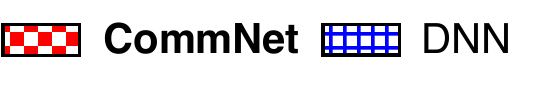}\\
    \includegraphics[width=0.9\linewidth]{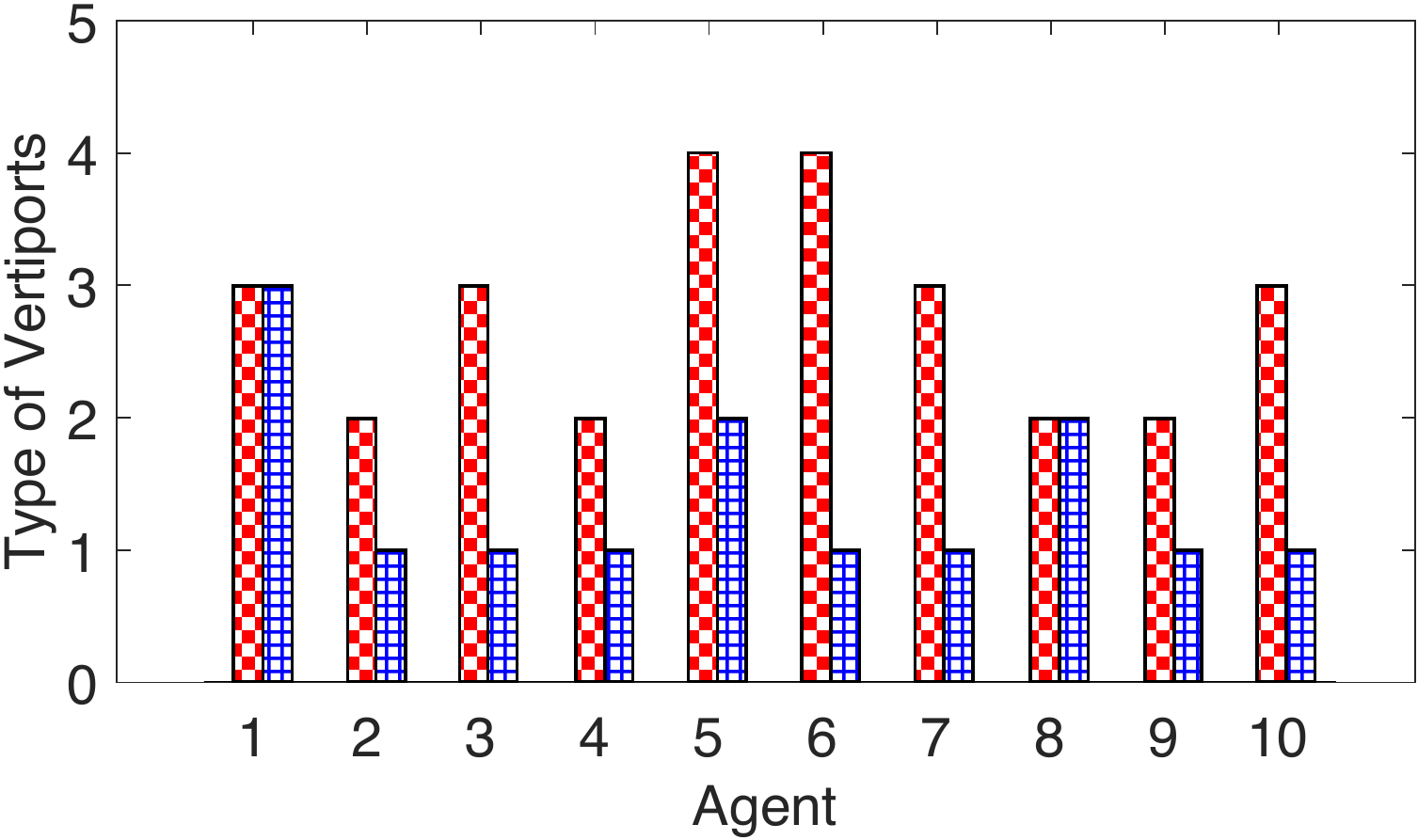}
    \caption{Type of vertiport where each UAM lands in Fig.~\ref{fig:Trajectory}.}
    \label{fig:train_bar}
\end{figure}

More exact values are organized in Fig.~\ref{fig:train_bar}, where the total type of vertiport CommNet-based UAMs landed on is twice as DNN-based UAMs. In addition, every CommNet-based UAM landed on various vertiports on average $2.3$ times more than a DNN-based UAM. It is also confirmed that all CommNet-based UAMs successfully transported passengers by visiting at least more than two different vertiports, with the maximum types of vertiports being visited even reaching four (Agents 5 and 6). However, in DNN, only three UAMs (Agents 1, 5, and 8) landed on more than two different vertiports, and the maximum type of visited vertiport is less than the CommNet.

\subsubsection{Equity in Policy Training}\label{sec:V-C-3}

A MARL algorithm is meaningless if at least one agent fails to learn the policy, even if it goes through the training process and achieves a high total reward value. Therefore, to ensure that all UAMs have equally well-trained their policies, Fig.~\ref{fig:Agentwise_reward} provides the convergence behavior of every UAM trained by the proposed CommNet/CTDE-based MADRL algorithm. All UAMs show similar tendencies in learning policies, starting with an average reward value of $0.0203$ ranging from $0.0119$ to $0.0274$. At the end of the training, all UAMs get reward values of $0.0348$--$0.0765$. Here, the average value is $0.0564$, which is $53.01\,\%$ higher than the average reward value of \textit{Monte Carlo} ($\approx0.0265$) summarized in Tables~\ref{table:reward} and~\ref{table:CTDE_reward}. As a result, it is confirmed that all UAMs trained by the proposed algorithm have equitable training performance.

\subsection{Feasibility of the Proposed Air Transportation System}\label{sec:V-D}
This section evaluates the feasibility of the proposed air transportation network in versatile aspects with $100$ inference times in the POMDP environment. For feasibility studies, this paper adopts the quality of air transportation service, UAMs' unbiased performance, and energy management as evaluation indicators.

\subsubsection{Service Quality}\label{sec:V-D-1}

\begin{figure}[t!]
    \centering
    \includegraphics[width=\columnwidth]{Figure/Comp1.pdf}\\
    \subfigure[Number of services.]
    {
        \includegraphics[width=0.295\columnwidth]{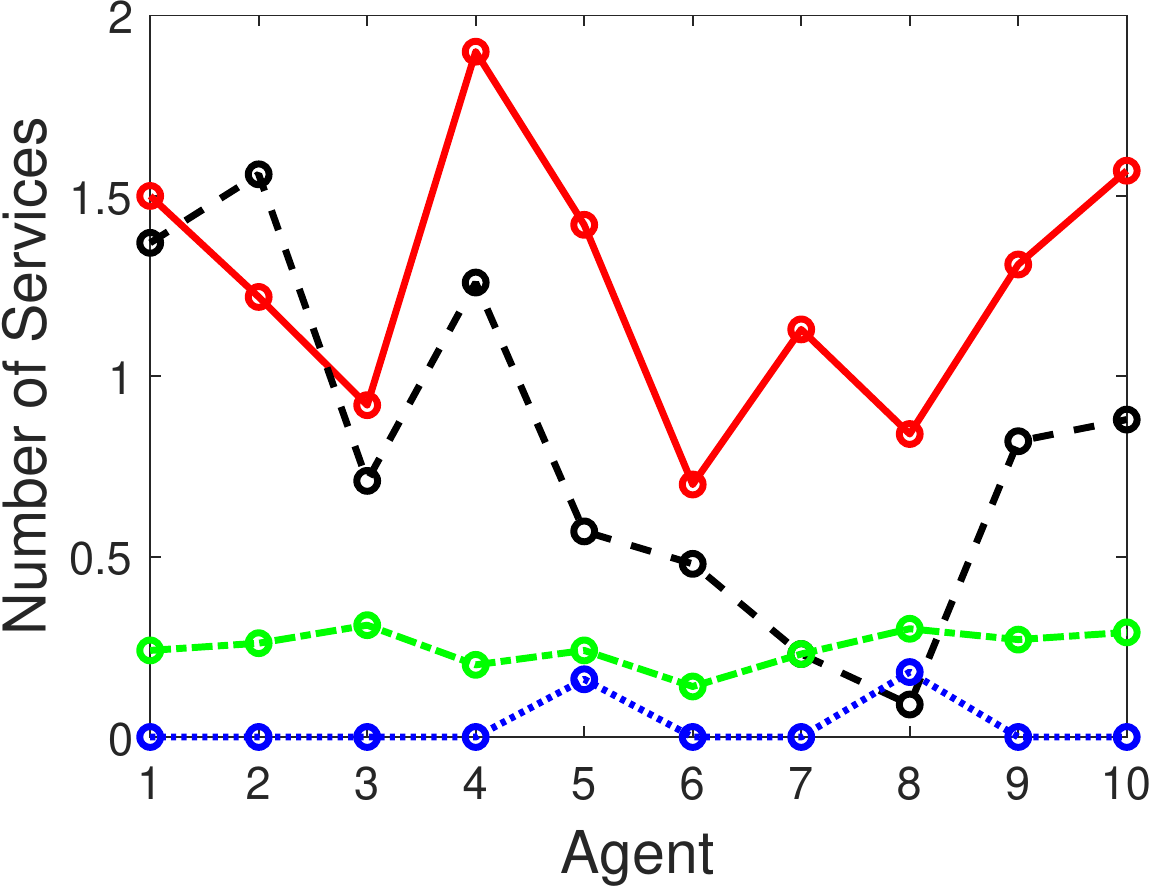}
        \label{fig:NumUsers}
    }
    \subfigure[Number of landings.]
    {
        \includegraphics[width=0.295\columnwidth]{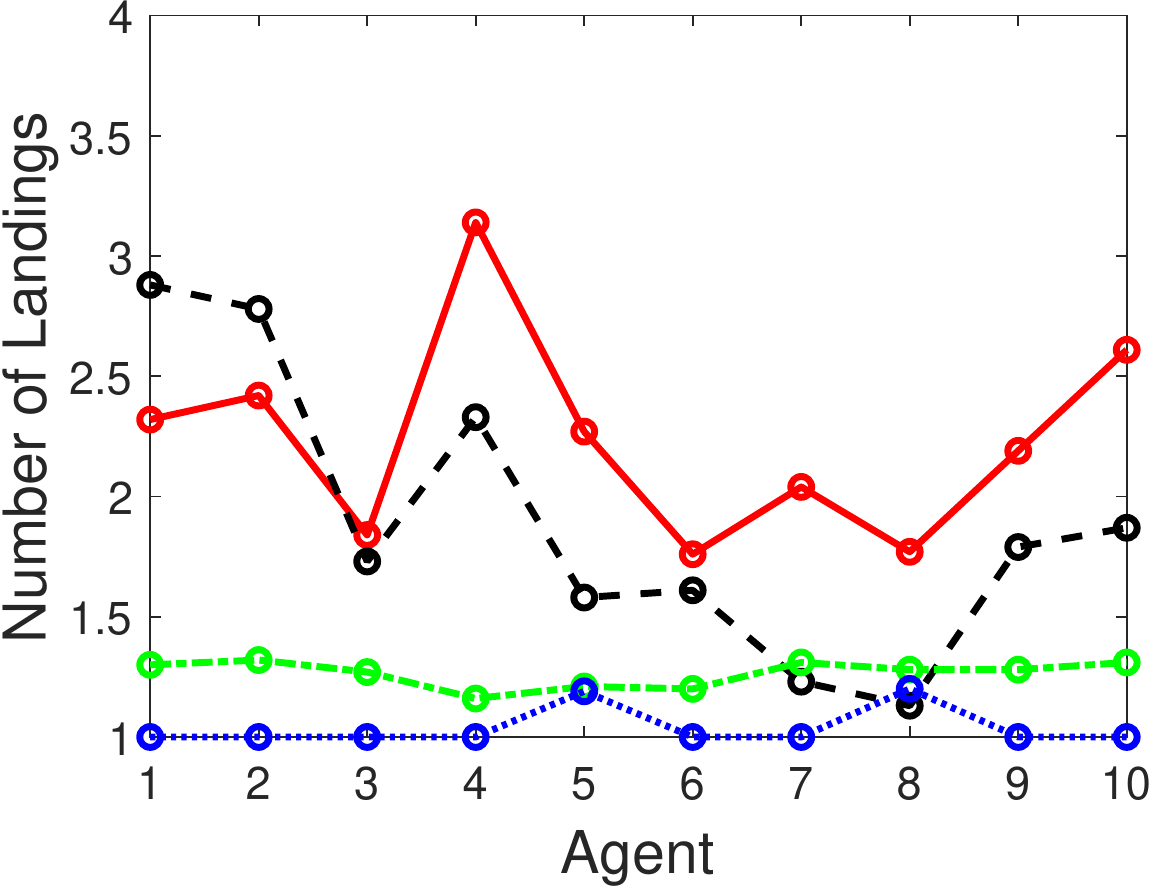}
        \label{fig:NumVertiport}
    }
    \subfigure[Type of vertiports.]
    {
        \includegraphics[width=0.295\columnwidth]{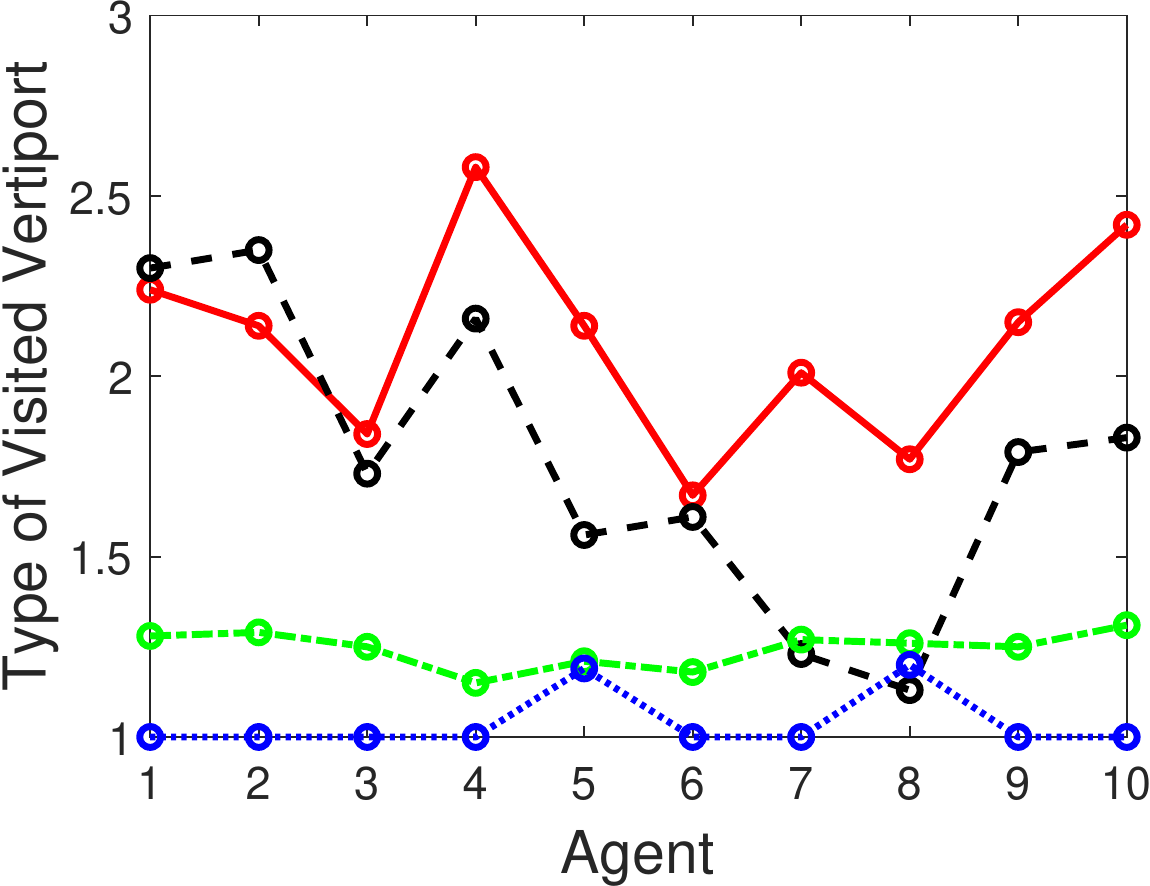}
        \label{fig:TypeVertiport}
    }\\
    \caption{Average validation performance in $100$ inference processes of the first benchmark group.}
    \label{fig:Inference}
\end{figure}

\begin{figure}[t!]
    \centering
    \includegraphics[width=0.9\columnwidth]{Figure/Comp2.pdf}\\
    \subfigure[Number of services.]
    {
        \includegraphics[width=0.295\columnwidth]{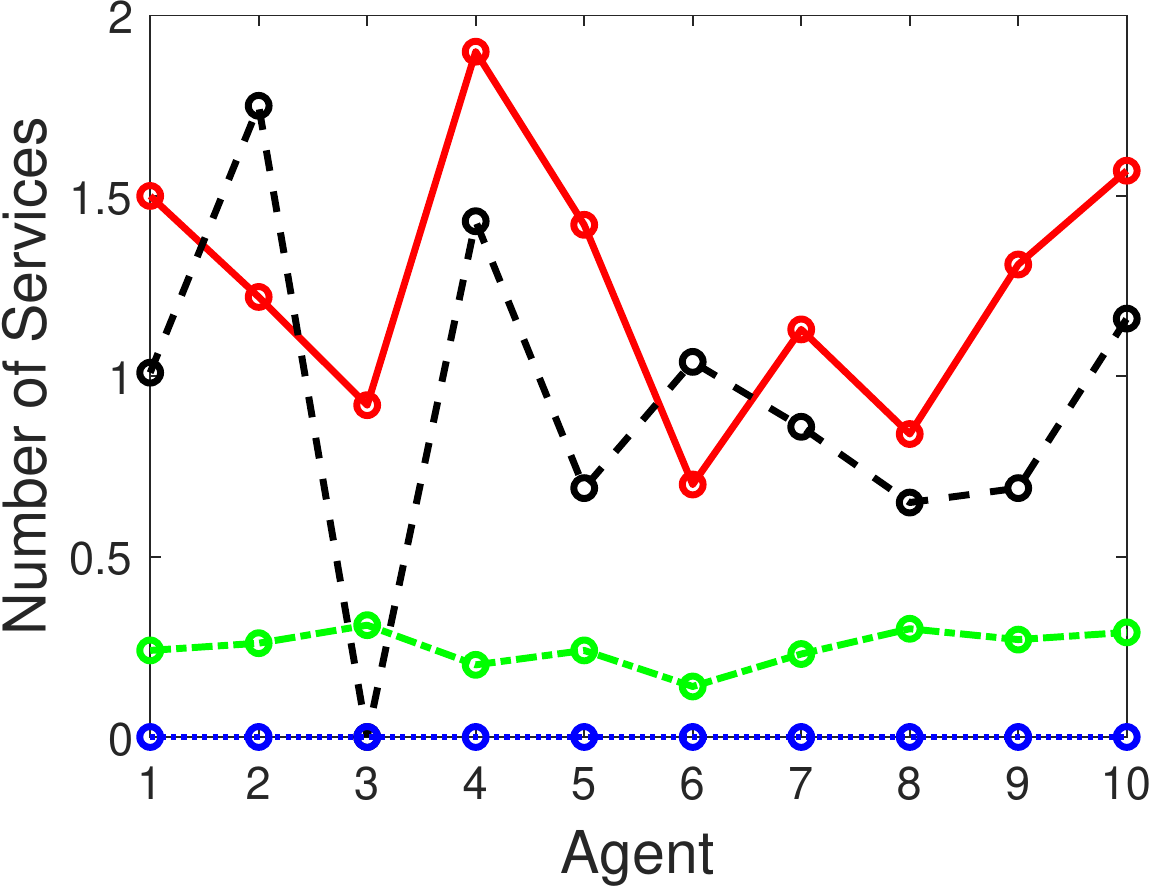}
        \label{fig:CTDE_NumUsers}
    }
    \subfigure[Number of landings.]
    {
        \includegraphics[width=0.295\columnwidth]{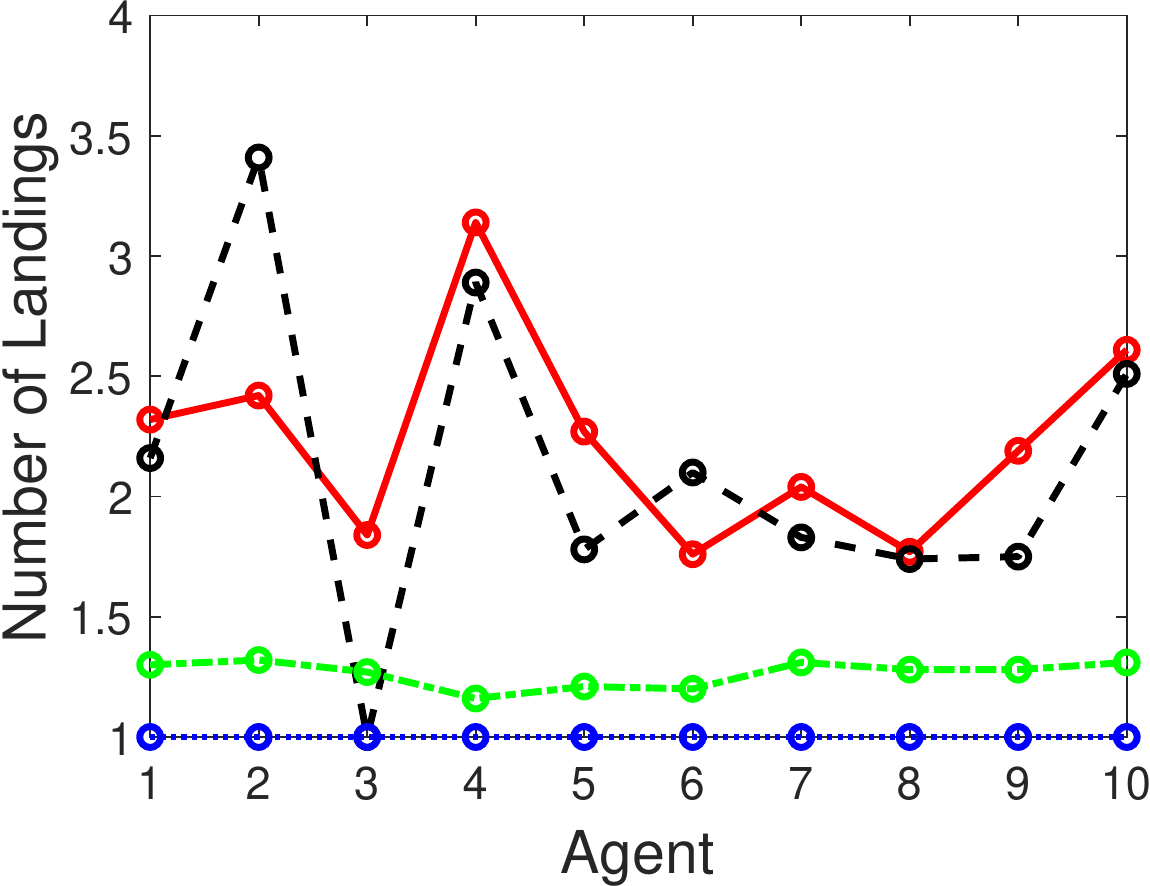}
        \label{fig:CTDE_NumVertiport}
    }
    \subfigure[Type of vertiport.]
    {
        \includegraphics[width=0.295\columnwidth]{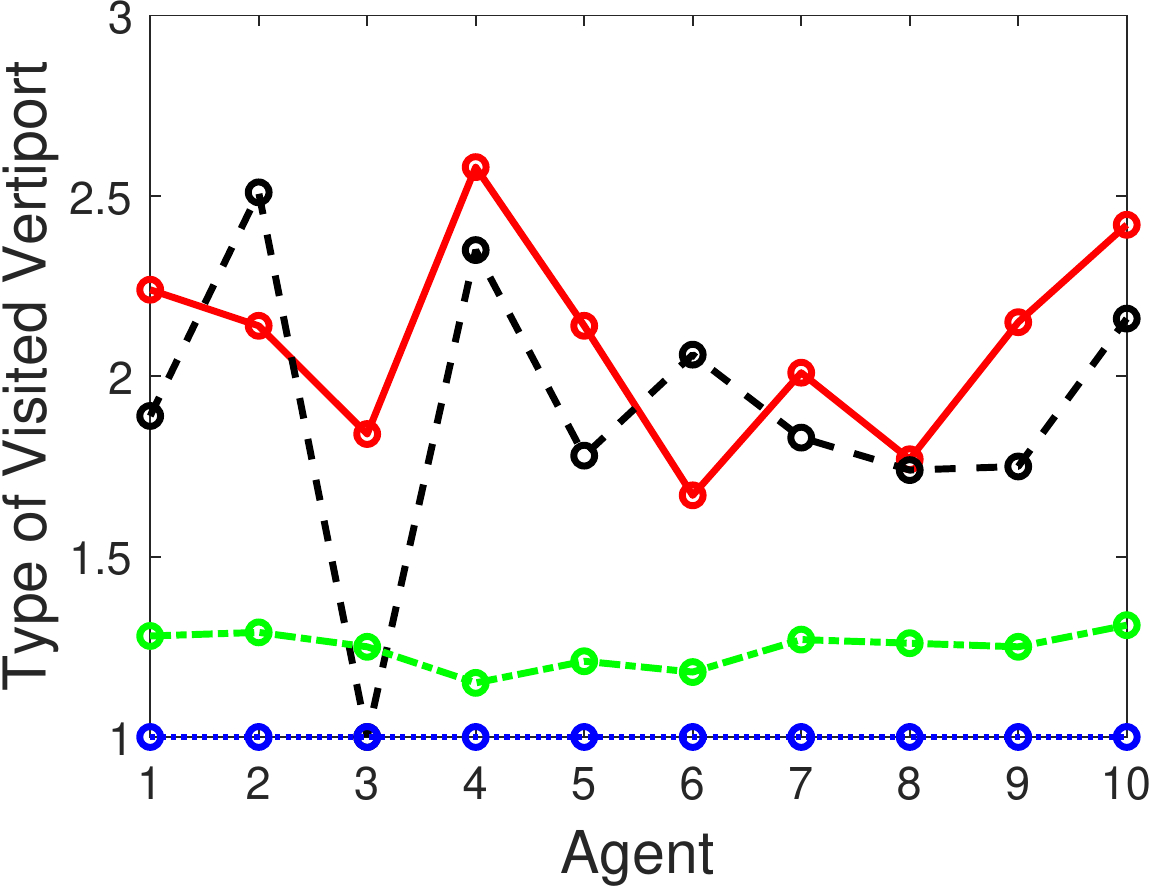}
        \label{fig:CTDE_TypeVertiport}
    }\\
    \caption{Average validation performance in $100$ inference processes of the second benchmark group.}
    \label{fig:CTDE_Inference}
\end{figure}

\begin{table}[t!]
    \centering          
    \caption{Performance Comparison for Validation of Trained Policy.}
    \resizebox{1\columnwidth}{!}{\begin{minipage}[h]{1\columnwidth}
    \centering
    \label{tab:validation}
    \newcolumntype{R}{>{\raggedleft\arraybackslash}X}
\begin{tabularx}{1\linewidth}{l l l l}
    \toprule[1pt]
    \multicolumn{4}{c}{\textbf{\circled{3} : Transportation Service Quality in Inference Phase in POMDP}} \\
    \midrule[.5pt]
    \ \textit{Benchmark} & \ \ \ \ \ Fig.\ref{fig:NumUsers} & \ \ \ \ \ Fig.\ref{fig:NumVertiport} & \ \ \ \ \ Fig.\ref{fig:TypeVertiport} \\     
    \cmidrule(lr){1-1} \cmidrule(lr){2-2} \cmidrule(lr){3-3} \cmidrule(lr){4-4}
    \textbf{Proposed} & {$12.51$} \hspace{1pt}  \tikz{
        \fill[fill=color1] (0.0,0) rectangle (1,0.2);
        \fill[pattern=north west lines, pattern color=black!30!color1] (0.0,0) rectangle (1,0.2);
    
    } & {$2.236$} \hspace{1pt}  \tikz{
        \fill[fill=color1] (0.0,0) rectangle (1,0.2);
        \fill[pattern=north west lines, pattern color=black!30!color1] (0.0,0) rectangle (1,0.2);
    } & {$2.096$} \hspace{1pt}  \tikz{
        \fill[fill=color1] (0.0,0) rectangle (1,0.2);
        \fill[pattern=north west lines, pattern color=black!30!color1] (0.0,0) rectangle (1,0.2);
    } \\
    Hybrid & {$7.970$} \hspace{1pt}  \tikz{
        \fill[fill=color2] (0.0,0) rectangle (0.637,0.2);
        \fill[pattern=north west lines, pattern color=black!30!color2] (0.0,0) rectangle (0.637,0.2);
    } & {$1.893$} \hspace{1pt}  \tikz{
        \fill[fill=color2] (0.0,0) rectangle (0.8465,0.2);
        \fill[pattern=north west lines, pattern color=black!30!color2] (0.0,0) rectangle (0.8465,0.2);
    } & {$1.769$} \hspace{1pt}  \tikz{
        \fill[fill=color2] (0.0,0) rectangle (0.844,0.2);
        \fill[pattern=north west lines, pattern color=black!30!color2] (0.0,0) rectangle (0.844,0.2);
    } \\
    DNN & {$0.340$} \hspace{1pt}  \tikz{
        \fill[fill=color3] (0.0,0) rectangle (0.027,0.2);
        \fill[pattern=north west lines, pattern color=black!30!color3] (0.0,0) rectangle (0.027,0.2);
    } & {$1.039$} \hspace{1pt}  \tikz{
        \fill[fill=color3] (0.0,0) rectangle (0.4645,0.2);
        \fill[pattern=north west lines, pattern color=black!30!color3] (0.0,0) rectangle (0.4645,0.2);
    } & {$1.039$} \hspace{1pt}  \tikz{
        \fill[fill=color3] (0.0,0) rectangle (0.4955,0.2);
        \fill[pattern=north west lines, pattern color=black!30!color3] (0.0,0) rectangle (0.4955,0.2);
    } \\
    IAC & {$9.280$} \hspace{1pt}  \tikz{
        \fill[fill=color4] (0.0,0) rectangle (0.7415,0.2);
        \fill[pattern=north west lines, pattern color=black!30!color4] (0.0,0) rectangle (0.7415,0.2);
    } & {$2.117$} \hspace{1pt}  \tikz{
        \fill[fill=color4] (0.0,0) rectangle (0.9465,0.2);
        \fill[pattern=north west lines, pattern color=black!30!color4] (0.0,0) rectangle (0.9465,0.2);
    } & {$1.000$} \hspace{1pt}  \tikz{
        \fill[fill=color4] (0.0,0) rectangle (0.9095,0.2);
        \fill[pattern=north west lines, pattern color=black!30!color4] (0.0,0) rectangle (0.9095,0.2);
    } \\
    DQN & {$0.000$} \hspace{1pt}  \tikz{
        \fill[fill=color5] (0.0,0) rectangle (0.005,0.2);
        \fill[pattern=north west lines, pattern color=black!30!color5] (0.005,0) rectangle (0,0.2);
    } & {$1.000$} \hspace{1pt}  \tikz{
        \fill[fill=color5] (0.0,0) rectangle (0.447,0.2);
        \fill[pattern=north west lines, pattern color=black!30!color5] (0.0,0) rectangle (0.447,0.2);
    } & {$1.000$} \hspace{1pt}  \tikz{
        \fill[fill=color5] (0.0,0) rectangle (0.477,0.2);
        \fill[pattern=north west lines, pattern color=black!30!color5] (0.0,0) rectangle (0.477,0.2);
    } \\
    \textit{Monte Carlo} & {$2.480$} \hspace{1pt}  \tikz{
        \fill[fill=color7] (0.0,0) rectangle (0.198,0.2);
        \fill[pattern=north west lines, pattern color=black!30!color7] (0.0,0) rectangle (0.198,0.2);
    } & {$1.264$} \hspace{1pt}  \tikz{
        \fill[fill=color7] (0.0,0) rectangle (0.565,0.2);
        \fill[pattern=north west lines, pattern color=black!30!color7] (0.0,0) rectangle (0.565,0.2);
    } & {$1.245$} \hspace{1pt}  \tikz{
        \fill[fill=color7] (0.0,0) rectangle (0.594,0.2);
        \fill[pattern=north west lines, pattern color=black!30!color7] (0.0,0) rectangle (0.594,0.2);
    } \\
    \bottomrule[1pt]
\end{tabularx}
    \end{minipage}}
    \vspace{-5mm}
\end{table}

\begin{figure*}[t!]
    \centering
    \includegraphics[width=\linewidth]{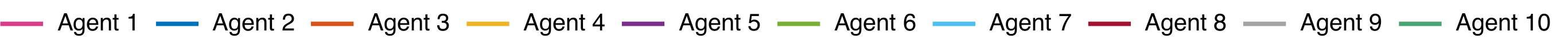}\\
    \subfigure[\textbf{Number of services (Proposed).}]{
        \includegraphics[width=0.3\linewidth]{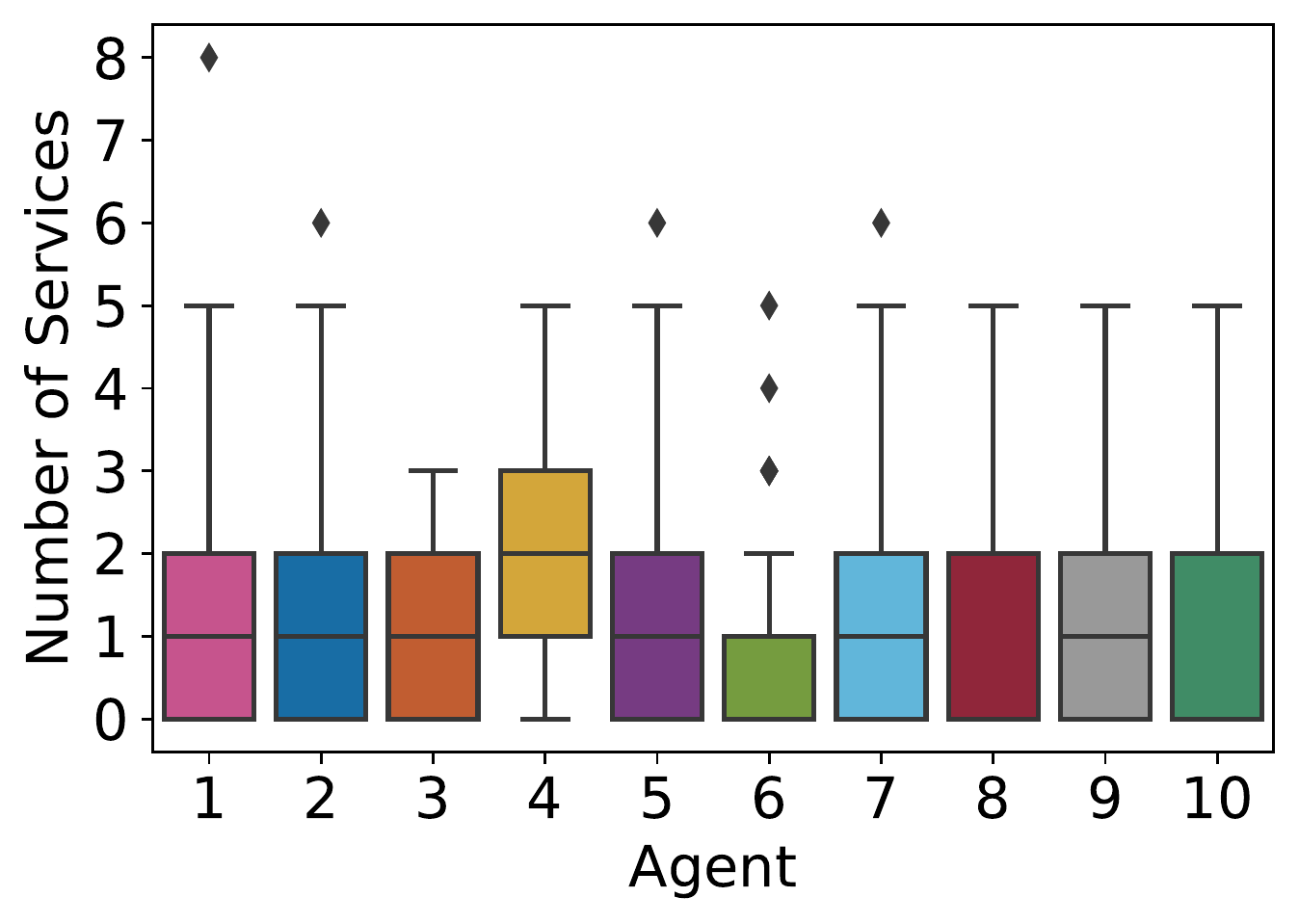}
        \label{fig:BoxPlot_Proposed_NumServices}
    }
    \subfigure[Number of services (Hybrid).]{
        \includegraphics[width=0.3\linewidth]{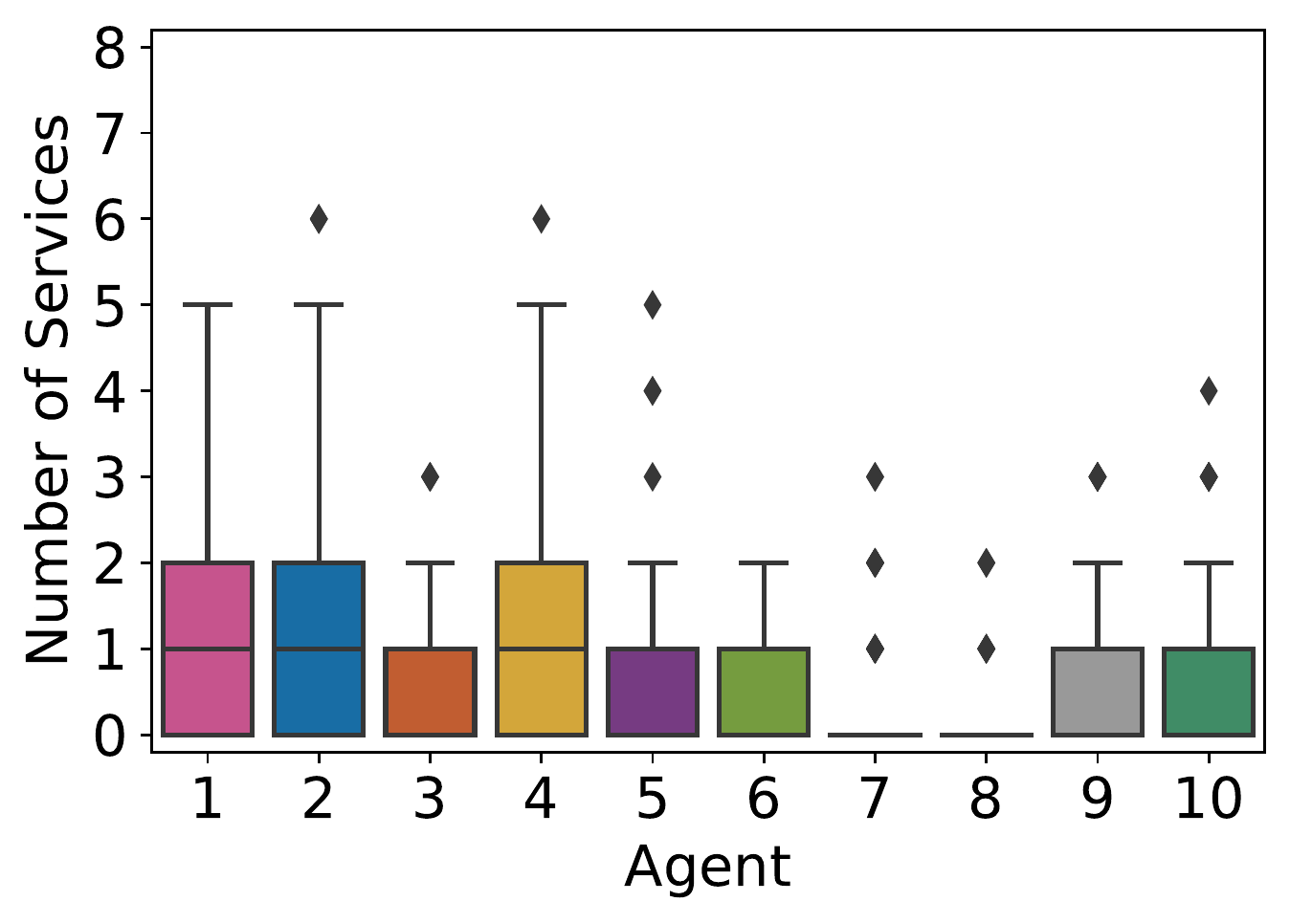}
        \label{fig:BoxPlot_Dual_NumServices}
    }
    \subfigure[Number of services (IAC).]{
        \includegraphics[width=0.3\linewidth]{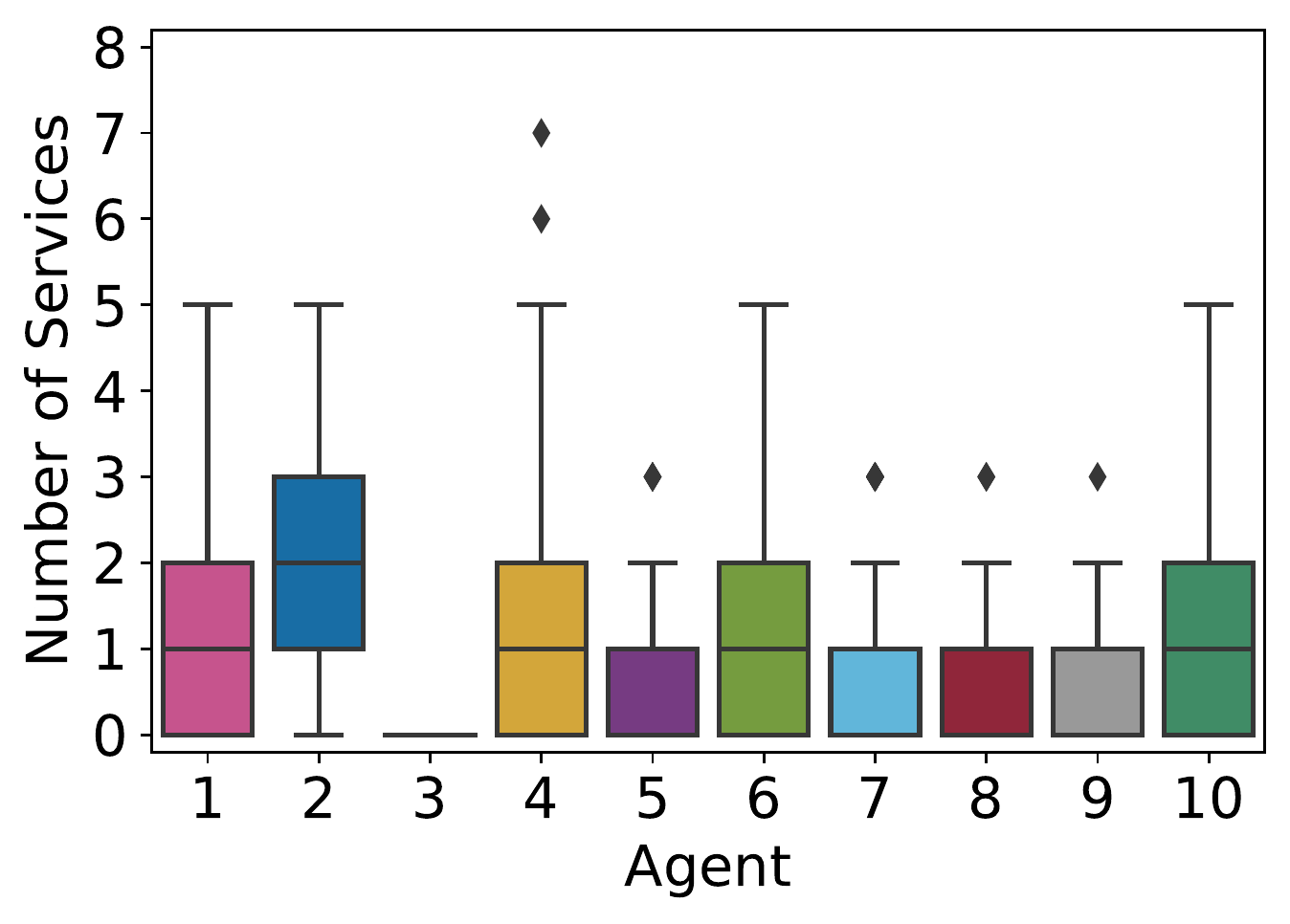}
        \label{fig:BoxPlot_IAC_NumServices}
    }
    \\
    \subfigure[\textbf{Number of landings (Proposed).}]{
        \includegraphics[width=0.3\linewidth]{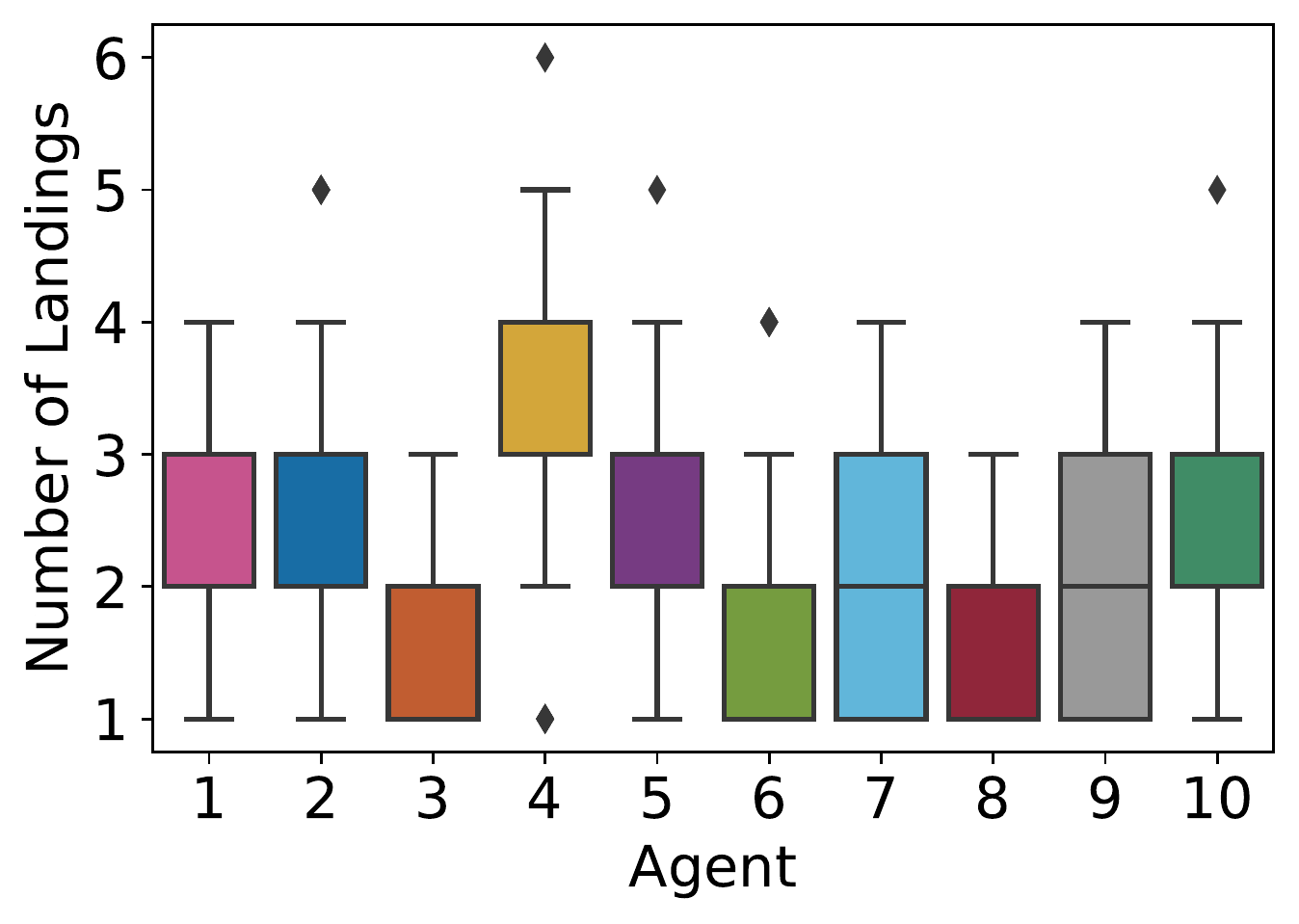}
        \label{fig:BoxPlot_Proposed_NumLandings}
    }
    \subfigure[Number of landings (Hybrid).]{
        \includegraphics[width=0.3\linewidth]{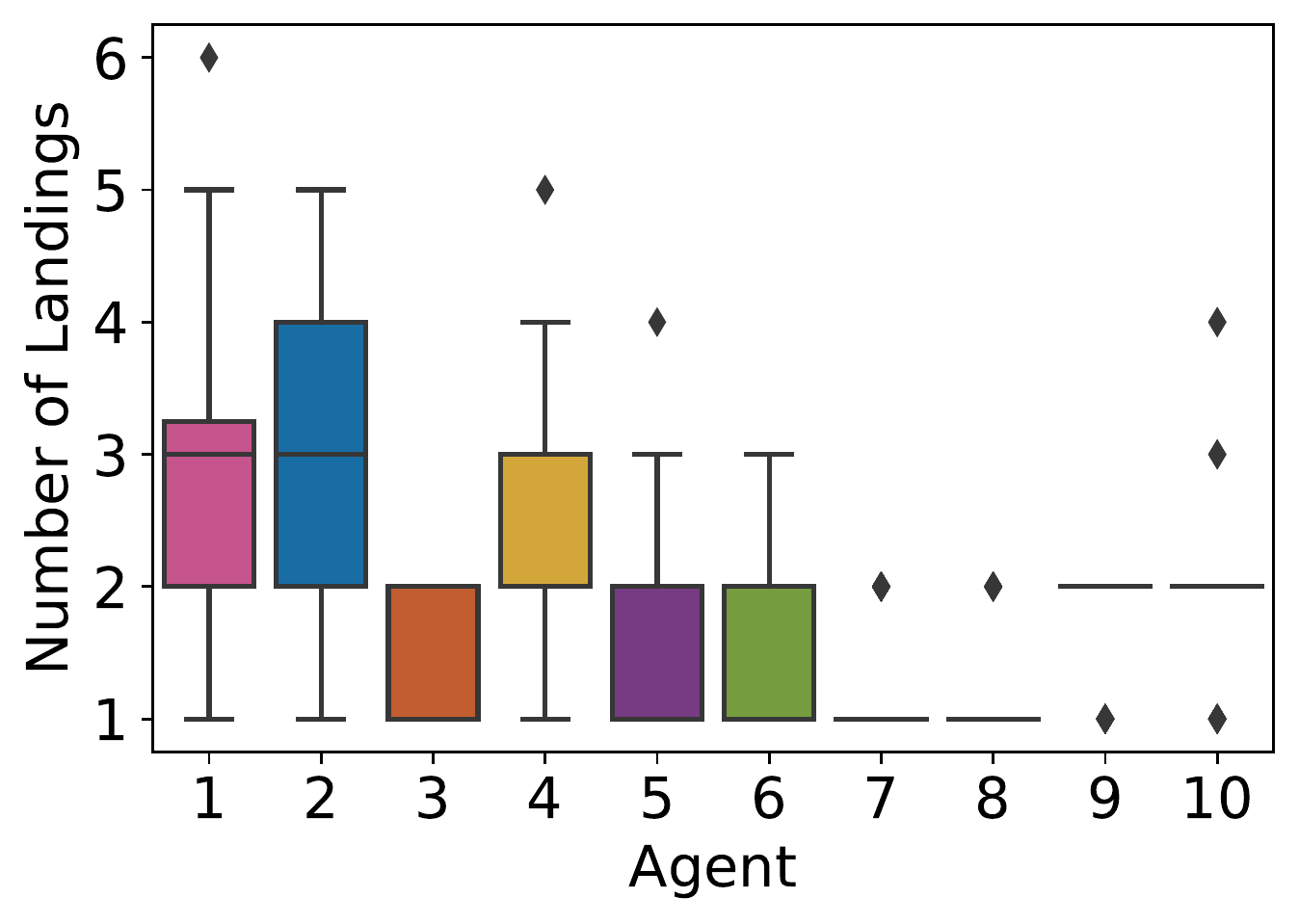}
        \label{fig:BoxPlot_Dual_NumLandings}
    }
    \subfigure[Number of landings (IAC).]{
        \includegraphics[width=0.3\linewidth]{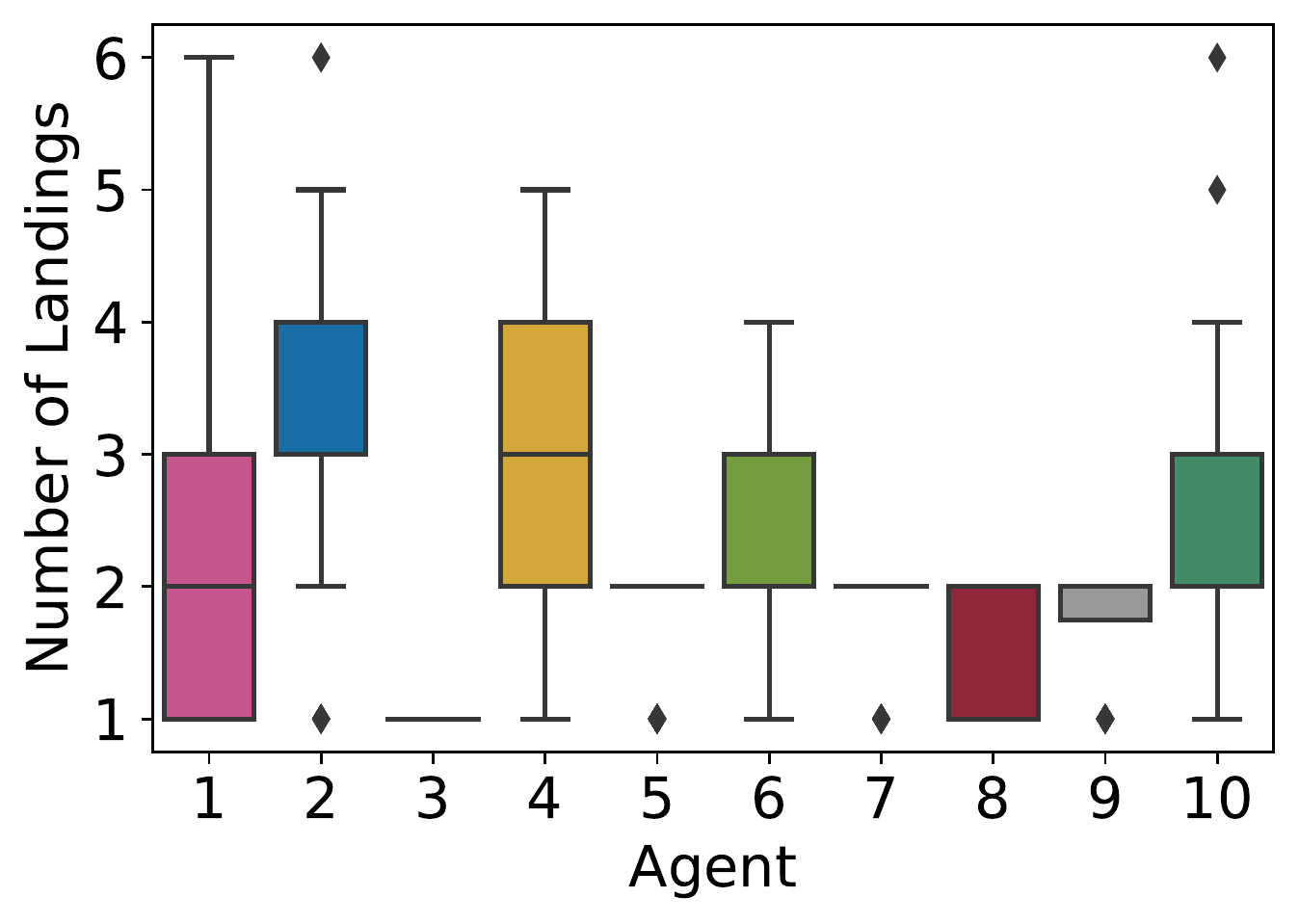}
        \label{fig:BoxPlot_IAC_NumLandings}
    }
    \\
    \subfigure[\textbf{Type of vertiport (Proposed).}]{
        \includegraphics[width=0.3\linewidth]{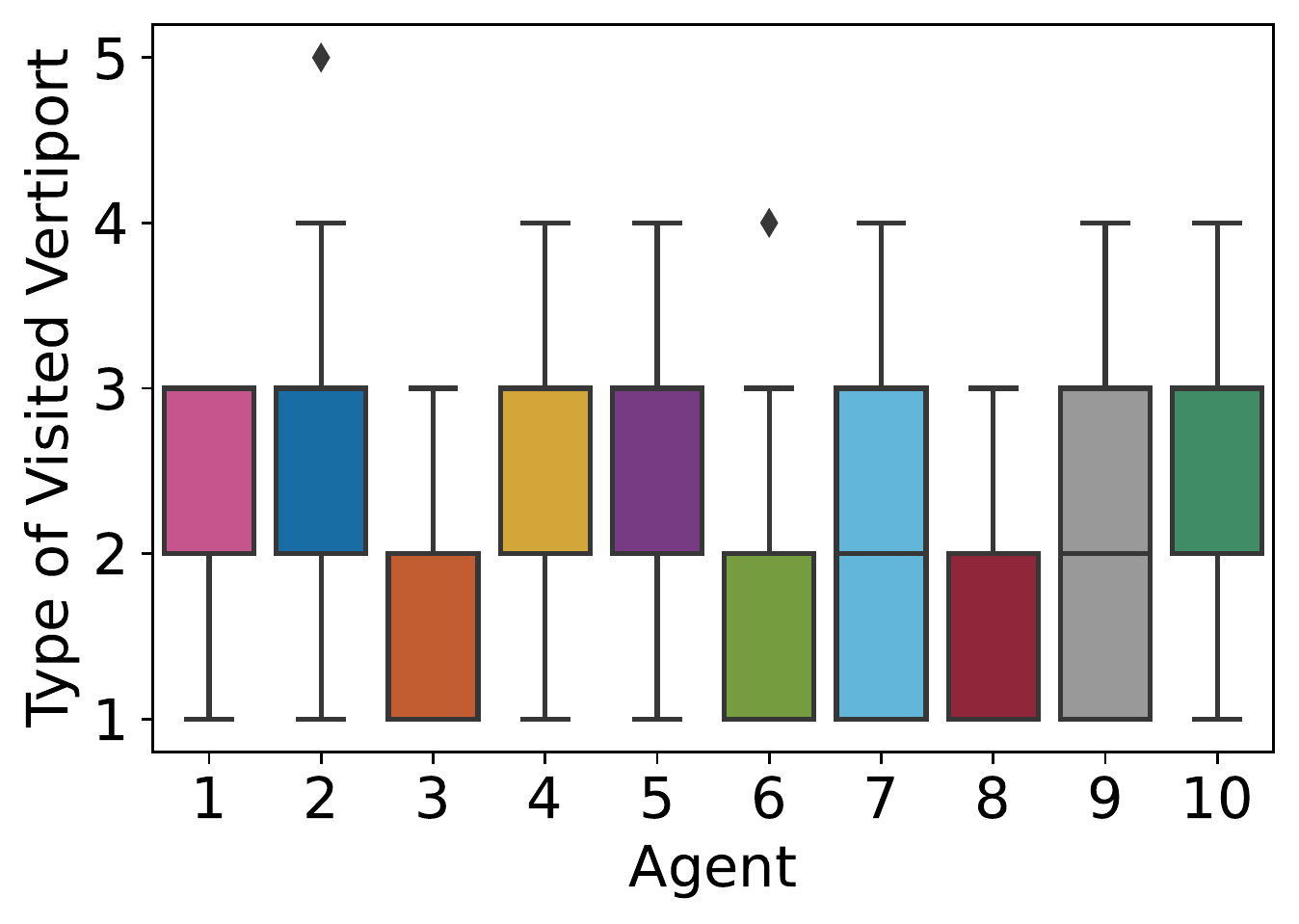}
        \label{fig:BoxPlot_Proposed_TypeVertiports}
    }
    \subfigure[Type of vertiport (Hybrid).]{
        \includegraphics[width=0.3\linewidth]{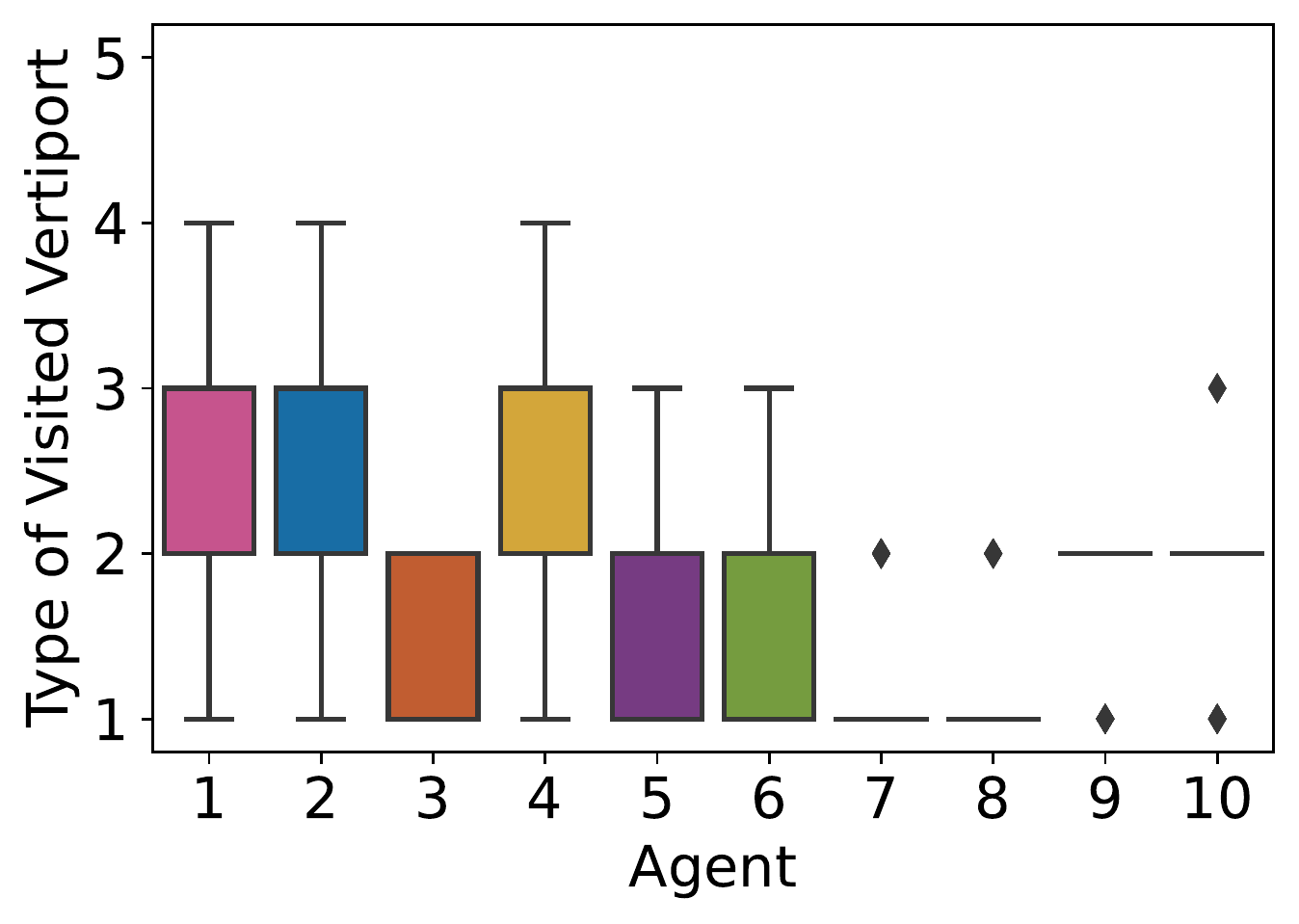}
        \label{fig:BoxPlot_Dual_TypeVertiports}
    }
    \subfigure[Type of vertiport (IAC).]{
        \includegraphics[width=0.3\linewidth]{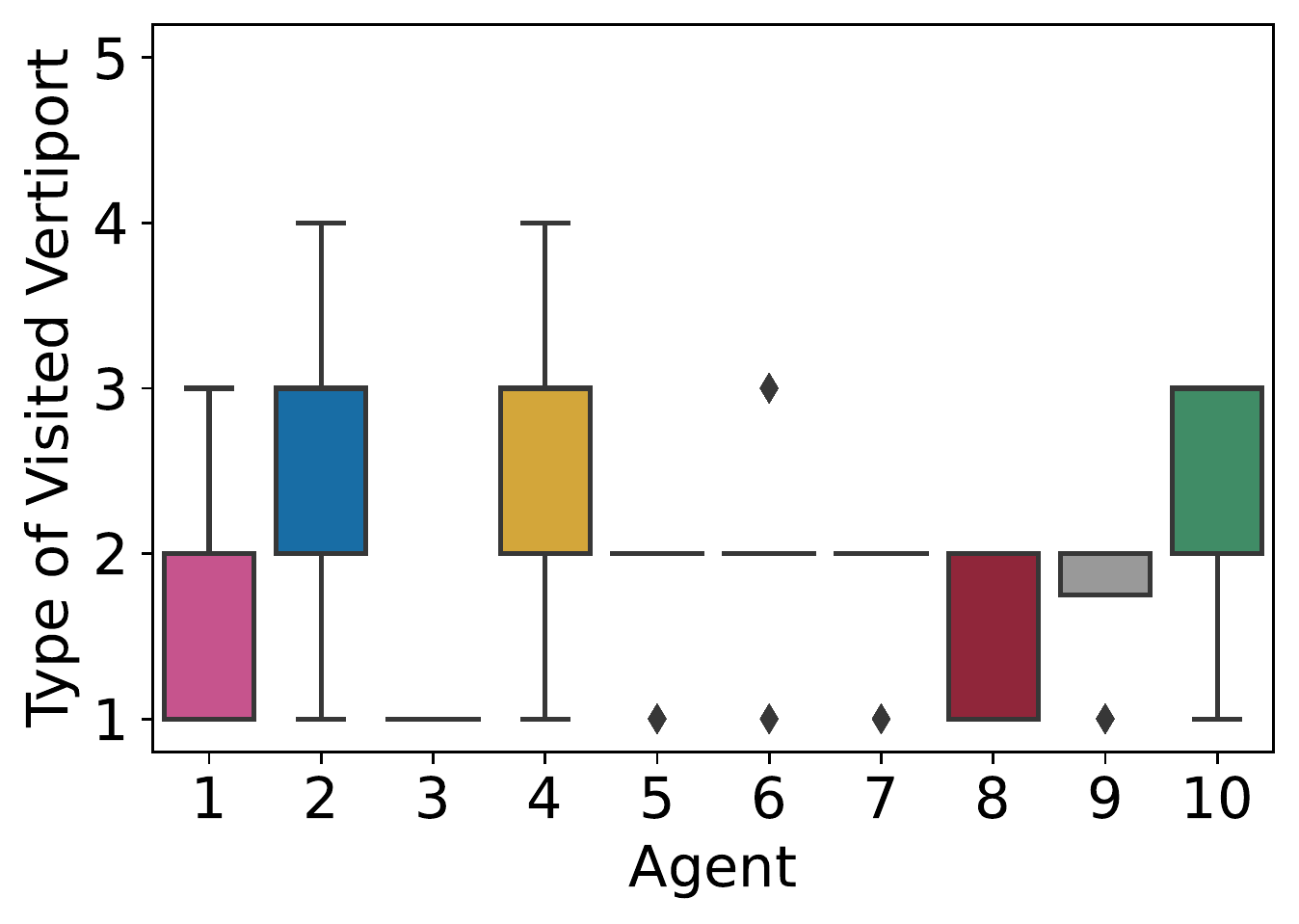}
        \label{fig:BoxPlot_IAC_TypeVertiports}
    }
    \caption{Total air transportation service quality of all UAMs trained with the proposed, Hybrid, and IAC algorithms in the inference process.}
    \label{fig:BoxPlot}
\end{figure*}

The air transportation service quality considered in this paper encompasses the number of serviced passengers and the number/types of landing vertiports.
Successfully transporting many passengers to their destination vertiports is the most crucial element of air transportation services. In addition, the above quality factors are correlated since the more diverse vertiports UAM visits, the more passengers it can provide air transportation services to.

Fig.~\ref{fig:Inference} presents the service quality of the first benchmark group. It can be seen that UAMs in CommNet outperform regarding air transportation service quality, except for less than two UAMs. Furthermore, it is noteworthy that among all UAMs trained by other benchmark algorithms that outperform UAMs in the proposed MADRL algorithm, CommNet-based UAMs in Hybrid are unique, corresponding to agent 2 in Fig.~\ref{fig:Inference}(a) and agents 1, 2 in Figs.~\ref{fig:Inference}(b)--(c).
The other DNN-based UAMs in Hybrid and DNN show inferior validation performance to CommNet-based UAMs in CommNet and Hybrid.
This result definitely verifies that inter-agent communications by CommNet can help UAMs learn optimal policies to serve autonomous air transportation services.

Regarding the second benchmark group's service quality, UAMs in CTDE outperform others in all service quality factors except for agents 2 and 6 according to Fig.~\ref{fig:CTDE_Inference}. Even though these two UAMs provided the second-highest quality of service, they served similar quality of air transportation service to other IAC agents. Even though these two UAMs provided the second highest quality of service, they served comparable air transportation services to other UAMs in IAC. Moreover, they outperformed UAMs in DQN and \textit{Monte Carlo}. In the case of DQN, as shown in Fig.~\ref{fig:CTDE_Inference}(a), UAMs cannot serve air transportation service to any passenger similar to the training process in Fig.~\ref{fig:CTDE_Reward}.

Finally, Table~\ref{tab:validation} summarizes the average validation performance of all UAMs trained with each benchmark represented in Figs.~\ref{fig:Inference}--\ref{fig:CTDE_Inference}. It can be seen that the proposed training scheme with CommNet and CTDE algorithms shows the best air transportation performance in terms of the number of serviced passengers and the number/types of vertiports on which UAMs land. The benchmark with the second-highest air transportation performance is IAC consisting entirely of CommNet-based UAMs. Next, Hybrid follows with the third highest air transportation performance. The other learning benchmarks, which correspond to DNN and DQN, failed to train parameterized policies of UAMs by serving inferior performance than \textit{Monte Carlo} which is not a learning algorithm. In summary, it is confirmed that environmental information sharing by CommNet and training strategy based on CTDE are suitable for building efficient autonomous air transportation networks.

\subsubsection{Service Fairness}\label{sec:V-D-2}


\begin{table}[t!]
    \centering          
    \caption{Variance Comparison for Validation of Policy Equity.}
    \resizebox{1\columnwidth}{!}{\begin{minipage}[h]{1\columnwidth}
    \centering
    \label{tab:Variance}
    \newcolumntype{R}{>{\raggedleft\arraybackslash}X}
\begin{tabularx}{1\linewidth}{l l l l}
    \toprule[1pt]
    \multicolumn{4}{c}{\textbf{\circled{4} : Variance of Air Transportation Service Quality in POMDP}} \\
    \midrule[.5pt]
    \ \textit{Benchmark} & \ \ Figs.\ref{fig:BoxPlot}(a)--(c) & \ \ Figs.\ref{fig:BoxPlot}(d)--(f) & \ \ Figs.\ref{fig:BoxPlot}(g)--(i) \\     
    \cmidrule(lr){1-1} \cmidrule(lr){2-2} \cmidrule(lr){3-3} \cmidrule(lr){4-4}
    \textbf{Proposed} & {$0.182$} \hspace{1pt}  \tikz{
        \fill[fill=color1] (0.0,0) rectangle (0.395,0.2);
        \fill[pattern=north west lines, pattern color=black!30!color1] (0.0,0) rectangle (0.395,0.2);
    
    } & {$0.081$} \hspace{1pt}  \tikz{
        \fill[fill=color1] (0.0,0) rectangle (0.471,0.2);
        \fill[pattern=north west lines, pattern color=black!30!color1] (0.0,0) rectangle (0.471,0.2);
    } & {$0.135$} \hspace{1pt}  \tikz{
        \fill[fill=color1] (0.0,0) rectangle (0.574,0.2);
        \fill[pattern=north west lines, pattern color=black!30!color1] (0.0,0) rectangle (0.574,0.2);
    } \\
    Hybrid & {$0.354$} \hspace{1pt}  \tikz{
        \fill[fill=color2] (0.0,0) rectangle (0.768,0.2);
        \fill[pattern=north west lines, pattern color=black!30!color2] (0.0,0) rectangle (0.768,0.2);
    } & {$0.172$} \hspace{1pt}  \tikz{
        \fill[fill=color2] (0.0,0) rectangle (1,0.2);
        \fill[pattern=north west lines, pattern color=black!30!color2] (0.0,0) rectangle (1,0.2);
    } & {$0.235$} \hspace{1pt}  \tikz{
        \fill[fill=color2] (0.0,0) rectangle (1,0.2);
        \fill[pattern=north west lines, pattern color=black!30!color2] (0.0,0) rectangle (1,0.2);
    } \\
    IAC & {$0.461$} \hspace{1pt}  \tikz{
        \fill[fill=color4] (0.0,0) rectangle (1,0.2);
        \fill[pattern=north west lines, pattern color=black!30!color4] (0.0,0) rectangle (1,0.2);
    } & {$0.172$} \hspace{1pt}  \tikz{
        \fill[fill=color4] (0.0,0) rectangle (1,0.2);
        \fill[pattern=north west lines, pattern color=black!30!color4] (0.0,0) rectangle (1,0.2);
    } & {$0.229$} \hspace{1pt}  \tikz{
        \fill[fill=color4] (0.0,0) rectangle (0.974,0.2);
        \fill[pattern=north west lines, pattern color=black!30!color4] (0.0,0) rectangle (0.974,0.2);
    }\\
    \bottomrule[1pt]
\end{tabularx}
    \end{minipage}}
\end{table}

This section inspects in detail the quality of air transportation services of all UAMs. Fig.~\ref{fig:BoxPlot} provides the records of serving air transportation services in POMDP of benchmarks that outperform  \textit{Monte Carlo} in Table~\ref{tab:validation}, which corresponds to the proposed, Hybrid, and IAC algorithms, with $100$ inference times. As shown in Fig.~\ref{fig:BoxPlot_Proposed_NumServices}, agent 1 in the proposed algorithm has served the largest number of services, providing air transportation service to 8 passengers. Additionally, agent 2 has visited all vertiports in the environment as shown in Fig.~\ref{fig:BoxPlot_Proposed_TypeVertiports}.
In addition to these best records, UAMs in the proposed algorithm have shown the highest and fairest air transportation service than the other benchmarks without an inferior UAM. However, there are UAMs with inferior performance to other benchmarks, such as agents 7-10 in Hybrid and agents 3, 5, and 7--9 in IAC.
Note that DNN-based UAMs (Agents 6--10) perform relatively inferior to CommNet-based UAMs (Agents 1--5) in Hybrid.

Table~\ref{tab:Variance} shows the variance of all service quality factors represented in Fig.~\ref{fig:BoxPlot}. Notably, it can be confirmed that UAMs trained with the proposed algorithm have the smallest variance than Hybrid and IAC while achieving the highest service quality as known in Fig.~\ref{fig:BoxPlot}. These results demonstrate that all UAMs in the proposed MADRL strategy cooperate unbiasedly to construct a high-quality autonomous air transportation network.

\subsubsection{Energy Management}\label{sec:V-D-3}

\begin{figure}
    \centering
    \includegraphics[width=\columnwidth]{Figure/Agents.pdf}\\
    \includegraphics[width=0.9\linewidth]{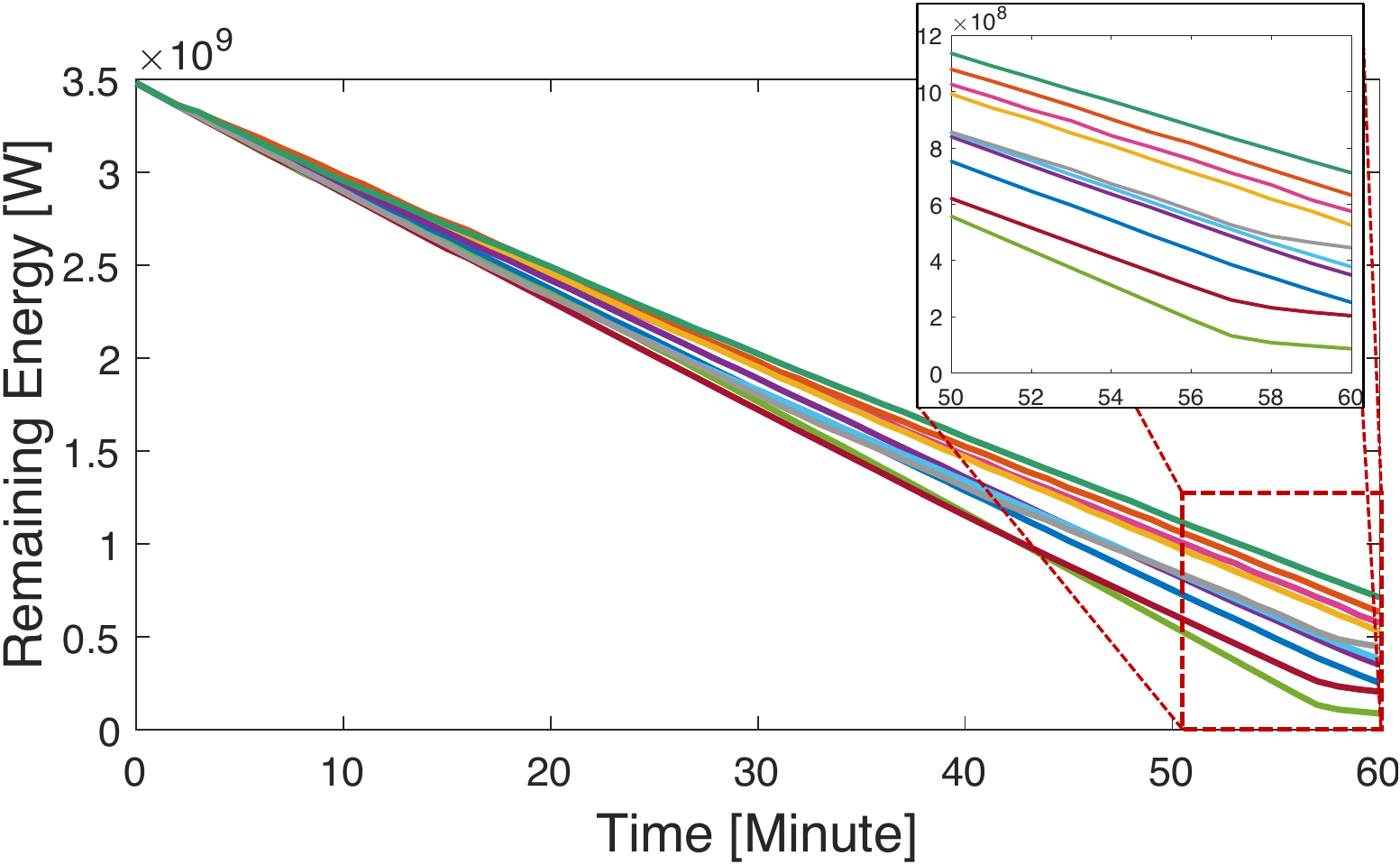}
    \caption{Energy state of all UAMs in the progress of the episode.}
    \label{fig:Energy}
\end{figure}

The reward function is designed in the direction of preventing the battery from being completely discharged with energy-related observation information when UAM learns its policy in Sec.~\ref{sec:IV-A-4}.
Fig.~\ref{fig:Energy} shows the energy state of all UAMs in the progress of the episode. It can be observed that UAMs trained by the proposed MADRL algorithm consume their energy within the maximum energy capacity limit for a given episode length $T$. In particular, agents 6 and 8 prevent complete discharging by reducing their energy consumption near the end of the episode. Hence, it is confirmed that UAMs successfully optimized their policies in both service quality and energy management.

\subsection{Discussions}
This section analyzes the above training and inference results in detail by describing the effects of CommNet and CTDE, the proposed training methods in this paper.

\subsubsection{Effect of CommNet}
As mentioned, the justification for using CommNet lies in achieving a common objective through mutual communication between UAMs. To evaluate the effect of CommNet, this paper conducted an ablation study according to the number of CommNet-based UAMs. The supremacy of this inter-communication scheme is confirmed in Fig.~\ref{fig:Reward} and Table~\ref{table:reward} by accomplishing the highest reward between all benchmarks in the training phase. In particular, it is meaningful for agent-to-agent communication through CommNet because UAMs in the proposed algorithm obtained a higher reward than FOMDP despite information loss in POMDP. Furthermore, CommNet-based information sharing helps UAMs learn out-of-range environmental information, allowing them to transport passengers efficiently to diversified vertiports as illustrated in Fig.~\ref{fig:Trajectory}. Next, the performance of the air transportation service quality provided after the learning phase is presented in Fig.~\ref{fig:Inference}. In the inference phase, the proposed strategy consisting of only CommNet-based UAMs served the best quality of service, as in the learning phase. Furthermore, parameter-sharing while training policies can make all policies optimal without any inferior UAMs as observed in Figs.~\ref{fig:Agentwise_reward} and~\ref{fig:BoxPlot}, and Table~\ref{tab:Variance}.

\subsubsection{Effect of CTDE}
In addition to CommNet, this subsection investigates the effect of utilizing a \textit{centralized critic} for CTDE. Since the \textit{critic} network evaluating the value of every state also needs to be trained to proceed with the episode, so using one \textit{centralized critic} can serve as a solid standard for multiple \textit{actors} to learn the policy with fewer episodes than when utilizing \textit{independent actor-critic} strategy~\cite{fujimoto2018addressing}. The advantage of CTDE in both training and inference processes is well represented in Fig.~\ref{fig:CTDE_Reward} and Table~\ref{table:CTDE_reward}, where training performance is different depending on the learning method if the same number of CommNet-based UAMs are employed.
Indeed, since the \textit{centralized critic} network can learn $J$-times more according to the progress of training epochs and establish more firm criteria for judging the state value than each \textit{independent critic} network, the fastest convergence was achieved with the highest reward as depicted in Fig.~\ref{fig:CTDE_Reward}. In addition, UAMs trained with a \textit{centralized critic} network outperform other benchmarks in the inference phase in Fig.~\ref{fig:CTDE_Inference}. Finally, the existence of a firm criterion for evaluating state values ensures that all multiple \textit{actors} have uniformly superior policies as shown in Figs.~\ref{fig:Agentwise_reward} and~\ref{fig:BoxPlot}, and Table~\ref{tab:Variance}.

\section{Conclusions and Future Work}\label{sec:VI}
This work aims to efficiently manage autonomous air transportation systems utilizing CommNet and CTDE-based MADRL algorithm. In particular, this paper conducts a realistic evaluation by adopting an actual vertiport map and considering the POMDP environment. Through extensive numerical results, it is demonstrated that the proposed MADRL algorithm has the most superior performance among conventional DRL algorithms in terms of air transportation service quality by helping multiple UAMs cooperatively learn their policies to coordinate their actions. In addition, all UAMs trained with the proposed strategy serve unbiased air transportation service without any inferior UAMs. In other words, the proposed MADRL algorithm has the most robust policy training performance for information loss in POMDP, effectively reflecting the physical limitations of UAM model. {Since UAM-based air transportation services are not yet commercialized, the scale outlined in this paper is suitable for preliminary air transportation service considerations. Nevertheless, once UAM materializes in the future, it would be advantageous to take into account a larger number of UAMs. Indeed, as demonstrated in~\cite{liu2020multi}, the MARL-based approach has the potential for scalability and the capacity to handle a considerable number of agents. Lastly, in order to improve the quality of the autonomous air transportation service, it is also a promising direction to consider incorporating factors that take passenger comfort into account in our reward function.}

\bibliographystyle{IEEEtran}
\bibliography{ref_tiv}

\begin{thebibliography}{10}
\providecommand{\url}[1]{#1}
\csname url@samestyle\endcsname
\providecommand{\newblock}{\relax}
\providecommand{\bibinfo}[2]{#2}
\providecommand{\BIBentrySTDinterwordspacing}{\spaceskip=0pt\relax}
\providecommand{\BIBentryALTinterwordstretchfactor}{4}
\providecommand{\BIBentryALTinterwordspacing}{\spaceskip=\fontdimen2\font plus
\BIBentryALTinterwordstretchfactor\fontdimen3\font minus
  \fontdimen4\font\relax}
\providecommand{\BIBforeignlanguage}[2]{{%
\expandafter\ifx\csname l@#1\endcsname\relax
\typeout{** WARNING: IEEEtran.bst: No hyphenation pattern has been}%
\typeout{** loaded for the language `#1'. Using the pattern for}%
\typeout{** the default language instead.}%
\else
\language=\csname l@#1\endcsname
\fi
#2}}
\providecommand{\BIBdecl}{\relax}
\BIBdecl

\bibitem{icc2023park}
C.~Park, S.~Park, G.~S. Kim, S.~Jung, J.-H. Kim, and J.~Kim, ``Multi-agent deep
  reinforcement learning for efficient passenger delivery in urban air
  mobility,'' in \emph{Proc. IEEE International Conference on Communications
  (ICC)}, Rome, Italy, May/June 2023.

\bibitem{Uber_2}
\BIBentryALTinterwordspacing
U.~Elevate, ``Fast-forwarding to a future of on-demand urban {Air}
  transportation,'' \emph{Uber Air}, October 2016. [Online]. Available:
  \url{https://www.uber.com/kr/ko/elevate/}
\BIBentrySTDinterwordspacing

\bibitem{goodrich2015demand}
K.~Goodrich and M.~Moore, ``On-demand mobility ({ODM}) technical pathway:
  enabling ease of use and safety,'' in \emph{Proc. American Institute of
  Aeronautics and Astronautics (AIAA) Aviation}, June 2015.

\bibitem{9447255}
A.~P. Cohen, S.~A. Shaheen, and E.~M. Farrar, ``Urban air mobility: History,
  ecosystem, market potential, and challenges,'' \emph{IEEE Transactions on
  Intelligent Transportation Systems}, vol.~22, no.~9, pp. 6074--6087,
  September 2021.

\bibitem{neto2021trajectory}
E.~C.~P. Neto, D.~M. Baum, J.~R. de~Almeida, J.~B. Camargo, and P.~S. Cugnasca,
  ``A trajectory evaluation platform for urban air mobility ({UAM}),''
  \emph{IEEE Transactions on Intelligent Transportation Systems}, vol.~23,
  no.~7, pp. 9136--9145, July 2022.

\bibitem{thipphavong2018urban}
D.~P. Thipphavong, R.~Apaza, B.~Barmore, V.~Battiste, B.~Burian, Q.~Dao,
  M.~Feary, S.~Go, K.~H. Goodrich, J.~Homola \emph{et~al.}, ``Urban air
  mobility airspace integration concepts and considerations,'' in \emph{Proc.
  Aviation Technology, Integration, and Operations Conference}, Atlanta,
  Georgia, June 2018, p. 3676.

\bibitem{kohlman2019urban}
L.~W. Kohlman, M.~D. Patterson, and B.~E. Raabe, ``Urban air mobility network
  and vehicle type-modeling and assessment,'' \emph{NASA, CA, USA, Tech. Rep.
  NASA/TM-2019-220072}, February 2019.

\bibitem{straubinger2020overview}
A.~Straubinger, R.~Rothfeld, M.~Shamiyeh, K.-D. B{\"u}chter, J.~Kaiser, and
  K.~O. Pl{\"o}tner, ``An overview of current research and developments in
  urban air mobility--setting the scene for {UAM} introduction,'' \emph{Journal
  of Air Transport Management}, vol.~87, p. 101852, August 2020.

\bibitem{americas2018new}
G.~Americas, ``New services \& applications with 5g ultra-reliable low latency
  communications,'' Technical Report, 5G Americas, Tech. Rep., November 2018.

\bibitem{ullah20195g}
H.~Ullah, N.~G. Nair, A.~Moore, C.~Nugent, P.~Muschamp, and M.~Cuevas, ``5g
  communication: An overview of vehicle-to-everything, drones, and healthcare
  use-cases,'' \emph{IEEE Access}, vol.~7, pp. 37\,251--37\,268, March 2019.

\bibitem{choi2022exploring}
J.~H. Choi and Y.~Park, ``Exploring economic feasibility for airport shuttle
  service of urban air mobility ({UAM}),'' \emph{Transportation Research Part
  A: Policy and Practice}, vol. 162, pp. 267--281, August 2022.

\bibitem{courtin2018feasibility}
C.~Courtin, M.~J. Burton, A.~Yu, P.~Butler, P.~D. Vascik, and R.~J. Hansman,
  ``Feasibility study of short takeoff and landing urban air mobility vehicles
  using geometric programming,'' in \emph{Proc. Aviation Technology,
  Integration, and Operations Conference (AIAA)}, Atlanta, Georgia, June 2018,
  p. 4151.

\bibitem{soltani2020eco}
M.~Soltani, S.~Ahmadi, A.~Akgunduz, and N.~Bhuiyan, ``An eco-friendly aircraft
  taxiing approach with collision and conflict avoidance,''
  \emph{Transportation Research Part C: Emerging Technologies}, vol. 121, p.
  102872, December 2020.

\bibitem{kim2022effects}
Y.~Kim, J.~Jo, D.~Kim, H.~Lee, and R.~Myong, ``Effects of lightning on {UAM}
  aircraft: Complex zoning and direct effects on composite prop-rotor blade,''
  \emph{Aerospace Science and Technology}, vol. 124, p. 107560, May 2022.

\bibitem{sukhbaatar2016learning}
S.~Sukhbaatar, R.~Fergus \emph{et~al.}, ``Learning multiagent communication
  with backpropagation,'' in \emph{Proc. Advances in Neural Information
  Processing Systems (NeurIPS)}, December 2016.

\bibitem{foerster2018counterfactual}
J.~Foerster, G.~Farquhar, T.~Afouras, N.~Nardelli, and S.~Whiteson,
  ``Counterfactual multi-agent policy gradients,'' in \emph{Proc. AAAI
  Conference on Artificial Intelligence (AAAI)}, vol.~32, no.~1, New Orleans,
  LA, USA, February 2018.

\bibitem{wang2022metavehicles}
F.-Y. Wang, ``Metavehicles in the metaverse: Moving to a new phase for
  intelligent vehicles and smart mobility,'' \emph{IEEE Transactions on
  Intelligent Vehicles}, vol.~7, no.~1, pp. 1--5, March 2022.

\bibitem{cao2022future}
D.~Cao, X.~Wang, L.~Li, C.~Lv, X.~Na, Y.~Xing, X.~Li, Y.~Li, Y.~Chen, and F.-Y.
  Wang, ``Future directions of intelligent vehicles: Potentials, possibilities,
  and perspectives,'' \emph{IEEE Transactions on Intelligent Vehicles}, vol.~7,
  no.~1, pp. 7--10, March 2022.

\bibitem{huang2022survey}
Y.~Huang, J.~Du, Z.~Yang, Z.~Zhou, L.~Zhang, and H.~Chen, ``A survey on
  trajectory-prediction methods for autonomous driving,'' \emph{IEEE
  Transactions on Intelligent Vehicles}, vol.~7, no.~3, pp. 652--674, September
  2022.

\bibitem{grigorescu2020survey}
S.~Grigorescu, B.~Trasnea, T.~Cocias, and G.~Macesanu, ``A survey of deep
  learning techniques for autonomous driving,'' \emph{Journal of Field
  Robotics}, vol.~37, no.~3, pp. 362--386, April 2020.

\bibitem{chen2017learning}
J.~Chen, U.~Yatnalli, and D.~Gesbert, ``Learning radio maps for {UAV}-aided
  wireless networks: A segmented regression approach,'' in \emph{Proc. IEEE
  International Conference on Communications (ICC)}, Paris, France, May 2017,
  pp. 1--6.

\bibitem{lu2020machine}
L.~Lu, Z.~Yang, M.~Chen, Z.~Zang \emph{et~al.}, ``Machine learning for
  predictive deployment of {UAVs} with multiple access,'' \emph{arXiv preprint
  arXiv:2003.02631}, 2020.

\bibitem{kong2022trajectory}
F.~Kong, J.~Li, B.~Jiang, H.~Wang, and H.~Song, ``Trajectory optimization for
  drone logistics delivery via attention-based pointer network,'' \emph{IEEE
  Transactions on Intelligent Transportation Systems}, April 2023.

\bibitem{mozaffari2016optimal}
M.~Mozaffari, W.~Saad, M.~Bennis, and M.~Debbah, ``Optimal transport theory for
  power-efficient deployment of unmanned aerial vehicles,'' in \emph{Proc. IEEE
  ICC}, Kuala Lumpur, Malaysia, May 2016, pp. 1--6.

\bibitem{kalantari2016number}
E.~Kalantari, H.~Yanikomeroglu, and A.~Yongacoglu, ``On the number and {3D}
  placement of drone base stations in wireless cellular networks,'' in
  \emph{Proc. IEEE VTC}, Montreal, QC, Canada, September 2016, pp. 1--6.

\bibitem{sayyadi2020integrated}
R.~Sayyadi and A.~Awasthi, ``An integrated approach based on system dynamics
  and {ANP} for evaluating sustainable transportation policies,''
  \emph{International Journal of Systems Science: Operations \& Logistics},
  vol.~7, no.~2, pp. 182--191, 2020.

\bibitem{bellman2010dynamic}
R.~E. Bellman, \emph{Dynamic programming}.\hskip 1em plus 0.5em minus
  0.4em\relax Princeton university press, 2010.

\bibitem{ijcai2019shin}
M.~Shin and J.~Kim, ``Randomized adversarial imitation learning for autonomous
  driving,'' in \emph{Proc. International Joint Conference on Artificial
  Intelligence (IJCAI)}, 2019, pp. 4590--4596.

\bibitem{kiran2021deep}
B.~R. Kiran, I.~Sobh, V.~Talpaert, P.~Mannion, A.~A. Al~Sallab, S.~Yogamani,
  and P.~P{\'e}rez, ``Deep reinforcement learning for autonomous driving: A
  survey,'' \emph{IEEE Transactions on Intelligent Transportation Systems},
  vol.~23, no.~6, pp. 4909--4926, June 2021.

\bibitem{rashid2020monotonic}
T.~Rashid, M.~Samvelyan, C.~S. De~Witt, G.~Farquhar, J.~Foerster, and
  S.~Whiteson, ``Monotonic value function factorisation for deep multi-agent
  reinforcement learning,'' \emph{The Journal of Machine Learning Research},
  vol.~21, no.~1, pp. 7234--7284, January 2020.

\bibitem{icte202103yun}
W.~J. Yun, S.~Jung, J.~Kim, and J.-H. Kim, ``Distributed deep reinforcement
  learning for autonomous aerial {eVTOL} mobility in drone taxi applications,''
  \emph{ICT Express}, vol.~7, no.~1, pp. 1--4, March 2021.

\bibitem{foerster2016learning}
J.~Foerster, I.~A. Assael, N.~de~Freitas, and S.~Whiteson, ``Learning to
  communicate with deep multi-agent reinforcement learning,'' in \emph{Proc.
  NeurIPS}, vol.~29, Barcelona, Spain, December 2016, pp. 2137--2145.

\bibitem{hou2021bicnet}
R.~Hou, H.~Chang, B.~Ma, R.~Huang, and S.~Shan, ``{BiCNet-TKS}: {L}earning
  efficient spatial-temporal representation for video person
  re-identification,'' in \emph{Proc. IEEE CVPR}, Virtual, June 2021, pp.
  2014--2023.

\bibitem{zhu2022survey}
C.~Zhu, M.~Dastani, and S.~Wang, ``A survey of multi-agent reinforcement
  learning with communication,'' \emph{arXiv preprint arXiv:2203.08975}, 2022.

\bibitem{tii202005shin}
M.~Shin, D.-H. Choi, and J.~Kim, ``Cooperative management for {PV/ESS}-enabled
  electric vehicle charging stations: A multiagent deep reinforcement learning
  approach,'' \emph{IEEE Transactions on Industrial Informatics}, vol.~16,
  no.~5, pp. 3493--3503, May 2020.

\bibitem{tvt202106jung}
S.~Jung, W.~J. Yun, M.~Shin, J.~Kim, and J.-H. Kim, ``Orchestrated scheduling
  and multi-agent deep reinforcement learning for cloud-assisted multi-{UAV}
  charging systems,'' \emph{IEEE Transactions on Vehicular Technology},
  vol.~70, no.~6, pp. 5362--5377, June 2021.

\bibitem{tii202210yun}
W.~J. Yun, S.~Park, J.~Kim, M.~Shin, S.~Jung, D.~A. Mohaisen, and J.-H. Kim,
  ``Cooperative multiagent deep reinforcement learning for reliable
  surveillance via autonomous multi-{UAV} control,'' \emph{IEEE Transactions on
  Industrial Informatics}, vol.~18, no.~10, pp. 7086--7096, October 2022.

\bibitem{Uber_1}
\BIBentryALTinterwordspacing
CORGAN, ``Connect evolved uber elevate 2019,'' \emph{Uber Air}, June 2019.
  [Online]. Available: \url{https://www.corgan.com/}
\BIBentrySTDinterwordspacing

\bibitem{mobisys2010paek}
J.~Paek, J.~Kim, and R.~Govindan, ``Energy-efficient rate-adaptive {GPS}-based
  positioning for smartphones,'' in \emph{Proc. ACM International Conference on
  Mobile Systems, Applications, and Services (MobiSys)}, 2010, pp. 299--314.

\bibitem{equation_sjung}
S.~Jung, W.~J. Yun, M.~Shin, J.~Kim, and J.-H. Kim, ``Orchestrated scheduling
  and multi-agent deep reinforcement learning for cloud-assisted multi-{UAV}
  charging systems,'' \emph{IEEE Transactions on Vehicular Technology},
  vol.~70, no.~6, pp. 5362--5377, 2021.

\bibitem{zeng2017energy}
Y.~Zeng and R.~Zhang, ``Energy-efficient {UAV} communication with trajectory
  optimization,'' \emph{IEEE Transactions on Wireless Communications}, vol.~16,
  no.~6, pp. 3747--3760, June 2017.

\bibitem{jobyaviation}
\BIBentryALTinterwordspacing
P.~Sciarra, ``Joby aviation analyst day,'' \emph{Joby aviation}, June 2021.
  [Online]. Available: \url{https://ir.jobyaviation.com/about-us/presentations}
\BIBentrySTDinterwordspacing

\bibitem{pajarinen2011periodic}
J.~Pajarinen and J.~Peltonen, ``Periodic finite state controllers for efficient
  {POMDP} and {DEC-POMDP} planning,'' in \emph{Proc. Advances in neural
  information processing systems (NeurIPS)}, vol.~24, Granada, Spain, December
  2011, pp. 2636--2644.

\bibitem{cui2022multi}
Q.~Cui, X.~Zhao, W.~Ni, Z.~Hu, X.~Tao, and P.~Zhang, ``Multi-agent deep
  reinforcement learning-based interdependent computing for mobile edge
  computing-assisted robot teams,'' \emph{IEEE Transactions on Vehicular
  Technology}, December 2022.

\bibitem{lowe2017multi}
R.~Lowe, Y.~Wu, A.~Tamar, J.~Harb, O.~Pieter~Abbeel, and I.~Mordatch,
  ``Multi-agent actor-critic for mixed cooperative-competitive environments,''
  in \emph{Proc. Advances in Neural Information Processing Systems (NeurIPS)},
  Long Beach, CA, USA, December 2017, pp. 6379--6390.

\bibitem{fujimoto2018addressing}
S.~Fujimoto, H.~Hoof, and D.~Meger, ``Addressing function approximation error
  in actor-critic methods,'' in \emph{proc. International Conference on Machine
  Learning (ICML)}, July 2018, pp. 1587--1596.

\bibitem{mnih2013playing}
V.~Mnih, K.~Kavukcuoglu, D.~Silver, A.~Graves, I.~Antonoglou, D.~Wierstra, and
  M.~Riedmiller, ``Playing atari with deep reinforcement learning,''
  \emph{arXiv preprint arXiv:1312.5602}, 2013.

\bibitem{malysheva2018deep}
A.~Malysheva, T.~T. Sung, C.-B. Sohn, D.~Kudenko, and A.~Shpilman, ``Deep
  multi-agent reinforcement learning with relevance graphs,'' \emph{arXiv
  preprint arXiv:1811.12557}, 2018.

\bibitem{weibel2011establishing}
R.~Weibel, M.~Edwards, and C.~Fernandes, ``Establishing a risk-based separation
  standard for unmanned aircraft self separation,'' in \emph{11th AIAA Aviation
  Technology, Integration, and Operations (ATIO) Conference, including the AIAA
  Balloon Systems Conference and 19th AIAA Lighter-Than}, 2011, p. 6921.

\bibitem{ATC}
M.~S. Nolan, ``Fundamentals of air traffic control-5th edition,'' \emph{Delmar
  Cengage Learning}, 2010.

\bibitem{liu2020multi}
Y.~Liu, W.~Wang, Y.~Hu, J.~Hao, X.~Chen, and Y.~Gao, ``Multi-agent game
  abstraction via graph attention neural network,'' in \emph{Proc. AAAI
  Conference on Artificial Intelligence (AAAI)}, vol.~34, no.~5, New York, NY,
  USA, February 2020, pp. 7211--7218.

\end{thebibliography}

\begin{IEEEbiography}[{\includegraphics[width=1in,height=1.25in,clip,keepaspectratio]{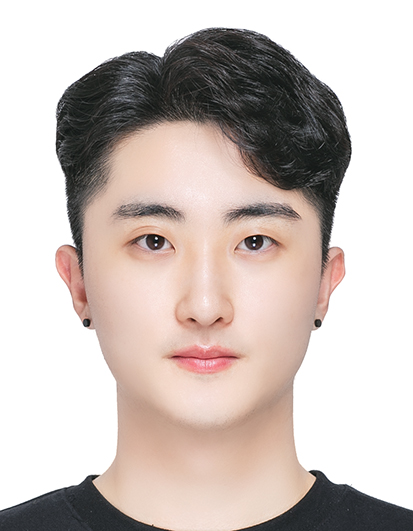}}]{Chanyoung Park} 
is currently a Ph.D. student at the Department of Electrical and Computer Engineering, Korea University, Seoul, Republic of Korea, since September 2022. He received the B.S. degree in electrical and computer engineering from Ajou University, Suwon, Republic of Korea, in 2022, with honor (early graduation). His research focuses include deep learning algorithms and their applications to communications and networks. 
\end{IEEEbiography}

\begin{IEEEbiography}[{\includegraphics[width=1in,height=1.25in,clip,keepaspectratio]{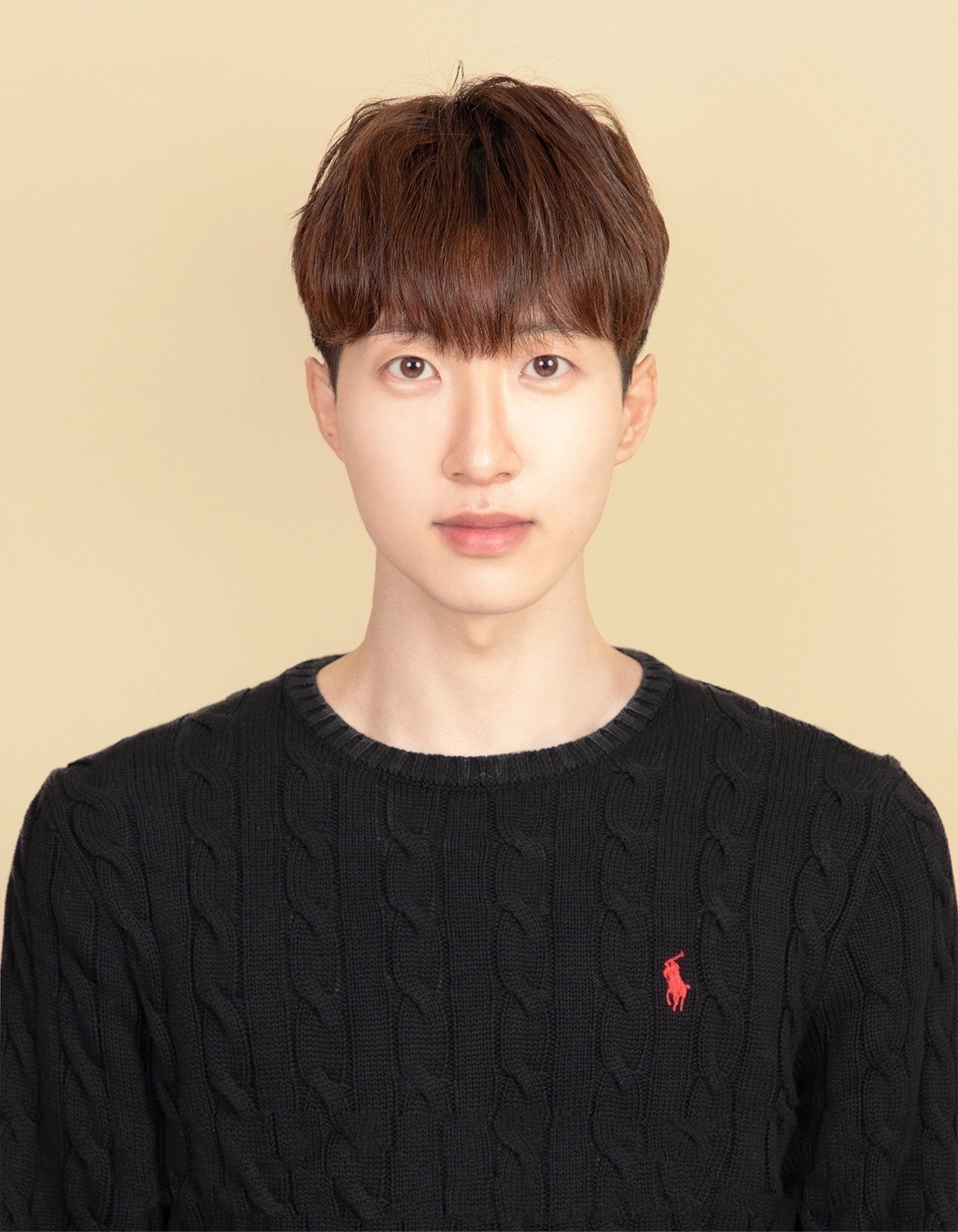}}]{Gyu Seon Kim} 
is currently a Ph.D. student at the Department of Electrical and Computer Engineering, Korea University, Seoul, Republic of Korea. He received the B.S. degree in aerospace engineering from Inha University, Incheon, Republic of Korea. 

His research focuses include deep reinforcement learning algorithms and their applications to autonomous mobility systems. 
\end{IEEEbiography}

\begin{IEEEbiography}[{\includegraphics[width=1in,height=1.25in,clip,keepaspectratio]{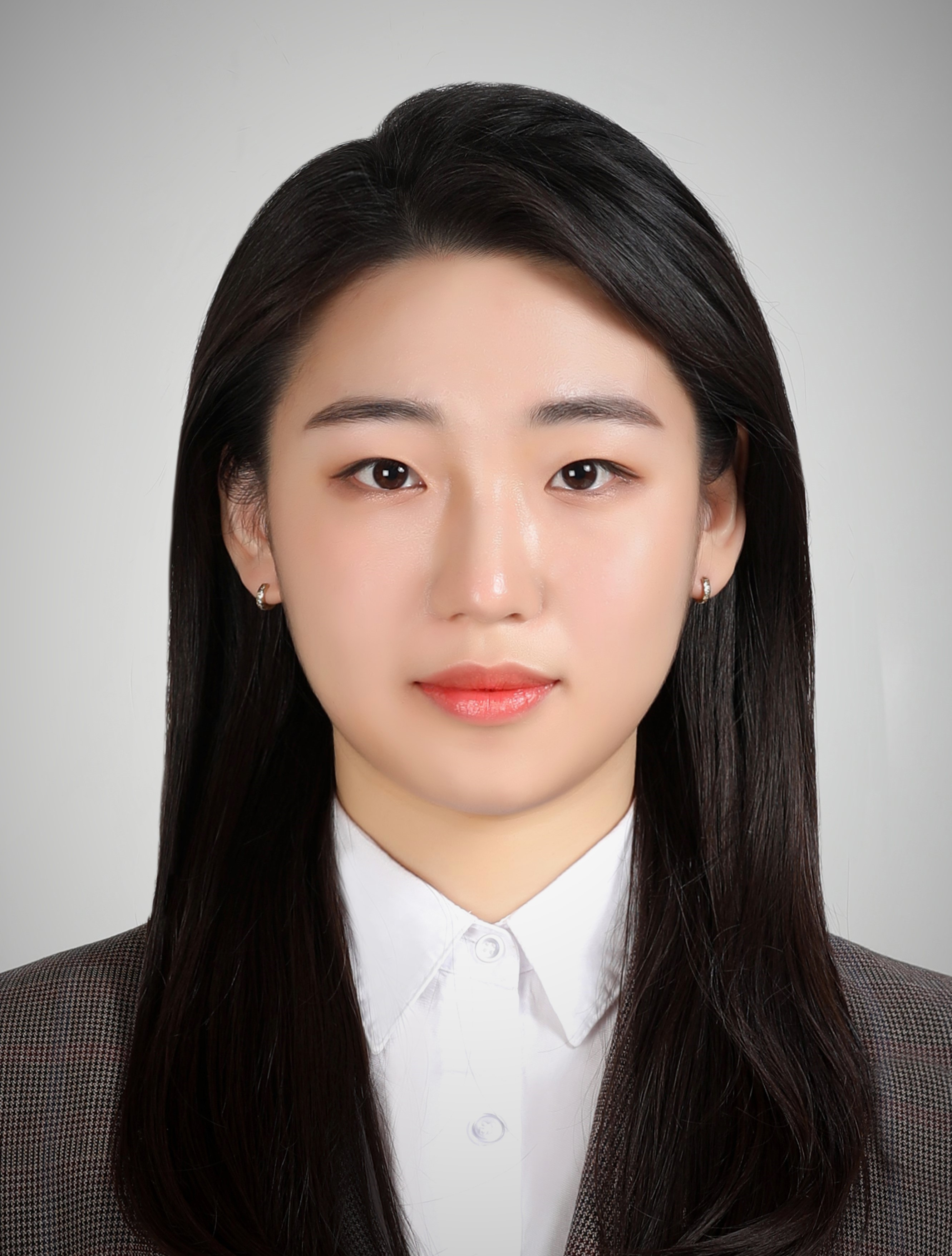}}]{Soohyun Park} has been a postdoctoral scholar at the Department of Electrical and Computer Engineering, Korea University, Seoul, Republic of Korea, since September 2023. She received the Ph.D. degree in electrical and computer engineering from Korea University, Seoul, Republic of Korea, in August 2023. She received the B.S. degree in computer science and engineering from Chung-Ang University, Seoul, Republic of Korea, in February 2019. Her research focuses include deep learning applications, quantum machine learning algorithms and their software engineering methodologies, big-data platforms, and autonomous networking. 

She was a recipient of the IEEE Vehicular Technology Society (VTS) Seoul Chapter Award (2019), IEEE Seoul Section Student Paper Content Award (2020), and \textit{ICT Express (Elsevier)} Best Reviewer Award (2021).\end{IEEEbiography}

\begin{IEEEbiography}[{\includegraphics[width=1in,height=1.25in,clip,keepaspectratio]{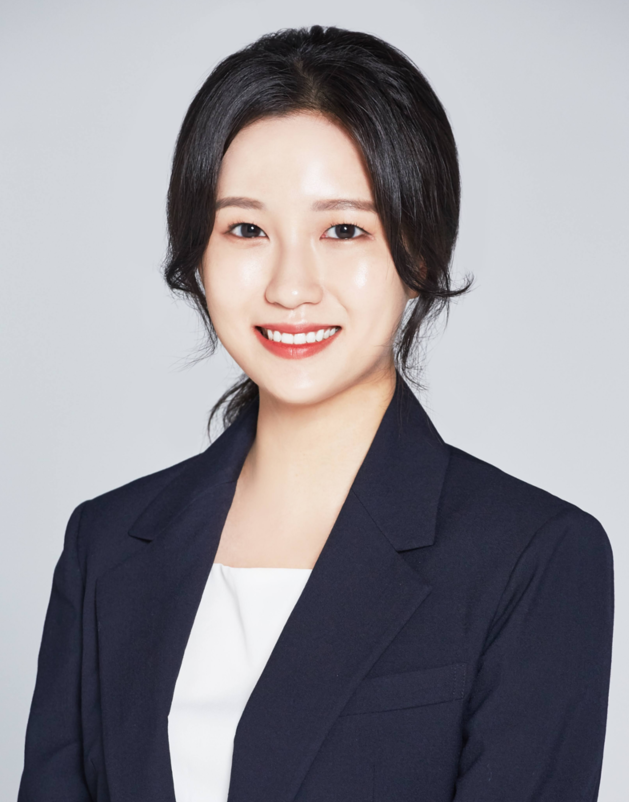}}]{Soyi Jung} 
(Member, IEEE) has been an assistant professor at the Department of Electrical of Computer Engineering, Ajou University, Suwon, Republic of Korea, since September 2022. Before joining Ajou University, she was an assistant professor at Hallym University, Chuncheon, Republic of  Korea, from 2021 to 2022; a visiting scholar at Donald Bren School of Information and Computer Sciences, University of California, Irvine, CA, USA, from 2021 to 2022; a research professor at Korea University, Seoul, Republic of Korea, in 2021; and a researcher at Korea Testing and Research (KTR) Institute, Gwacheon, Republic of Korea, from 2015 to 2016. She received her B.S., M.S., and Ph.D. degrees in electrical and computer engineering from Ajou University, Suwon, Republic of Korea, in 2013, 2015, and 2021, respectively. 

Her current research interests include network optimization for autonomous vehicles communications, distributed system analysis, big-data processing platforms, and probabilistic access analysis. 
She was a recipient of Best Paper Award by KICS (2015), Young Women Researcher Award by WISET and KICS (2015), Bronze Paper Award from IEEE Seoul Section Student Paper Contest (2018), ICT Paper Contest Award by Electronic Times (2019), and IEEE ICOIN Best Paper Award (2021).
\end{IEEEbiography}

\begin{IEEEbiography}[{\includegraphics[width=1in,height=1.25in,clip]{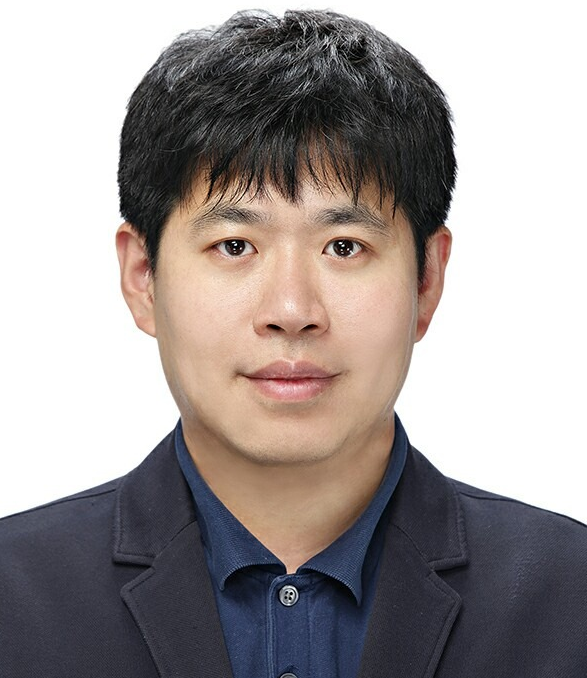}}]{Joongheon Kim} 
(M'06--SM'18) has been with Korea University, Seoul, Korea, since 2019, and he is currently an associate professor. He received the B.S. and M.S. degrees in computer science and engineering from Korea University, Seoul, Korea, in 2004 and 2006, respectively; and the Ph.D. degree in computer science from the University of Southern California (USC), Los Angeles, CA, USA, in 2014. Before joining Korea University, he was with LG Electronics (Seoul, Korea, 2006--2009), InterDigital (San Diego, CA, USA, 2012), Intel Corporation (Santa Clara in Silicon Valley, CA, USA, 2013--2016), and Chung-Ang University (Seoul, Korea, 2016--2019). 

He serves as an editor for \textsc{IEEE Transactions on Vehicular Technology}, \textsc{IEEE Transactions on Machine Learning in Communications and Networking}, \textsc{IEEE Communications Standards Magazine}, \textit{Computer Networks (Elsevier)}, and \textit{ICT Express (Elsevier)}. He is also a distinguished lecturer for \textit{IEEE Communications Society (ComSoc)} (2022-2023) and \textit{IEEE Systems Council} (2022-2024).

He was a recipient of Annenberg Graduate Fellowship with his Ph.D. admission from USC (2009), Intel Corporation Next Generation and Standards (NGS) Division Recognition Award (2015), \textsc{IEEE Systems Journal} Best Paper Award (2020), IEEE ComSoc Multimedia Communications Technical Committee (MMTC) Outstanding Young Researcher Award (2020), IEEE ComSoc MMTC Best Journal Paper Award (2021), \textit{ICT Express (Elsevier)} Best Special Issue Guest Editor Award (2022), and \textit{ICT Express (Elsevier)} Best Editor Award (2023). He also received numerous awards from IEEE conferences including IEEE ICOIN Best Paper Award (2021), IEEE Vehicular Technology Society (VTS) Seoul Chapter Awards for APWCS (2019, 2021, 2022), and IEEE ICTC Best Paper Award (2022). 
\end{IEEEbiography}

\end{document}